\documentclass[twocolumn]{aastex7}

\newcommand\jc[1]{\textcolor{orange}{#1}}

\newcommand\Msun{{\rm M}_\odot}
\newcommand\Mearth{{\rm M}_\oplus}

\newcommand\water{${\rm H_2O}$}
\newcommand\CtwoHtwo{${\rm C_2H_2}$}
\newcommand\COtwo{${\rm CO_2}$}

\received{April 6, 2025}
\accepted{July 8, 2026}

\begin{document}

\title{Probing Planet Formation with JWST Spectroscopy of IC348:\\
Sustained Diversity but Limited Chemical Evolution and Pebble Drift}

\author[orcid=0000-0002-5538-6954,gname=John,sname=Carr]{John S. Carr}
\affiliation{Department of Astronomy, University of Maryland, College Park, MD 20742, USA}
\email{jscarr108@gmail.com}  

\author[orcid=0000-0002-5758-150X,gname=Joan,sname=Najita]{Joan R. Najita}
\affiliation{NOIRLab, 950 North Cherry Avenue, Tucson, AZ 85719, USA}
\email{joan.najita@noirlab.edu}


\begin{abstract}

Inner disk chemistry offers a valuable window onto disk evolution and planet formation. Key planet formation processes, including pebble drift, planetesimal and planet formation, dust traps, and snow lines, determine the delivery rates of oxygen and carbon to the inner disk. As a result, measurements of molecules sensitive to the gas-phase C/O ratio, in disks over a range of ages, can constrain the relative importance and time evolution of these processes. Here we report JWST/MIRI observations of T Tauri disks in the young cluster IC 348, 
extending to older ages previous studies of disks. 
Like younger disks, IC 348 sources show a broad range of molecular ratios, indicating diverse planet formation histories. 
However, the similar   HCN/\water\ and \CtwoHtwo/\water\ flux ratio distributions of IC 348 (2--5 Myr old) and Taurus (1--2 Myr old) sources imply little to no evolution in the inner disk C/O ratio over this time interval, even as disk masses decline by a factor of 5 on average. This result contradicts the general prediction of increasing C/O ratio with time in models of disk chemical evolution, indicating the need to better understand the interplay between (and efficiencies of) planet formation and disk evolution processes. 
We also find that the cold-to-warm water flux ratios show no evolution and 
do not correlate with 
the hydrocarbon-to-water ratios. These results suggest that most inner T Tauri disks are not dominated by rapid
icy pebble drift and avoid the ``meter-size barrier problem'' once thought to be an obstacle to planet formation.
\end{abstract}



\section{Introduction} \label{sec:intro}

The inner regions of protoplanetary disks 
show remarkable diversity in their mid-infrared molecular line emission, indicating that their inner regions are chemically diverse. 
The discovery, made with the {\it Spitzer Space Telescope}, that disks display a rich emission spectrum of water and simple organic molecules \citep[e.g., HCN, \CtwoHtwo, and \COtwo;][]{CN08} opened the door to broader {\it Spitzer} studies of 1--2 Myr old classical T Tauri stars (CTTS), which revealed marked variety in the relative strengths of their molecular emission features
\citep[][hereafter N13] {Pontoppidan10, CN11, Salyk08, Salyk11, N13}.
In particular, the relative strengths of simple hydrocarbons to water suggested  that inner disks,
the region within the water snow line at $\sim 1$ au, sport a range of inner disk C/O ratios (N13). 

Chemical diversity of this kind is expected, in principle, from familiar 
ideas about the processes involved in disk evolution and planet formation, e.g., the inward drift of pebbles and larger solids, the formation of planetesimals and protoplanets, and pebble trapping in pressure bumps created by forming giant planets; \citep[see e.g.,][]
{Ciesla06, Rice06, Pinilla12, Booth17}. 
In brief, as icy grains grow into pebbles, they are expected to sink toward the disk midplane and, as a consequence of gas drag, drift inward rapidly relative to the accreting gas, evaporating their  water ice when they pass the snow line. If drifting solids reach the inner disk in abundance, they {\it superhydrate} it, lowering its C/O ratio. Pebble inflow is expected to dominate inner disk chemistry at early times and in disks that fail to 
lock up their icy solids in 
planetesimals and protoplanets
and/or in pebble traps within pressure bumps
\citep[e.g.,][]{Ciesla06, Banzatti20, Kalyaan21, Mah24}.

In contrast, disks that engage in core accretion (the preferred theory of planet formation) efficiently convert their icy solids into large objects, with processes like the streaming instability and pebble accretion converting pebbles into icy planetesimals and protoplanets \citep[e.g.,][]{Johansen17}, 
the building blocks of small and giant planets. These objects are so large they 
drop out of the inward flow of gas and pebbles, 
and sequester ice (water and oxygen) beyond the snow line. If giant planets then form, they are expected to create pressure bumps in the gas disk that can inhibit the inward flow of drifting pebbles. If pressure bumps are efficient in trapping the solids, they further sequester ices and can induce additional planetesimal and planet formation 
\citep[e.g.,][]{Pinilla12,Carrera21,Morbidelli20,Jiang23, Danti2023, Houge2026leaky}. 
If that occurs, material reaching the inner disk is {\it dehydrated} (oxygen-poor) and enhances the C/O ratio of the inner disk. 
Hence, the general expected chemical signature of planet formation processes is an evolution toward a C-enhanced inner disk; 
conversely, an O-rich inner disk would suggest a system that has not engaged efficiently in planet formation.

These ideas are supported by the wide range in molecular flux ratios, of simple hydrocarbons to water, seen in {\it Spitzer} spectra.
In earlier work \citep[][N13]{CN11,N11},
we described how 
the large range in observed molecular flux ratios could be understood in terms of a range in inner disk C/O ratios
produced by the competing effects of rapid inward migration of icy solids vs.\ the sequestration of icy solids (via planet formation processes) in the outer disk.
Since disks are
likely to have diverse histories, each experiencing different degrees of pebble inflow and core accretion over time,
the expected timescales for these processes imply that inner disks as a population 
can develop diverse C/O ratios by $\sim 1$ Myr age \citep[e.g.,][]{Ciesla06, Booth17, Mah24, Houge25carbon}. 
These disks would show dramatically different MIR molecular spectra, because modest changes in the C/O ratio dramatically affect inner disk chemistry.
Thermal-chemical models of inner disk atmospheres find that the observed order of magnitude range in molecular flux ratios can be explained by a small increase in C/O (factor of $\sim 2$), which drives, for example, a large enhancement (factor of $\sim 10$) in the relative abundance of HCN or \CtwoHtwo\ to water, because HCN and \CtwoHtwo\ are enhanced and \water\ is reduced in abundance \citep[e.g.,][]{N11,Anderson21,Arabhavi26}.

These results suggest that understanding how inner disk molecular ratios evolve over time can potentially lend insights into the efficiencies and timescales of planet formation and disk evolution processes, constraints that are valuable given the difficulty of directly observing  these processes in action. Neither pebble drift nor planetesimal formation is known to produce a directly observable signature. 
Even giant planet formation, which can be inferred from the presence of disk substructure (e.g., gaps and rings), is typically a challenge to diagnose at distances within 10\,au. 
Observing inner disks over a range of ages can test current theories of disk chemical evolution, which generally predict that their C/O ratio grows over time under a broad range of conditions \citep[e.g.,][]{Mah24, Williams25,Sellek_vanDishoeck_25}.

With the much higher sensitivity of MIRI on JWST compared to {\it Spitzer/IRS},
it is now possible to obtain good signal-to-noise spectra of disks beyond the nearest star forming regions ($> 140$ pc) and  probe populations in clusters with a greater range of ages. Here we describe JWST observations of disks in the $\sim 3$ Myr old cluster IC 348 located at a distance of 320 pc. 

Nearly all of the classical T Tauri (CTTS) disks measured with {\it Spitzer}/IRS at high spectral resolution are located in younger associations $\sim$1--2~Myr old. 
Because disks are expected to dissipate within $\sim 5$ Myr (
\citealt{Hernandez07, Fedele10}),  
an intermediate age ($\sim 3$ Myr) 
population is an evolutionary sweet spot, well suited to probing the chemical evolution of inner disks,
because a large fraction of  stars still have disks.
In contrast, in older clusters, only a small subset of stars retain disks, a situation that can challenge our ability to connect observed systems with younger populations.
As one example, in the $\sim$5--10 Myr old Upper Sco association, $\sim 20$\% of $0.2-1 \Msun$ sources have disks \citep[][]{LuhmanEsplin20,Pfalzner26}, and of these only about half (or $\sim 10$\%) have detectable MIR molecular emission (C.\ Xie et al.\ 2026, accepted). Thus, observations of intermediate age disks provide a pathway to connect older and younger disk populations.

The cluster IC 348 is well studied and characterized \citep{Luhman03}, with an estimated age of 2--5 Myr 
\citep{Muench07, Luhman24}.
A full census of  IC 348 cluster members has been carried out with {\it Spitzer} IRAC and MIPS photometry \citep{Lada06, Muench07}. The cluster was also the subject of an ALMA 1.3 mm continuum survey  \citep{Rodriguez18}. 

To place the IC 348 results in context with younger disk populations, we also make a direct comparison to a $\sim$1--2 Myr old disk population from a star-forming region over the same range stellar mass.  
We selected the extensively studied Taurus star-forming region, 
which allowed us to construct the largest possible comparison sample by drawing upon readily available
{\it Spitzer} IRS-SH spectra. 
Isochronal ages, cluster disk fractions, stellar accretion rates, and disk masses argue that IC 348 is evolutionarily older than Taurus by $\sim$1--3~Myr (see Section 5 and Appendix~\ref{sec:appendix:ages} for details). 

In sections~\ref{sec:sample} and \ref{sec:obs}, we describe the properties of the IC 348 and Taurus samples and the MIRI observations. The results are analyzed in section~\ref{sec:results} and discussed in section~\ref{sec:discussion}. We present our conclusions in section~\ref{sec:conclusions}.

\section{Sample Selection} \label{sec:sample}

We selected IC 348 targets from the ALMA 1.3 mm continuum survey of
\citet{Rodriguez18}.
{Ru{\'\i}z-Rodr{\'\i}guez} et al.\  observed 136 Class II members in IC 348, 
selecting optically thick disks based on the {\it Spitzer} studies of \citet{Lada06}
and \citet{Muench07}. They detected 40 of the targets, but with a beam size of 0.8 arcsec
(250 au), only a handful were marginally resolved.
The inferred dust masses range from $19\,\Mearth$ down to 3$\sigma$ limits of $\sim 1\,\Mearth$.

\begin{figure}[b]
\centering
\includegraphics[trim = 1.0cm 0cm 0.5cm 1cm, clip, angle=270, 
width = 1.0\columnwidth]
{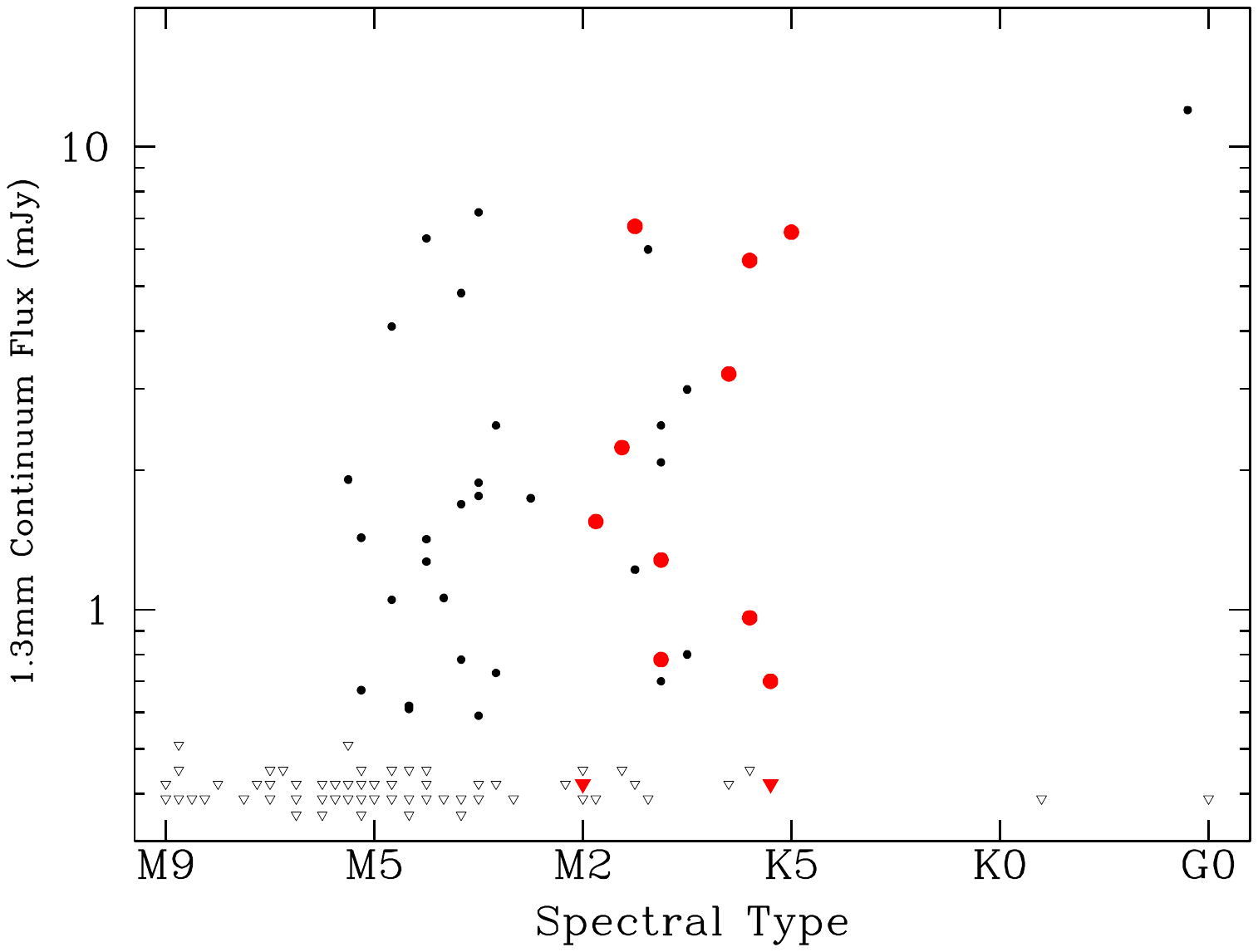}
\caption{The 1.3\,mm continuum flux as a function of stellar spectral type for disks in IC 348  \citep[from][]{Rodriguez18}.  Inverted triangles are upper limits. The sources selected for study with JWST/MIRI are indicated in red. The sample covers the full range in disk mass (1.3\,mm flux) within the spectral type interval M2 to K5. 
}
\label{fig:sample_selection_submm}
\end{figure}

Figure~\ref{fig:sample_selection_submm} plots the 1.3 mm continuum flux against spectral type for the IC 348 sources. Our MIRI sample, which is limited to spectral types K5--M2 ($0.4$–-$0.8\, \Msun$), is plotted in red.
The ALMA survey is complete in this spectral range in that it includes all of the 29 full disk
K5--M2 stars (plus one transition disk) in the studies of
\citet{Lada06} and \citet{Muench07}.
Our targets were chosen to cover a range in disk luminosity, including two upper limits.
Because the disks in the ALMA survey are optically thick in the infrared,
targets with upper limits are likely low-mass disks. 
The 12 disks in our MIRI program sample 41\% of the full disks between K5--M2, although
lower mass disks with upper limits on the 1.3 mm flux are underrepresented.

Table~\ref{tab:IC 348_sources} collects some basic information on the 12 sources from the literature (e.g., spectral types, distance, and 1.3\,mm flux). 
In this paper, we use the source identifier LRL (abbreviated as L in the plots), 
following the numbering system used in
\citet{Luhman03}, \citet{Lada06} and \citet{Muench07}.
We adopted a distance to IC 348 of 320 pc \citep{OritzLeon18}. This is consistent with the average {\it Gaia} distance of 319 +/- 8 pc for the 9 sample stars with good parallaxes.

\begin{figure}[b]
\centering
\includegraphics[trim = 0.6cm 0.5cm 0.5cm 1.3cm, clip, width=0.4\textwidth]{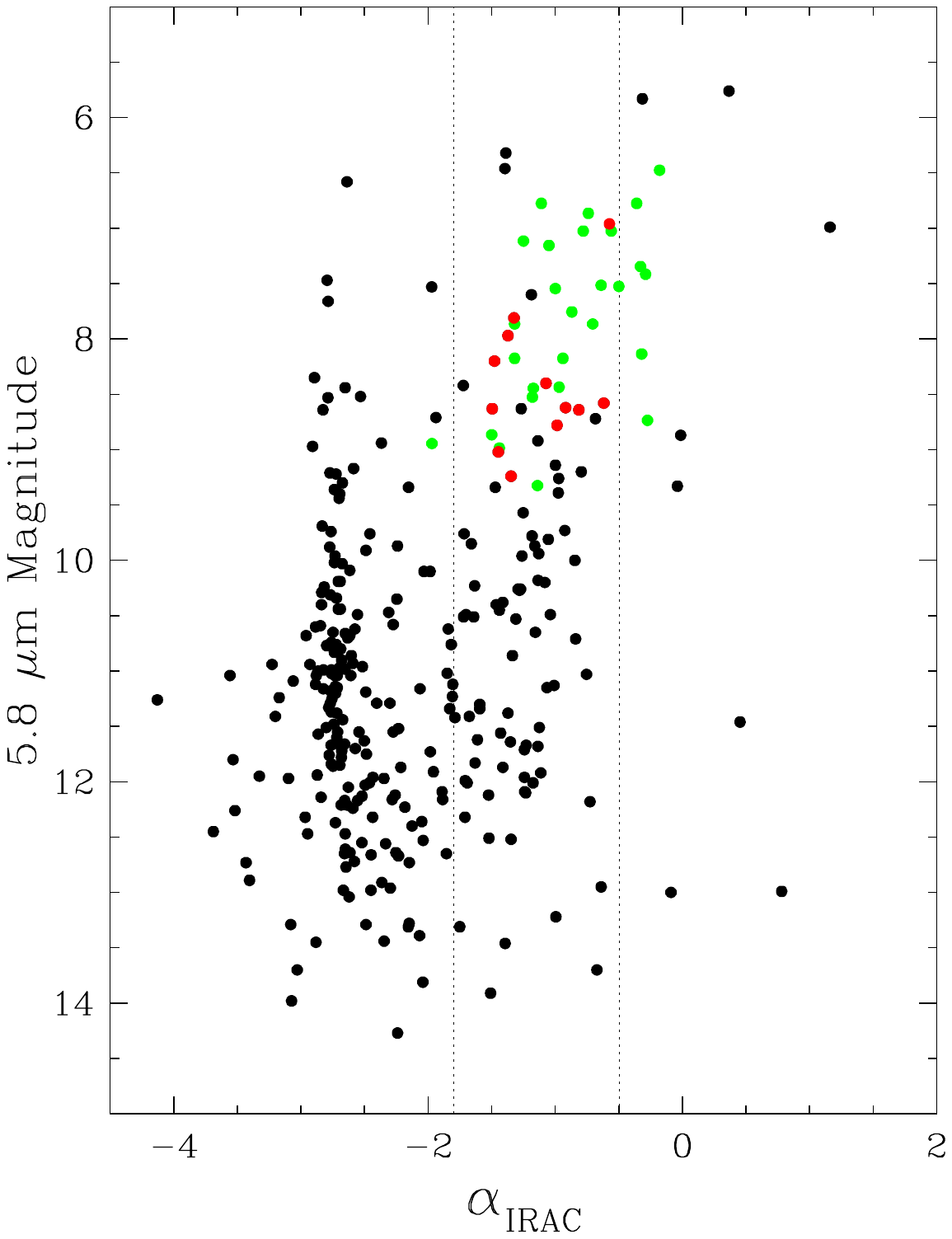}
\caption{
{\it Spitzer} color-magnitude diagram for IC 348 plotting the IRAC 5.8~\micron\ magnitude vs.\ the IRAC 3.6--8.0~\micron\ spectral slope \citep{Muench07}. The sample selected for MIRI spectroscopy is marked as red points.
The Taurus comparison sample is plotted as green points \citep{Luhman10},
with magnitudes scaled from 140 pc to the 320 pc distance of IC 348.
The vertical dotted lines are the values of 
$\alpha_{IRAC}$ used to separate Classes I, II, and III+anemic disks \citep[right to left;][] {Muench07}. }
\label{fig:sample_selection_IRAC}
\end{figure}

\begin{deluxetable}{lllccccccccc}
\tabletypesize{\scriptsize}
\tablecaption{Stellar and Disk Properties for IC 348 Sources \label{tab:IC 348_sources}}
\tablehead{
\colhead{LRL} & \colhead{RA} & \colhead{Dec} & \colhead{SpT} & \colhead{$d$} & \colhead{err} & \colhead{RR18} & \colhead{$F_{1.3\mathrm{mm}}$} & \colhead{$F_\mathrm{err}$} 
}
\startdata
15    & 56.1863174 & 32.067421  & M0.5  & 620 & 104 & 4   & 1.28 & 0.15 \\
32    & 56.1578712 & 32.134491  & K7    & 318 & 6   & 7   & 5.68 & 0.45 \\
36    & 56.1602783 & 32.1265945 & K6    & 335 & 5   & 8   & $<$0.42 \\
37    & 56.1582985 & 32.0582771 & K6    & 317 & 2   & 9   & 0.70 & 0.14 \\
40    & 56.1238441 & 32.1777344 & K8    & 320 & 7   & 10  & 3.23 & 0.29 \\
41    & 56.0900612 & 32.1771164 & K7    & 326 & 11  & 11  & 0.96 & 0.14 \\
55    & 56.1306458 & 32.0040855 & M0.5  & \nodata & \nodata & 14 & 0.78 & 0.22 \\
58    & 56.1606255 & 32.1335258 & M1.25 & 308 & 6   & 15  & 2.24 & 0.22 \\
63    & 55.9954376 & 32.1908531 & M1.75 & 313 & 9   & 17  & 1.55 & 0.17 \\
1933  & 56.318125  & 32.105542  & K5    & 315 & 5   & 126 & 6.54 & 0.52 \\
9060  & 56.1065941 & 32.1918106 & M2    & \nodata & \nodata & 89 & $<$0.42 \\
10352 & 56.3352432 & 32.1095581 & M1    & 315 & 4   & 91  & 6.73 & 0.58 \\
\enddata
\tablecomments{The LRL source names follow the numbering system used in
\citet{Luhman03}, \citet{Lada06} and 
\citet{Muench07}. SpT is the stellar spectral type. $d$ is the {\it Gaia} distance in pc. RR18 is the source ID from \citet{Rodriguez18}. $F_{1.3\mathrm{mm}}$ is the 1.3 mm flux density in mJy. Upper limits are $3\sigma$. 
} 
\end{deluxetable}

\begin{deluxetable*}{llccccccccccccccccccc}
\tablecaption{Taurus Sample and IRS Fluxes and Ratios\label{tab:taurus_irs}}
\tablehead{
\colhead{Name} & \colhead{SpT} & 
\colhead{\water} &  
\colhead{HCN} &  
\colhead{\CtwoHtwo} &  
\colhead{HCN/\water} &  
\colhead{\CtwoHtwo/\water} &  
\colhead{Cont} & \colhead{Fmm} & 
\colhead{D} & 
\colhead{\ion{H}{1} Flux} & 
\colhead{log(L$_{\rm acc}$)}  
}
\startdata
AA Tau & M0.6 &  4.93 (0.18) & 2.51 (0.08) & 1.38 (0.11) & 0.509 (0.025) & 0.279 (0.024) & 0.32 & 65.0  & 135 & 0.35 (0.12) & -2.11 (0.3) \\
BP Tau & M0.5 &  7.25 (0.19) & 2.32 (0.09) & $<$0.35     & 0.320 (0.015) & $<$0.048      & 0.37 & 45.2   & 127 & 1.31 (0.16) & -1.02 (0.11) \\
CI Tau & K5.5 &  6.80 (0.18) & 2.48 (0.09) & 1.66 (0.14) & 0.365 (0.017) & 0.244 (0.021) & 0.5 & 142.4  & 160 & 2.49 (0.18) & -0.03 (0.07) \\
CW Tau & K3   & 19.48 (0.42) & 1.64 (0.21) & $<$0.71     & 0.084 (0.011) & $<$0.036      & 1.11 & 52.3  & 132 & 5.81 (0.38) & 0.39  (0.06) \\
CY Tau & M2.3 &  1.05 (0.07) & 0.68 (0.04) & 1.31 (0.05) & 0.731 (0.071) & 1.411 (0.117) & 0.14 & 79.4  & 126 & 0.61 (0.06) & -1.74 (0.08) \\
DE Tau & M2.3 &  1.84 (0.13) & 0.79 (0.05) & 0.78 (0.05) & 0.430 (0.042) & 0.422 (0.043) & 0.37 & 31.1   & 128 & 1.08 (0.08) & -1.19 (0.06) \\
DG Tau & K7   & 10.68 (1.23) & $<$0.99     & $<$0.97     & $<$0.093      & $<$0.091      & 5.2 & 344.2  & 125 & 8.41 (0.53) & 0.63  (0.06) \\
DH Tau & M2.3 &  2.90 (0.1)  & $<$0.10     & $<$0.11     & $<$0.034      & $<$0.038      & 0.14 & 26.7   & 133 & 0.65 (0.06) & -1.58 (0.08) \\
DK Tau & K8.5 & 16.25 (0.32) & 2.91 (0.15) & $<$0.51     & 0.179 (0.01)  & $<$0.031      & 0.75 & 30.1  & 132 & 2.08 (0.34) & -0.54 (0.15) \\
DL Tau & K5.5 &  3.08 (0.21) & 2.54 (0.1)  & 4.63 (0.12) & 0.824 (0.065) & 1.505 (0.111) & 0.88 & 170.7 & 160 & 3.89 (0.17) & 0.38  (0.04) \\
DN Tau & M0.3 &  1.91 (0.1)  & 1.50 (0.06) & 1.03 (0.07) & 0.786 (0.053) & 0.540 (0.048) & 0.3 & 88.6   & 129 & 0.62 (0.09) & -1.67 (0.14) \\
DO Tau & M0.3 &  5.46 (0.58) & 1.88 (0.25) & 1.34 (0.28) & 0.345 (0.059) & 0.246 (0.058) & 2.01 & 123.8 & 139 & 3.34 (0.35) & -0.02 (0.1) \\
DP Tau & M0.8 &  9.70 (0.28) & 1.41 (0.13) & $<$0.41     & 0.146 (0.014) & $<$0.042      & 0.67 & $<$3.8    & 140*& 2.13 (0.22) & -0.41 (0.09) \\
DQ Tau & M0.5 &  6.79 (0.28) & 1.80 (0.16) & $<$0.38     & 0.292 (0.029) & $<$0.056      & 0.64 & 69.3 & 195 & 2.23 (0.22) & 0.23  (0.09) \\
DR Tau & K6   & 26.67 (0.72) & 7.27 (0.29) & 2.03 (0.46) & 0.272 (0.013) & 0.076 (0.017) & 1.93 & 127.2 & 193 & 7.14 (0.6)  & 1.26  (0.08) \\
DS Tau & M0.4 &  5.42 (0.15) & 3.56 (0.07) & 0.83 (0.11) & 0.658 (0.022) & 0.153 (0.021) & 0.24 & 22.2  & 158 & 1.14 (0.13) & -0.76 (0.1) \\
FZ Tau & M0.5 & 21.78 (0.36) & 3.09 (0.25) & 1.57 (0.46) & 0.142 (0.012) & 0.072 (0.021) & 0.94 & 13.7  & 129 & 5.32 (0.57) & 0.27  (0.1) \\
GI Tau & M0.4 &  7.01 (0.24) & 2.83 (0.12) & 1.19 (0.15) & 0.404 (0.022) & 0.170 (0.023) & 0.76 & 17.7  & 129 & 1.05 (0.19) & -1.19 (0.16) \\
GK Tau & K6.5 &  5.24 (0.35) & 0.57 (0.13) & $<$0.31     & 0.108 (0.026) & $<$0.059      & 0.79 & 5.2   & 129 & 1.24 (0.21) & -1.04 (0.15) \\
Haro 6-13 & M0 & 7.04 (0.67) & $<$0.38     & $<$0.37     & $<$0.054      & $<$0.050      & 1.34 & 137.1  & 129 & 3.02 (0.22) & -0.24 (0.07) \\
HK Tau & M1.1 &  0.65 (0.12) & $<$0.07     & $<$0.08     & $<$0.110      & $<$0.120      & 0.23 & 33.2  & 131 & 0.56 (0.06) & -1.73 (0.09) \\
HN Tau & K5.5 &  3.95 (0.32) & 0.47 (0.12) & 0.57 (0.15) & 0.119 (0.031) & 0.145 (0.04)  & 0.97 & 12.3  & 144*& 3.79 (0.19) & 0.16  (0.05) \\
HQ Tau & K2   & $<$0.56      & $<$0.22     & $<$0.2      & nan           & nan           & 1.0 & 4.0 & 161 & 1.53 (0.14) & -0.46 (0.09) \\
IQ Tau & M1.1 & 3.43  (0.29) & 3.17 (0.15) & 2.10 (0.16) & 0.922 (0.089) & 0.612 (0.07)  & 0.38 & 64.1  & 131 & 0.79 (0.23) & -1.42 (0.26) \\
RW Aur & K2   & 23.28 (0.57) & 4.99 (0.31) & 3.78 (0.59) & 0.214 (0.014) & 0.162 (0.026) & 1.47 & 35.6   & 156*& 6.17 (0.57) & 0.75  (0.08) \\
UY Aur & K7   & 13.10 (1.0)  & 3.66 (0.39) & $<$0.92     & 0.279 (0.036) & $<$0.016      & 2.64 & 20.0  & 152 & 1.49 (0.6)  & -0.59 (0.36) \\
V836 Tau   & M0.8 &  1.35 (0.07) & 0.39 (0.04) & $<$0.08     & 0.287 (0.033) & $<$0.059      & 0.1 & 26.2 & 167 & $<$0.13     & $<$-2.62  \\
04303+2240 & M0.5 & 28.46 (0.63) & 5.77 (0.32) & 1.62 (0.76) & 0.203 (0.012) & 0.057 (0.027) & 1.75 & $<$6. & 161*& 3.41 (0.77) & 0.27  (0.2) \\
04385+2550 & M0   &  4.12 (0.2)  & 0.92 (0.08) & 0.85 (0.09) & 0.224 (0.023) & 0.206 (0.023) & 0.42 & 25.5 & 140*& 0.34 (0.11) & -2.06 (0.28) \\
\enddata
\tablecomments{Flux units are $10^{-17}$ W\,m$^{-2}$.
Cont is 14\,\micron\ continuum in Jy.
$F_{\rm mm}$ is 1.3 mm flux in mJy from \citet{Long19}, where available, and \citet{Andrews13} for the remainder.
D is Gaia distance in pc; * indicates the mean distance of its associated stellar group \citep{Luhman23} was adopted.
\ion{H}{1} is the combined flux of the (7-6) and (11-8) transitions after correction for H$_2$O contamination.
$L_{\rm acc}$ is in $L_\odot.$
$1\sigma$ uncertainties are given in parentheses.
}
\label{Table_TaurusIRS_Results}
\end{deluxetable*}

To create a large comparison sample of younger disks from a single star-forming region, we adopted the high signal-to-noise {\it Spitzer} IRS-SH archival spectra of Taurus CTTS.  
These data were reduced uniformly using  procedures established to produce high S/N spectra \citep{CN11}.
In order to compare similar spectral types with IC 348, we excluded stars with spectral types M2.5 and later or earlier than K, as well as known  transition disks (GM Aur, DM Tau, LkCa 15, UXTau A, and CokuTau/4). 
This left 29 Taurus disks for direct comparison to IC 348. The names, spectral types, and distances of the Taurus sample are listed in Table~\ref{Table_TaurusIRS_Results}.

The IC 348 and Taurus samples are compared in Figure~\ref{fig:sample_selection_IRAC}, which plots the 
Spitzer IRAC 5.8~\micron\ magnitude vs. the IRAC 3.6--8.0~\micron\ spectral slope \citep{Muench07}. 
Our IC 348 targets (red points) have IRAC slopes characteristic of optically thick Class II disks. By design, the infrared magnitudes 
of our sample fall in the brighter half of K5--M2 stars in IC 348,
in order to optimize the number of targets in a small JWST program. The IC 348 and Taurus samples overlap in 
the plot, although the Taurus disks extend to higher fluxes and redder slopes. 
This may be due to some Taurus disks 
having higher accretion rates  than those found in IC 348, 
consistent with the younger age and earlier evolutionary state of Taurus compared to IC 348 (see Appendix~\ref{sec:appendix:ages}).

\section{Observations}  \label{sec:obs}

The targets were observed through the JWST GO program 2826 using the Medium-Resolution Spectrometer module \citep[MRS;][]{Wells15, Argyriou23} 
of the Mid-Infrared Instrument
\citep[MIRI;][]{Rieke15, Wright15,Wright23}.
Data were obtained with all three MRS grating settings to give the full wavelength coverage of 4.9\,$\micron$--28\,$\micron$,
with a spectral resolving power of R = 1500--3500.
Targets were acquired with the standard 4-point dithering pattern.
The exposure times were set to achieve a minimum S/N = 250 at $14\,\micron$,
based on continuum fluxes or {\it Spitzer} IRS (low-res) spectra, when available.
 These JWST spectra can be obtained from the Mikulski Archive for Space Telescopes (MAST) at the Space Telescope Science Institute. The specific observations analyzed can be accessed via \dataset[doi: 10.17909/2vdw-5160]{https://doi.org/10.17909/2vdw-5160}.

The spectra were processed by the standard JWST pipeline (version 1.12.5).
The only additional processing was the removal of residual fringing in some wavelength regions of interest. This was performed using a sine fitting routine \citep[IRSFRINGE;][] {Lahuis03, irsfringe} 
to remove a single prominent fringe frequency.
Applying this to the 17--18\,$\micron$ region of the spectrum,
the measured continuum S/N increased from a typical value of $\sim 130$ to a median value of 230. 
Residual fringing was also removed from the 24\,$\micron$ region, which improved the S/N of the lines measured at this wavelength by a factor of 1.5 times, on average.

\begin{figure*}[htb]
\centering
\includegraphics[trim = 0.8cm 0cm 0.5cm 1cm, clip, angle=0, width=0.8\textwidth]{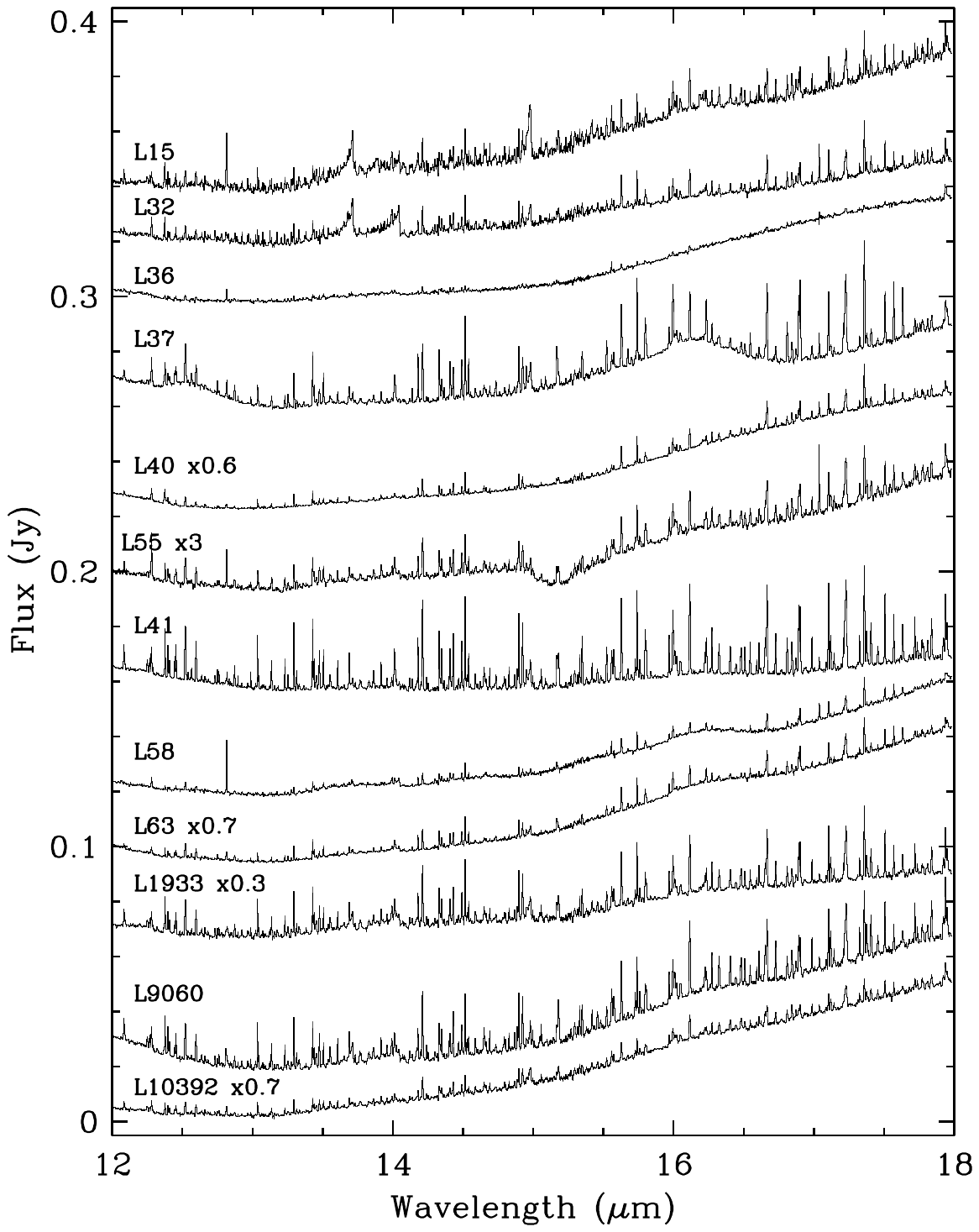}
\caption{JWST/MIRI spectra of the IC 348 sample in the 12--18 \micron\ region.  The spectra are shifted arbitrarily for plotting, and in some cases scaled in flux by the indicated factor.} 
\label{fig:spectra}
\end{figure*}

\section{Results and Analysis}  \label{sec:results}

Figure~\ref{fig:spectra} shows an overview of the IC 348 spectra in the 12--18 \micron\ region. 
The spectra 
show the rich mid-infrared emission typical of younger classical T Tauri
stars, with emission features of \water, HCN, \CtwoHtwo, OH, \COtwo, and \ion{H}{1}.
Emission lines of H$_2$  and \ion{Ne}{2} are also present in the reduced spectra. However, because both are present in the nebular gas
emission \citep{Dhariwal24} and are widespread in the MIRI spectral images, their presence in Figure~\ref{fig:spectra} does not imply that the emission is intrinsic to the source.
In the two disks with the strongest \CtwoHtwo\ emission (LRL 15 and LRL 32), 
a weak emission feature is present at the wavelength of ${\rm C_4H_2}$ (15.92 \micron),
but spectral modeling is required to firmly establish the detection of ${\rm C_4H_2}$ emission.
One of the targets (LRL 55) shows weak \COtwo\ ice absorption. 
This absorption is unlikely to be intrinsic to LRL 55, as shown by the presence of the same \COtwo\ ice absorption in the spectrum of an unnamed star 
in the same MIRI field.
LRL 55 has high extinction \citep{Muench07}, is not visible in the POSS image, and hence may be located behind gas and dust at the cluster's southern boundary \citep{Muench07}.

\begin{figure*}[ht]
\centering
\includegraphics[trim = 0.5cm 1.0cm 0.5cm 1cm, clip, angle=0, 
width=0.6\textwidth]
{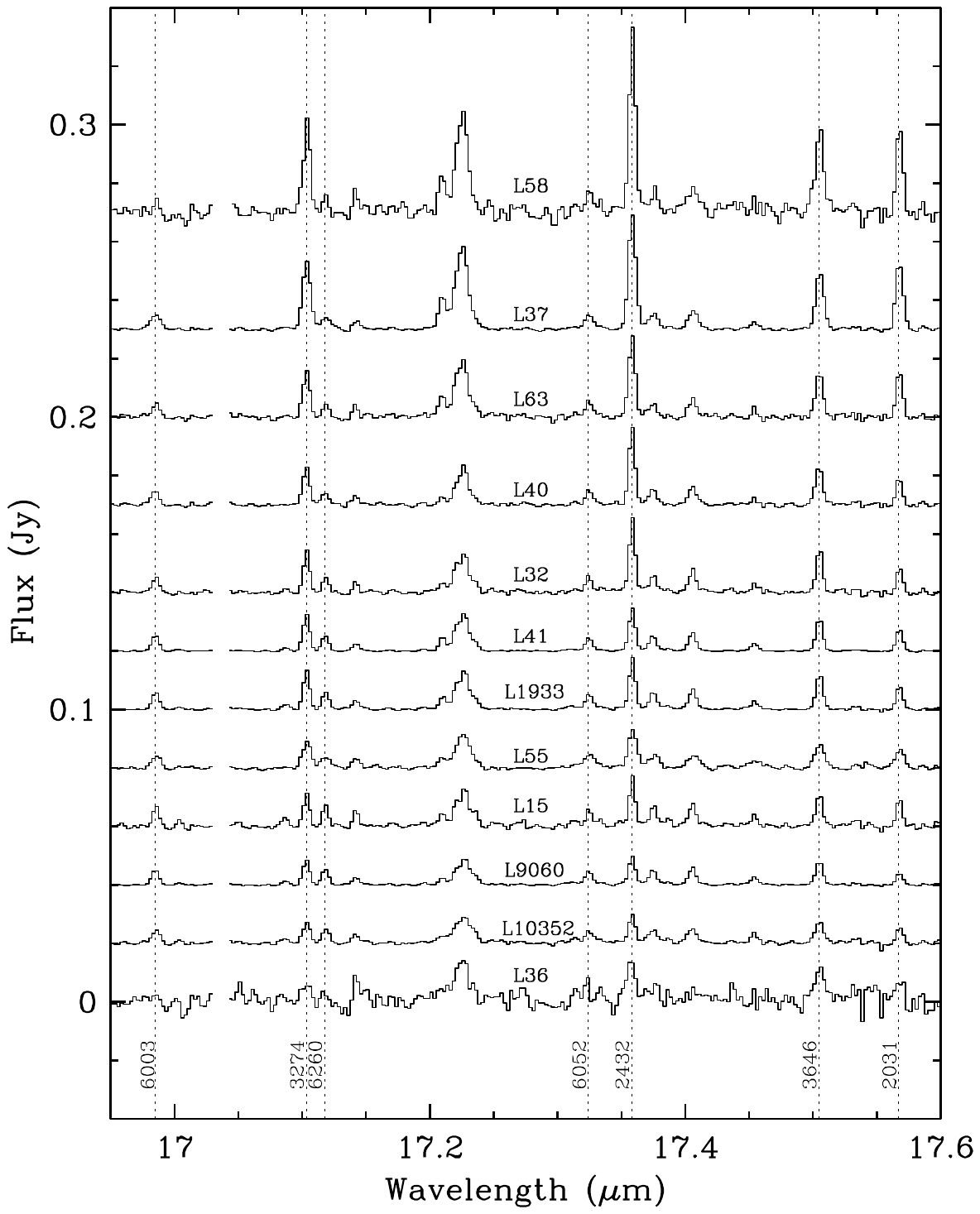}
\caption{Close-up of the 17--17.6 \micron\ region, which includes \water\ lines with a range of upper level energy. The upper level energy of the dominant transition for several \water\ features is indicated at the bottom of the plot. Each continuum-subtracted spectrum is scaled such that the high-energy water lines are similar in strength, then shifted vertically for plotting. The spectra are ordered by the ratio of cool to hot water, from high (top) to low (bottom). Spectra are not plotted at the wavelength of the 17 \micron\ H$_2$ line. \vspace{0.4cm}} 
\label{fig:spectra_H2O}
\end{figure*}

\begin{figure*}[ht]
\centering
\includegraphics[trim = 0.5cm 1.0cm 0.5cm 1cm, clip, angle=0, width=0.48\textwidth]{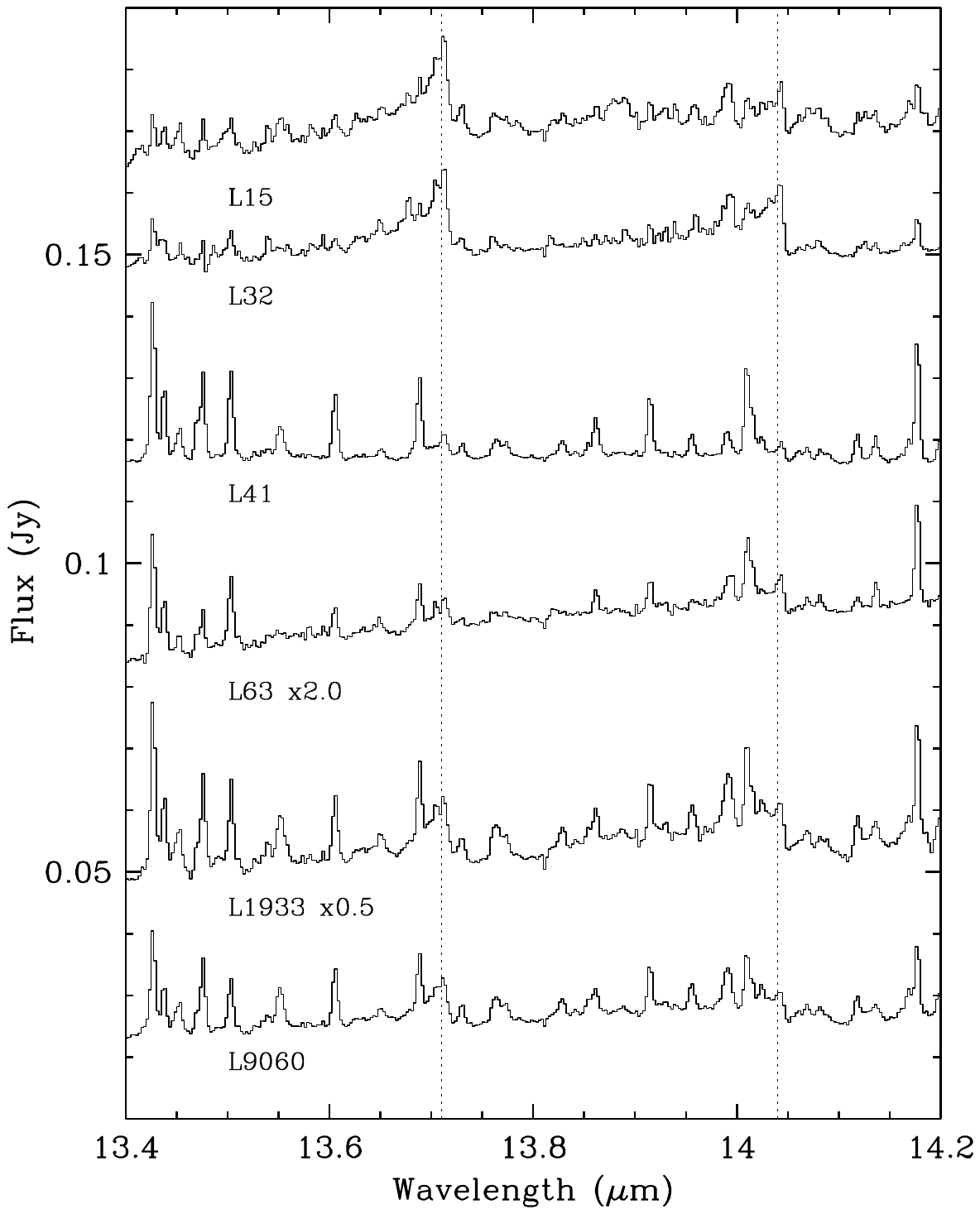}
\includegraphics[trim = 0.5cm 1.0cm 0.5cm 1cm, clip, angle=0, width=0.48\textwidth]{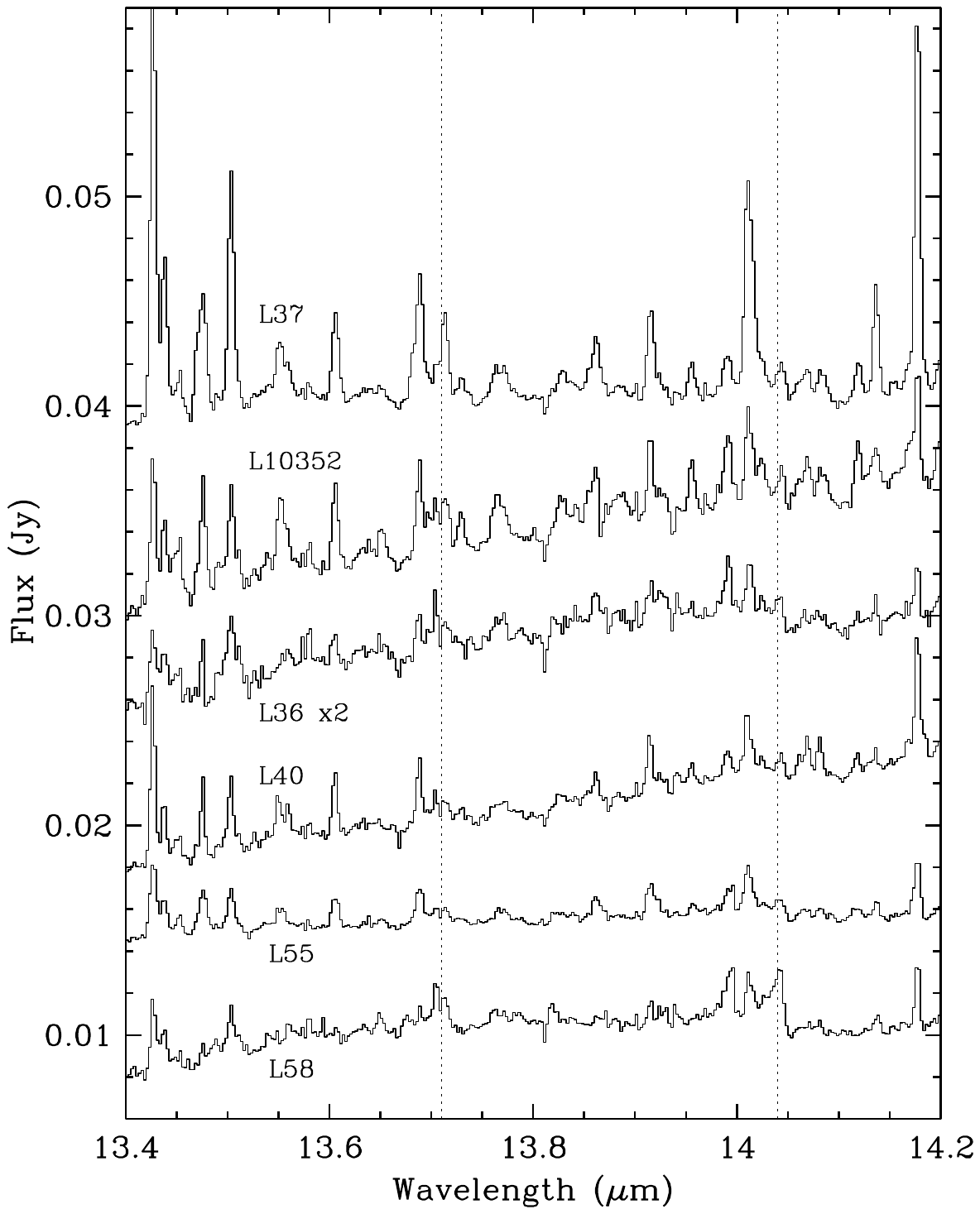}
\caption{Close-up of the 13.4--14.2 \micron\ region, showing the range in strength of the HCN and \CtwoHtwo\ features. 
The spectra are shifted vertically for plotting, and
in some cases scaled in flux by the indicated factor.
Spectra with stronger (weaker) continua are shown in the left (right) panels. The positions of the first bandhead of \CtwoHtwo\ and HCN are marked by the dotted lines near 13.7 and 14 \micron, respectively.
\vspace{0.5cm}} 
\label{fig:spectra_HCN}
\end{figure*}

The analysis presented here focuses on the \water\ lines near 17.3 \micron\ and 
the Q-branches of HCN at 14 \micron\ and \CtwoHtwo\ at 13.7 \micron.
The analysis also includes 
lower excitation \water\ lines ($\sim$ 1500 K) near 24 \micron\ 
(Figure~\ref{fig:cold_water}).
Water emission is clearly present in all 12 of the IC 348 disks.
Continuum subtracted \water\ spectra are shown in Figure~\ref{fig:spectra_H2O}.
Several \water\ features are marked, with upper level
energy ranging from 2000 to 6300 K.
An obvious characteristic of the \water\ spectra is the range in the flux ratios of transitions with different upper level energy.
This fact is emphasized in Figure~\ref{fig:spectra_H2O} by scaling the spectra to the
strength of the higher excitation ($\sim$ 6000 K) lines and ordering them by the ratio of lower to higher excitation lines.
The diversity of \water\ line emission ratios indicates a range in the ratio of hot and cold water emission that contributes to the spectra.

Spectra in the vicinity of the HCN and \CtwoHtwo\ bands are shown in Figure~\ref{fig:spectra_HCN}.
According to our measurements, HCN was detected in all 12 spectra,
although the equivalent width is low in many cases. Emission from \CtwoHtwo\ is
detected only in 5 spectra. Figure~\ref{fig:spectra_HCN} illustrates the large range in the ratio of HCN and \CtwoHtwo\ emission strengths relative to \water\ emission in the sample.

\subsection{Molecular Flux Measurements}

To characterize the molecular emission, we measured  fluxes in defined bands, using an approach analogous to that taken by N13 \citep[also][]{ Banzatti20}.
The wavelength intervals used to define the continuum and measure the fluxes,
for both the MIRI and IRS spectra,
are described in Appendix~\ref{sec:appendix:features}.

Model slab spectra were calculated as an aid in establishing the intervals and determining corrections for \water\ emission in the \CtwoHtwo\ and \ion{H}{1} bands. 
For the synthetic \water\ spectra, we used a line list based on HITRAN2020
\citep{Gordon22},  augmented by transitions from HITEMP  
\citep{Rothman10} in order to 
include higher energy transitions observed in the MIRI spectra that are missing in HITRAN
(see Appendix~\ref{sec:appendix:features} for details).
This augmented line list can be important for \CtwoHtwo, where a number of high energy water transitions make 
an observable contribution to the combined flux 
(as illustrated in Figure~\ref{fig:FZTau.c2h2}).

The bands used to measure the \water\ flux 
are close to those adopted in N13 and \citet{Banzatti20}.
In the current study, the HCN flux is the sum of 3 sub-bands 
that are largely free of other emission, 
in contrast to the method adopted by N13 and Banzatti et al., which first summed over this entire wavelength range and then corrected for the \water\ emission. 
For \CtwoHtwo, a correction was applied for the \water\ and HCN emission within the \CtwoHtwo\  band.

The molecular flux measurements for the IC 348 sample are reported, along with useful flux ratios, in 
Table~\ref{Table.IC 348.Results}.
To help characterize the warm and cool water emission components, 
we also measured the flux of the individual \water\ features marked in Figure~\ref{fig:spectra_H2O}, which are either single lines, close ortho-para pairs, or features dominated by a single \water\ transition.
These include two lines defined and used in \citet{Banzatti25} as measures of hot and
warm water emission.  We also measured the two lines near 24\,\micron\ 
(Fig.~\ref{fig:cold_water})
that were used by
\citet{Banzatti25} as a measure of cold water emission. Fluxes for the individual \water\ features are listed in Table~\ref{Table.IC 348.water.Results}.

\begin{deluxetable*}{lrrcrrcrcc}
\tablecaption{IC 348 Measured  Fluxes and Ratios}
\tablehead{
\colhead{LRL} &
\multicolumn{1}{c}{H$_2$O} &
\multicolumn{1}{c}{HCN} &
\multicolumn{1}{c}{C$_2$H$_2$} &
\multicolumn{1}{c}{C$_2$H$_2$band} &
\multicolumn{1}{c}{HCN/H$_2$O} &
\multicolumn{1}{c}{C$_2$H$_2$/H$_2$O} &
\colhead{Cont} 
}
\startdata
15    &  71.23 (1.15) & 29.15 (0.52) &   67.66 (1.15) & 77.69 (0.45) & 0.409 (0.010) &    0.950 (0.022) & 100 \\
32    &  53.70 (0.82) & 43.53 (0.32) &   57.58 (1.24) & 68.59 (0.68) & 0.811 (0.014) &    1.072 (0.028) &  67 \\
36    &   9.56 (0.56) &  5.97 (0.19) & $<$1.00        &  2.99 (0.34) & 0.625 (0.042) & $<$0.105         &  44 \\
37    & 117.44 (0.85) &  5.28 (0.20) & $<$3.56        & 17.72 (0.41) & 0.045 (0.002) & $<$0.030         &  46 \\
40    &  63.11 (0.84) &  7.97 (0.19) & $<$2.00        &  6.86 (0.34) & 0.126 (0.003) & $<$0.032         &  43 \\
41    & 146.50 (0.90) & 10.62 (0.30) & $<$5.01        & 22.55 (0.40) & 0.072 (0.002) & $<$0.034         &  33 \\
55    &  30.61 (0.49) &  5.14 (0.17) & $<$1.07        &  5.00 (0.21) & 0.168 (0.006) & $<$0.034         &  14 \\
58    &  24.58 (0.68) &  9.20 (0.17) &    4.57 (0.52) &  7.29 (0.25) & 0.374 (0.012) &    0.186 (0.022) &  44 \\
63    &  53.37 (1.08) & 10.51 (0.24) & $<$1.96        &  9.60 (0.24) & 0.197 (0.006) & $<$0.036         &  49 \\
1933  & 338.11 (2.55) & 91.88 (1.64) &   61.35 (5.50) &113.33 (1.10) & 0.272 (0.005) &    0.181 (0.016) & 235 \\
9060  & 123.75 (1.09) & 27.42 (0.34) &   23.90 (1.71) & 39.80 (0.44) & 0.222 (0.003) &    0.193 (0.014) &  63 \\
10352 &  70.89 (0.93) &  7.70 (0.37) & $<$4.10        & 12.45 (0.39) & 0.109 (0.005) & $<$0.058         &  79 \\
\enddata
\tablecomments{
Fluxes are in units of $10^{-19}$~W~m$^{-2}$.  
``C$_2$H$_2$band'' refers to the measured C$_2$H$_2$ flux before correction for H$_2$O and HCN.  
``Cont'' indicates continuum values measured at 14~$\mu$m in mJy.
$1\sigma$ uncertainties are given in parentheses. Upper limits are $2\sigma$.
\vspace{-0.3cm}}
\label{Table.IC 348.Results}
\end{deluxetable*}

\begin{deluxetable}{lccccccccc}
\tablecaption{Fluxes for Individual H$_2$O Features in IC 348 \label{tab:waterlines}}
\tablehead{
\colhead{LRL} &
\colhead{6003 K} &
\colhead{3274 K} &
\colhead{6260 K} &
\colhead{6052 K} &
\colhead{2433 K} &
\colhead{3646 K} &
\colhead{2031 K} &
\colhead{1448 K} &
\colhead{1615 K} \\
\colhead{} & \colhead{16.98 $\mu$m} & \colhead{17.10 $\mu$m} & \colhead{17.12 $\mu$m} & \colhead{17.32$^{\rm a}$ $\mu$m} &
\colhead{17.36$^{\rm a}$ $\mu$m} & \colhead{17.50$^{\rm a}$ $\mu$m} & \colhead{17.57 $\mu$m} & \colhead{23.82$^{\rm a}$ $\mu$m} & \colhead{23.90$^{\rm a}$ $\mu$m}}
\startdata
15   & 4.58 (0.30) & 6.99 (0.27) & 4.13 (0.27) & 3.58 (0.23) & 9.84 (0.26) & 6.90 (0.28) & 5.55 (0.25) & 8.31 (1.35) & 6.99 (1.54) \\
32   & 2.91 (0.21) & 6.80 (0.19) & 2.34 (0.19) & 2.61 (0.16) &11.41 (0.18) & 6.85 (0.20) & 3.87 (0.18) &15.91 (0.75) &15.40 (0.86) \\
36   & 0.19 (0.16) & 0.65 (0.14) & 0.15 (0.14) & 0.52 (0.12) & 1.40 (0.14) & 1.33 (0.15) & 0.50 (0.13) & $<$1.28        & $<$1.46 \\
37   & 4.92 (0.19) &18.09 (0.17) & 3.55 (0.17) & 3.75 (0.15) &27.89 (0.17) &14.61 (0.18) &15.97 (0.16) &18.84 (1.03) &17.90 (1.18) \\
40   & 3.77 (0.25) & 8.70 (0.22) & 2.87 (0.22) & 3.41 (0.20) &16.24 (0.22) & 9.03 (0.23) & 5.18 (0.21) &21.51 (0.79) &21.48 (0.91) \\
41   & 9.54 (0.19) &20.69 (0.17) & 8.75 (0.17) & 7.45 (0.15) &23.23 (0.17) &17.86 (0.18) &11.91 (0.17) &15.37 (0.79) &18.34 (0.91) \\
55   & 1.82 (0.13) & 3.76 (0.11) & 1.68 (0.11) & 1.90 (0.10) & 5.17 (0.11) & 3.69 (0.12) & 2.41 (0.11) & 6.12 (0.76) & 7.79 (0.87) \\
58   & 0.66 (0.17) & 3.66 (0.16) & 0.42 (0.16) & 0.97 (0.14) & 6.75 (0.15) & 3.66 (0.16) & 3.12 (0.15) & 4.61 (0.67) & 7.42 (0.77) \\
63   & 2.37 (0.26) & 7.27 (0.23) & 1.72 (0.23) & 2.57 (0.20) &11.92 (0.23) & 6.92 (0.24) & 6.30 (0.22) & 9.71 (1.18) & 9.75 (1.35) \\
1933 &21.00 (0.56) &43.29 (0.51) &20.11 (0.50) &18.07 (0.44) &53.87 (0.49) &39.42 (0.53) &25.08 (0.48) &32.56 (3.02) &46.80 (3.47) \\
9060 & 9.41 (0.26) &14.24 (0.23) & 8.98 (0.23) & 8.08 (0.20) &15.75 (0.23) &13.60 (0.24) & 6.66 (0.22) &12.41 (0.99) &13.62 (1.14) \\
10352& 4.72 (0.25) & 6.64 (0.23) & 4.35 (0.23) & 3.92 (0.20) & 7.99 (0.22) & 7.00 (0.24) & 4.52 (0.21) & 7.06 (1.12) &11.11 (1.28) \\
\enddata
\tablecomments{Individual \water\ features labeled by upper level energy (K) and wavelength in microns. Flux units are in 10$^{-19}$ W m$^{-2}$. $1\sigma$ uncertainties are given in parentheses. Upper limits are 2$\sigma$.}
\tablenotetext{a}{Line used in \citet{Banzatti25}.
\vspace{-0.3cm}}
\label{Table.IC 348.water.Results}
\end{deluxetable}

In order to compare the IC 348 molecular fluxes with those of the younger Taurus sample, we made analogous measurements from the {\it Spitzer} IRS spectra of the Taurus sources (Table~\ref{Table_TaurusIRS_Results}). 
Because of the lower spectral resolution of IRS, the wavelength regions adopted for the IRS spectra
sometimes differ from the wavelength regions used for the IC 348 MIRI spectra, and certain corrections were applied to account for the effects of different resolution or flux bandpass (see Appendix~\ref{sec:appendix:features} for details). 
As a test of the equivalence of fluxes measured from the IRS and MIRI spectra, we smoothed and binned the IC 348 MIRI spectra to match the resolution of the IRS data and then measured  fluxes in the same manner as for the Taurus IRS data. 
The fluxes from the smoothed and binned spectra recovered the values measured directly from the MIRI spectra to better than 10\% without systematic offsets.

\subsection{\ion{H}{1} Fluxes and Accretion Luminosity}

To provide a contemporaneous and uniform measure of the stellar accretion rates in our samples, we measured the mid-infrared \ion{H}{1} lines at 12.37 \micron\ in the MIRI and IRS spectra. 
The hydrogen emission at this wavelength includes both the \ion{H}{1}(7--6) and \ion{H}{1}(11--8) transitions. 
In MIRI spectra, these two transitions overlap to some extent, but the stronger (7--6) line can be largely separated from the (11--8) line \citep{Tofflemire25}, although both transitions are contaminated by \water\ emission.
At IRS resolution, the two \ion{H}{1} transitions are blended with three \water\ lines into a single feature.
We measured the combined \ion{H}{1}(7--6) and \ion{H}{1}(11--8) flux in order to use a consistent diagnostic for the MIRI and IRS based samples.
The wavelength intervals for the flux measurements, continuum definition, and the relation to the \water\ spectrum are described in Appendix~\ref{sec:appendix:features}.

\begin{deluxetable}{lccccc}
\tablecaption{\ion{H}{1} Line Fluxes and Accretion Luminosities \label{tab:HI_Lacc}}
\tablehead{
\colhead{LRL} & 
\colhead{Observed (\ion{H}{1}+\water)\tablenotemark{a}} & 
\colhead{\ion{H}{1}\tablenotemark{b}} & 
\colhead{log($L_{\mathrm{line}}$)\tablenotemark{c}} &
\colhead{log($L_{\mathrm{acc}}$)\tablenotemark{c}} &
\colhead{log($M_{\mathrm{acc}}$)\tablenotemark{d}} \\
\colhead{} & 
\colhead{($10^{-17}$ W m$^{-2}$)} &
\colhead{($10^{-17}$ W m$^{-2}$)} &
\colhead{($L_\odot$)} &
\colhead{($L_\odot$)} &
\colhead{($M_\odot\,{\rm yr}^{-1}$)}
}
\startdata
15    & 0.250 (0.007) & 0.144 (0.011) & -5.337 (0.033) & $-1.350$ (0.069) &  -8.05\\
32    & 0.186 (0.004) & 0.118 (0.008) & -5.423 (0.029) & $-1.530$ (0.061) &  -8.41\\
36    & 0.039 (0.002) & 0.015 (0.004) & -6.319 (0.116) & $-3.393$ (0.241) &  -10.25\\
37    & 0.222 (0.007) & 0.095 (0.011) & -5.517 (0.050) & $-1.726$ (0.105) &  -8.68\\
40    & 0.253 (0.004) & 0.176 (0.006) & -5.249 (0.015) & $-1.169$ (0.031) &  -8.15\\
41    & 0.307 (0.003) & 0.042 (0.007) & -5.872 (0.072) & $-2.463$ (0.151) &  -9.55\\
55    & 0.054 (0.002) & $<$0.008      & -6.592 (0.217) & $< -3.961$ &  $< -10.58$\\
58    & 0.031 (0.005) & $<$0.012      & -6.416 (0.217) & $< -3.595$ &  $< -10.41$\\
63    & 0.079 (0.005) & $<$0.015      & -6.319 (0.203) & $< -3.393$ &  $< -10.25$\\
1933  & 0.952 (0.016) & 0.372 (0.041) & -4.924 (0.048) & $-0.493$ (0.100) & -7.60\\
9060  & 0.320 (0.003) & 0.091 (0.007) & -5.536 (0.033) & $-1.765$ (0.069) &  -8.54\\
10352 & 0.124 (0.005) & 0.018 (0.009) & -6.240 (0.217) & $-3.228$ (0.452) &  -9.88\\
\enddata
\tablenotetext{a}{Flux of the blended \ion{H}{1}+H$_2$O feature.}
\tablenotetext{b}{Flux of the \ion{H}{1} (7–6)+(11–8) transitions after correction for H$_2$O contamination.}
\tablenotetext{c}{Log of line luminosity and accretion luminosity.}
\tablenotetext{d}{Log of accretion rate.}
\vspace{-0.2cm}
\end{deluxetable}

As was done for \CtwoHtwo, model slab spectra were used to determine corrections for the  contaminating \water\ transitions in the 12.37~\micron\ \ion{H}{1} feature (see Appendix~\ref{sec:appendix:features} for details). For the MIRI spectra, we used a linear combination of the flux of the 13.13\ \micron\ and 14.43\ \micron\ \water\ transitions to determine the correction for the two contaminating transitions. 
For the IRS spectra, the correction for the three contaminating \water\ transitions used a combination of blended water features at 11.64\ \micron\ and 11.72\ \micron.
The corrected \ion{H}{1} fluxes were converted to line luminosity, and accretion luminosities were derived using the relation in \citet{Rigliaco15}.\footnote{While \citet{Rigliaco15} specfically refer to the \ion{H}{1} (7--6) line luminosity,
their measurements and relation necessarily include both \ion{H}{1} transitions. Their relation is close to that of \citet{Tofflemire25} which is based solely on the (7--6) line. The direction of the offset between the two relations is consistent the fact that Rigliaco et al. measure both transitions, while the magnitude of the offset would imply that the  (7--6) line comprises about 70\% of the combined \ion{H}{1} flux on average.}
The \ion{H}{1} line fluxes and accretion luminosities are given in Table~\ref{tab:HI_Lacc} for IC 348
and in Table~\ref{Table_TaurusIRS_Results} for Taurus. 
Note that the line luminosities used to calibrate the relation in Rigliaco et al.\ do not extend below log($L_{\mathrm{HI}}$) of -6.0,
while several of the IC 348 line luminosities are lower;
hence, some of our values are extrapolated beyond the calibrated region.
The last column of Table~\ref{tab:HI_Lacc} gives the accretion rates for IC 348, calculated from the accretion luminosity using the stellar parameters from
\citet{Rodriguez18}.

\begin{figure*}[ht]
\centering
\includegraphics[trim = 0.5cm 4.5cm 1cm 3cm, clip, width=0.45\textwidth]{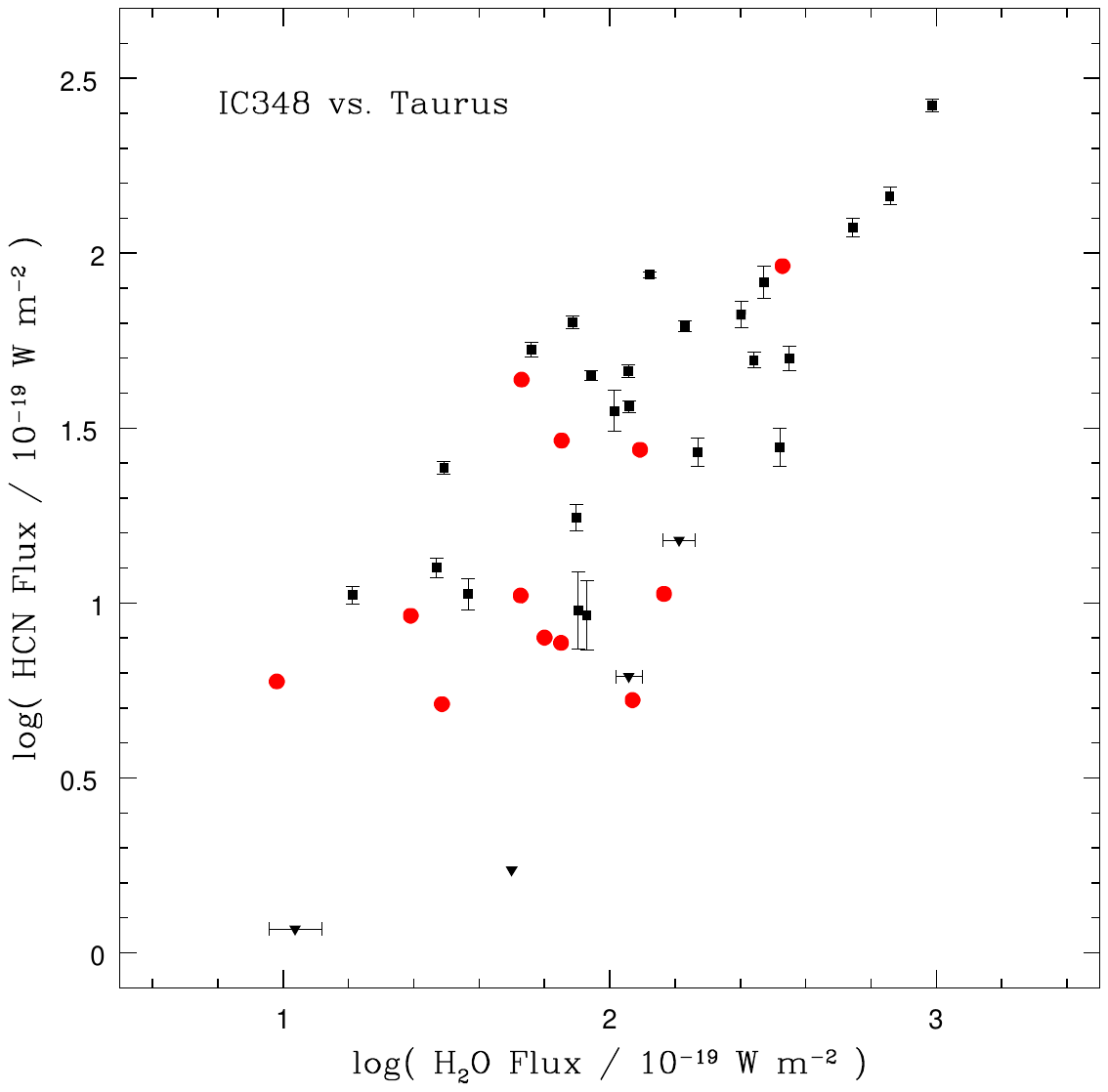}\includegraphics[trim = 0.5cm 4.5cm 1cm 3cm, clip, width=0.45\textwidth]{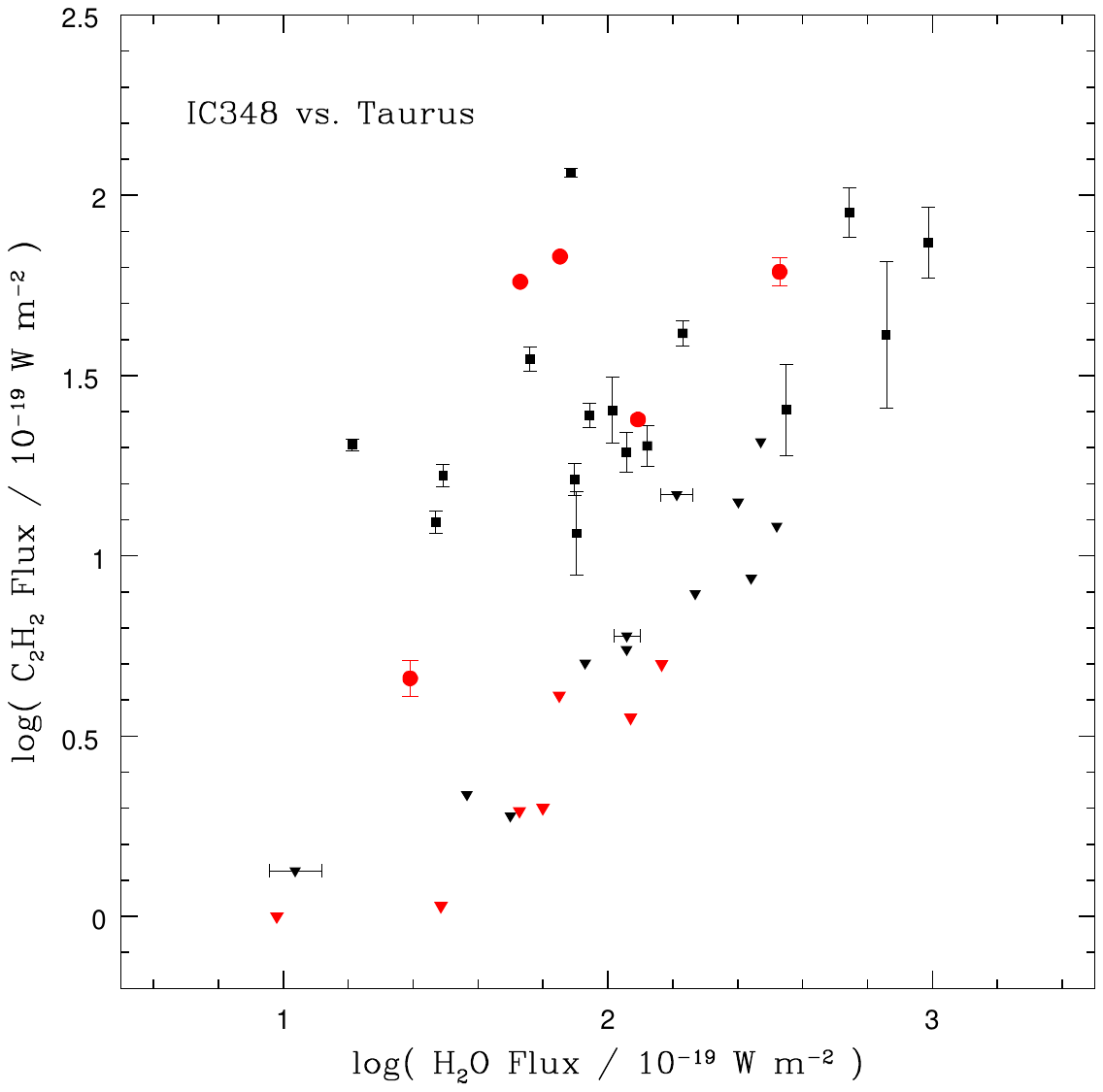}
\caption{Flux of HCN (left) and \CtwoHtwo\ (right) as a function of \water\ flux in the spectra of IC 348 (red points) and Taurus (black points) disks. The molecular flux values for the two samples overlap, with Taurus sources reaching higher values. In both panels, the Taurus fluxes are scaled to the IC 348 distance of 320 pc. Inverted triangles are upper limits. \vspace{0.7cm}}
\label{fig:HCN_C2H2_H2O_fluxes}
\end{figure*}

\subsection{Molecular Flux Ratios and Trends}
\label{sec:MolecularRatios}
\subsubsection{Molecular Fluxes}

The HCN, \CtwoHtwo, and \water\ fluxes for IC 348 are compared with those of Taurus in
Figure~\ref{fig:HCN_C2H2_H2O_fluxes}, by plotting the flux of HCN vs.\ \water\ (left panel) and \CtwoHtwo\ vs.\ \water\ (right panel).
 The absolute fluxes of \water, HCN and \CtwoHtwo\ overlap significantly, but those in IC 348 do not reach the highest values observed in the Taurus sample, and the fluxes for the IC 348 sample are weighted toward lower values.
The HCN and \water\ fluxes increase together in the same manner in both samples, with an order of magnitude spread. More dispersion is seen in the flux of \CtwoHtwo\ vs.\ \water, although there is still a general increase of \CtwoHtwo\ with \water.

In the mid-infrared spectra of CTTS, molecular emission features arise from the warm disk atmosphere, which is heated by FUV irradiation and X-rays  \citep[e.g.,][]{Adamkovics14,  Woitke16, Bosman22}.
Water has a special role to play in this process. Its absorption cross-section spans a broad swath of the FUV wavelength region at significant strength. As a result, UV absorption by water is an important heating process for disk atmospheres and is effective in putting disk molecular features (\water\ and other molecules) into emission 
\citep[e.g.,][]{Adamkovics16, NA17}. 
Since the FUV fluxes of CTTS are dominated by stellar accretion, the strength of molecular emission features is expected to rise with accretion rate diagnostics such as \ion{H}{1}. 
{\it In situ} accretion heating would have a related effect \citep{Glassgold04}.

\begin{figure}[htb]
\centering
\includegraphics[trim = 0.5cm 4.5cm 1cm 3cm, clip, width=0.46\textwidth]{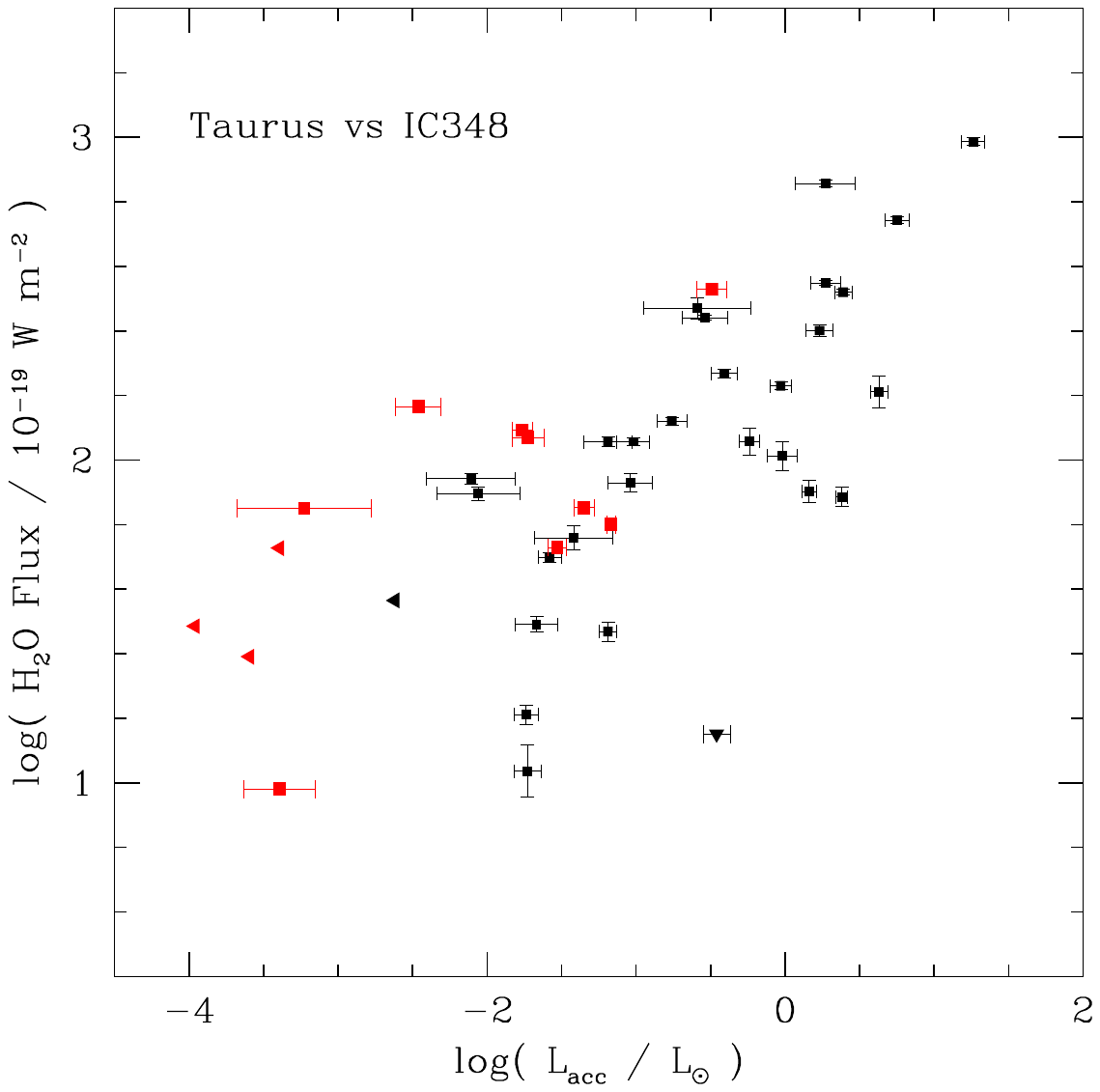}
\caption{\water\ flux vs.\ accretion luminosity, as measured by the 12.37 \micron\ \ion{H}{1} emission lines, for IC 348 (red points) and Taurus (black points) sources.
The  
quantities are correlated, as expected if  accretion heats the disk surface, 
putting molecular features into emission. 
The Taurus \water\ fluxes are scaled (on a star-by-star basis) to the IC 348 distance of 320 pc. Triangles indicate upper limits on accretion luminosity or \water\ flux.
}
\label{fig:H2O.v.HI}
\end{figure}

Figure~\ref{fig:H2O.v.HI} shows that, as expected, the \water\ flux is correlated
with accretion luminosity for both IC 348 and Taurus sources.
Equivalent trends have been presented in earlier work  \citep[e.g.,][]{Banzatti20,Banzatti25}. 
The two samples overlap in \water\ and accretion luminosities, but the Taurus disks extend to both higher accretion luminosities and water flux. The IC 348 sample is weighted to lower accretion rates, consistent with the interpretation that IC 348 is evolutionarily older than Taurus. 
In the larger Taurus sample, the linear trend between the log of \water\ flux and log of accretion  luminosity has a dispersion of about 0.3 dex.
While IC 348 shows a similar trend, a number of disks have much lower accretion rates compared to Taurus disks at similar \water\ luminosity.

\subsubsection{Molecular Flux Ratios}

\begin{figure*}[ht]
\centering
\includegraphics[trim = 0.5cm 5.0cm 1cm 3.5cm, clip, width=0.46\textwidth]{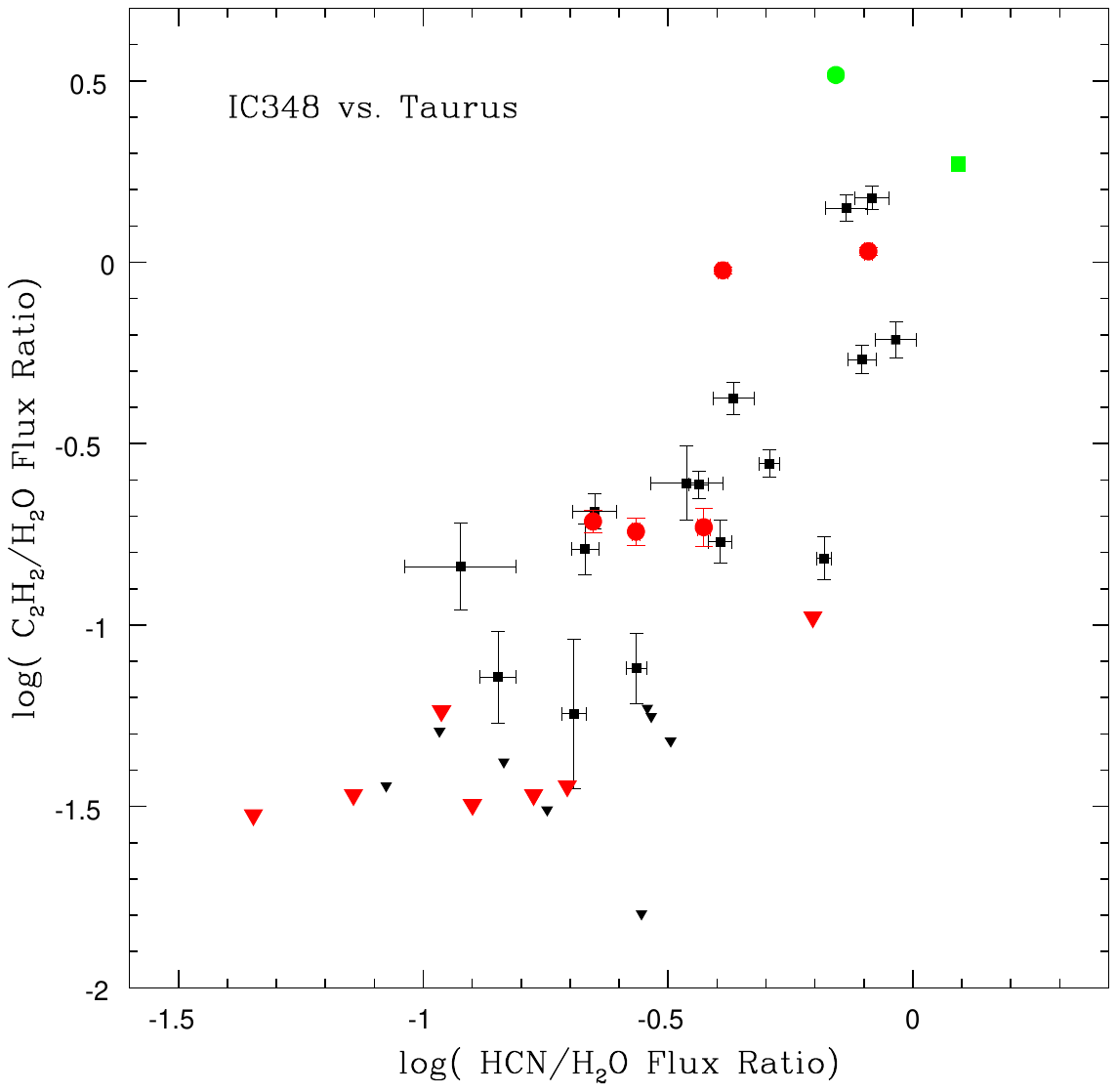}
\includegraphics[trim = 0.5cm 5.0cm 1cm 3.5cm, clip, width=0.46\textwidth]{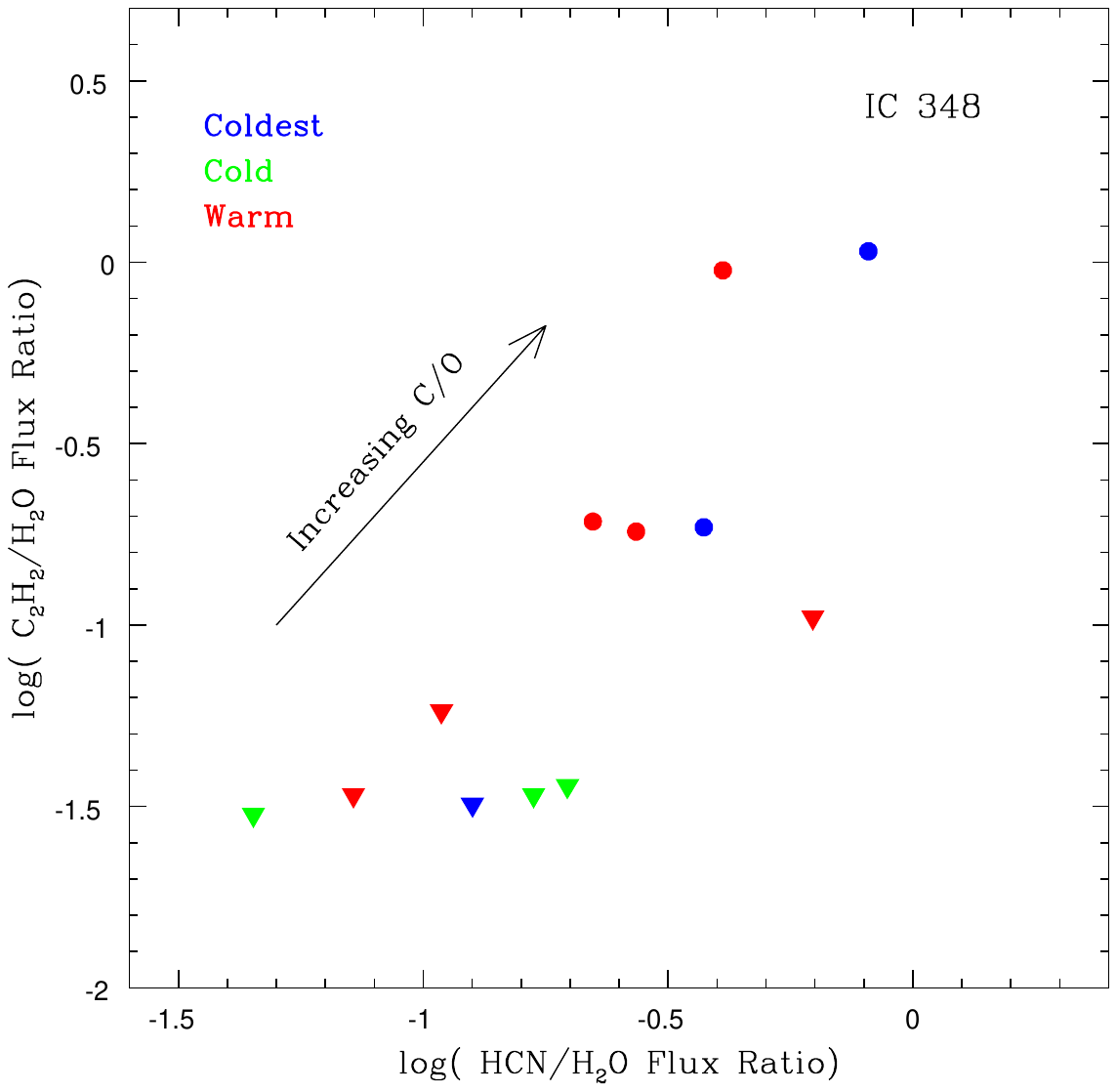}
\caption{ 
Left: 
\CtwoHtwo/\water\  vs.\ HCN/\water\ flux ratios for IC 348 (red) and Taurus (black). Despite their age difference, the two samples have similar flux ratio distributions. The log of the ratios increase together in a linear trend in both samples. 
Two disks from the literature which show enhanced C/O ratio, DoAr33 (green square) and GO Tau (green circle), are also shown. 
Inverted triangles indicate upper limits. 
Objects lacking both HCN and \CtwoHtwo\ detections are not plotted.
Error bars are  plotted for the IC 348 points only when large compared to the symbol size.
Right: The same plot for IC 348 alone, color-coded by the ratio of cool-to-warm water emission, as determined from the 1500/6000\,K and 1500/3500\,K water line ratios. 
The disks are divided into three groups, with Coldest denoting the largest cold-to-warm water ratios and Warm the lowest ratios. 
There is no correlation between the water ratio and the hydrocarbon-to-water ratio.
See also Figure~\ref{fig:C2H2.HCN.H2Oratios2}.
Downward triangles indicate upper limits for \CtwoHtwo. \vspace{0.5cm}}
\label{fig:C2H2.HCN.H2Oratios}
\end{figure*}

Figure~\ref{fig:C2H2.HCN.H2Oratios} (left panel) examines and compares the flux ratios of HCN and \CtwoHtwo\ to \water\ for IC 348 and Taurus.
A large range in the flux ratios is readily apparent, with a factor of 20 range in HCN/\water.
The \CtwoHtwo/\water\ ratio shows a factor of 30 and 50 range for IC 348 and Taurus, respectively.
In contrast to the lower median value of the absolute fluxes for IC 348 vs.\ Taurus,
the flux ratios of the two samples appear to have very similar distributions (also see Sec. 4.3.4).

Using the KS test to quantitatively compare the flux ratio distributions, 
we find that the HCN/\water\ ratios of the two samples are consistent with being drawn from the same population. 
The KS test is less robust for \CtwoHtwo/\water,
 due to the larger number of upper limits for both IC 348 and Taurus.
Therefore, as a more robust approach, we used the Kaplan Meier Estimator in the
{\it lifelines} python package to estimate the survival function of each sample and then
applied the logRank test to the functions. Here as well, the p-value 
says that the two samples are indistinguishable.
We conclude that the flux ratios of HCN/\water\ and \CtwoHtwo/\water\ in IC 348 and Taurus are consistent with having the same distribution. The possible implications of this similarity with regard to the chemical evolution of the inner disk are discussed in Sec. 5.2.

Interestingly, Figure~\ref{fig:C2H2.HCN.H2Oratios} (left)
also shows that the \CtwoHtwo/\water\ and HCN/\water\ flux ratios increase together in a linear fashion in a log-log plot. 
A fit to the Taurus sample (excluding upper limits) gives a slope of 1.6, while a fit to the five IC 348 points 
\jc{}
gives a slope of 1.5. Hence, \CtwoHtwo/\water\ increases faster than HCN/\water\ in both samples.
In other words, \CtwoHtwo/\water\ is more sensitive (varies more strongly) than HCN/\water\ in response to the processes that drive these ratios. 

We interpret the trend of increasing molecular ratios in Figure~\ref{fig:C2H2.HCN.H2Oratios} as due to increasing C/O ratio in the inner disk.
The two additional points (green symbols) plotted in Figure~\ref{fig:C2H2.HCN.H2Oratios} (left),
slightly beyond the upper end of this sequence, support this view.
Both are solar-mass T Tauri stars with evidence of particularly carbon-rich inner disks.
The K4 star DoAr 33 shows emission from ${\rm C_4H_2}$ and a high flux ratio  
of \CtwoHtwo\ relative to HCN \citep{Colmenares24}, which Colmenares et al.\  modeled and interpreted as the result of a high C/O ratio.
GO Tau (spectral type K5) has a similar spectrum \citep{Arulanantham25}, with a high \CtwoHtwo/HCN ratio and ${\rm C_4H_2}$ emission.
Similarly prominent ${\rm C_4H_2}$ is lacking in the IC 348 sample, although the two 
sources with the highest \CtwoHtwo/\water\ ratios show a 
tentative ${\rm C_4H_2}$ emission feature. 
Even higher molecular ratios are seen in disks around very low mass stars 
\citep{Pascucci09, Pascucci13, Grant25}, 
systems which also show emission from exotic hydrocarbons \citep[e.g., C$_2$H$_4$, C$_4$H$_2$, C$_6$H$_6$, HC$_3$N;][]{Tabone23, Arabhavi25, Long25}.
These properties are interpreted as signatures of very carbon-rich inner disks \citep[C/O $>1$;][]{Arabhavi25, Tabone23, Kanwar24}.
While the trend in
Figure~\ref{fig:C2H2.HCN.H2Oratios} 
can be understood to first order as due to the gas C/O ratio,
the actual values of the O/H and C/H abundances can also produce variation in the flux ratios, 
even at a given C/O ratio \citep[][]{Arabhavi26}.
In addition, other parameters could influence the flux ratios and contribute to their scatter,
including the UV flux, dust properties, and vertical mixing 
\citep[e.g.,][]{Walsh15,Woitke18,Greenwood19,Woitke22}.

\begin{figure}[htb]
\centering
\includegraphics[trim = 0.5cm 1.0cm 1cm 1.5cm, clip, width=0.45\textwidth]{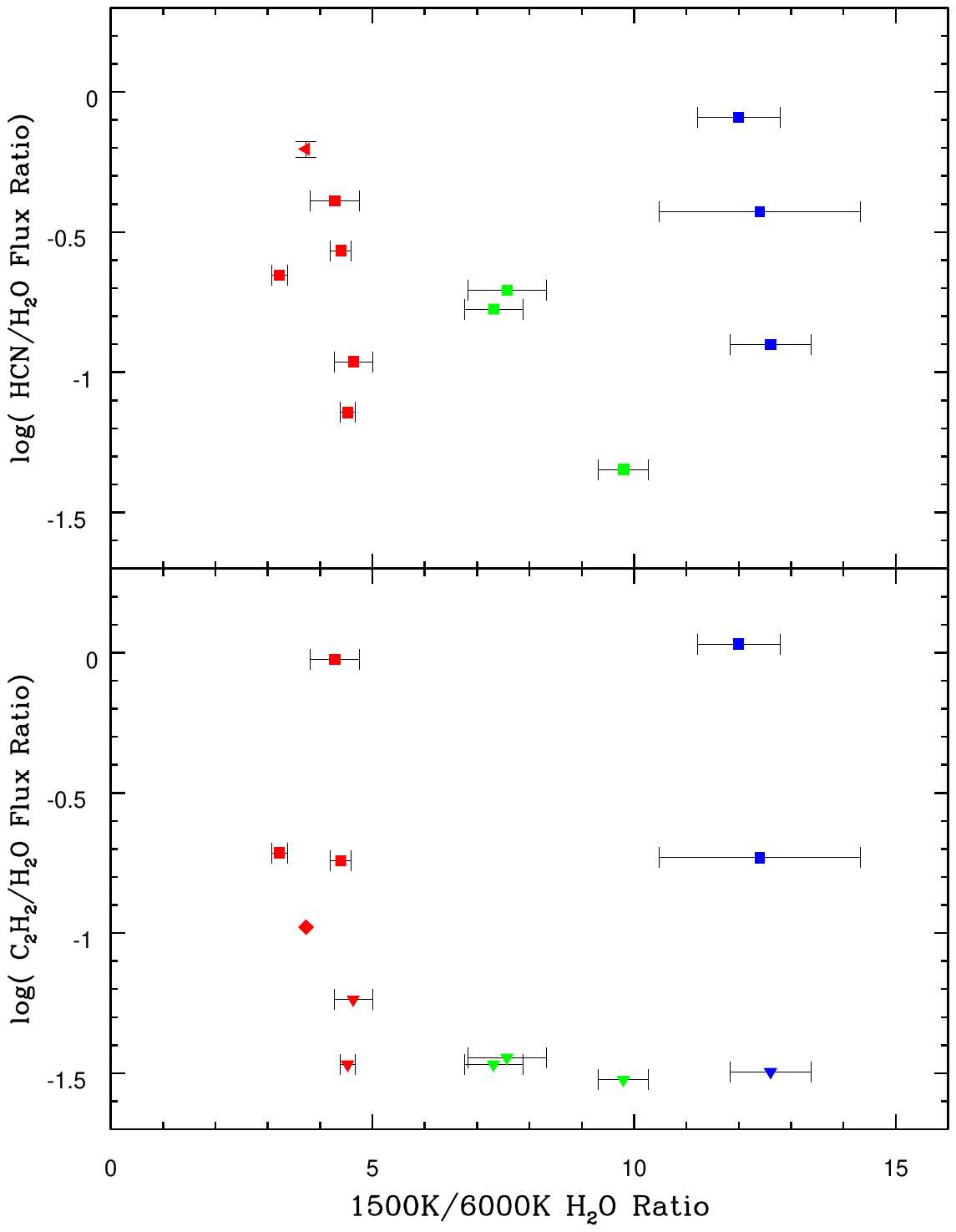}
\caption{ 
The flux ratios of HCN and \CtwoHtwo\ relative to \water\ in the spectra of IC 348 sources plotted against the 1500/6000\,K  water ratio. Neither HCN/\water\ nor \CtwoHtwo/\water\ are correlated with the relative amount of cold water. The color coding is the same as in Figure~\ref{fig:C2H2.HCN.H2Oratios} (right).
Triangles indicate upper limits, and the diamond an upper limit in both axes.
} 
\label{fig:C2H2.HCN.H2Oratios2}
\end{figure}

\subsubsection{Molecular Flux Ratios and Cold Water}

As seen in Figure~\ref{fig:spectra_H2O}, the disks observed in IC 348 show a significant range in  emission line ratios between low and high energy water transitions, suggesting a range in the relative flux (and hence emitting area and mass) from cold and warm \water.
To explore whether 
this diversity in cold to warm water emission ratios has any relation to the 
HCN/\water\ and \CtwoHtwo/\water\ ratios in IC 348,
we examined the specific water ratios defined by \citet{Banzatti25} based on the measurement of lines with upper energies of approximately 1500 K, 3600 K, and 6000 K
(see Table \ref{tab:waterlines}).
In particular, the two lines with $E_u$ near 1500 K are the most sensitive to cold \water\ (e.g., gas closer to the surface snowline)
within the MIRI bandpass,
with larger 1500/3600 K and 1500/6000 K ratios indicating a larger relative amount of cold water.

In the right panel of Figure~\ref{fig:C2H2.HCN.H2Oratios}, the molecular ratios of \CtwoHtwo/\water\ vs. HCN/\water\ are plotted for IC 348
(as in the left panel),
but now color-coded by the ratio of cool to warm water.
Based on the 1500/6000 K and 1500/3600 K ratios, the IC 348 disks can be readily divided into three groups (as shown in Figure~\ref{fig:C2H2.HCN.H2Oratios2}). 
Disks with the highest cold-to-warm water ratios are designated in the plot as ``Coldest'' (blue points), those with the lowest ratios as ``Warm'' (red points), and an intermediate group ``Cold'' (green points).
The three groups are spread throughout  Figure~\ref{fig:C2H2.HCN.H2Oratios}.
There is no preference for disks with high or low cold/warm ratios to have high or low flux ratios of HCN/\water\ and \CtwoHtwo/\water.

The absence of a trend 
in Figure~\ref{fig:C2H2.HCN.H2Oratios}
is illustrated in a different way in
Figure~\ref{fig:C2H2.HCN.H2Oratios2}, which plots the flux ratios HCN/\water\ and \CtwoHtwo/\water\ against the 1500/6000 K water ratio.
There is no correlation between the hydrocarbon-to-water flux ratios and the cold-to-hot water ratio, and
disks with high and low 1500/6000 water ratios show the same range in HCN/\water\ and \CtwoHtwo/\water.\footnote{An equivalent result is found when plotted against the 1500/3600 K water ratio.}
We discuss this result in connection with inner disk C/O ratios and pebble drift in Sec. 5.3.

\begin{figure*}[htb]
\centering
\includegraphics[trim = 0.5cm 2.0cm 10cm 1.5cm, clip, angle=270, width=0.97\textwidth]{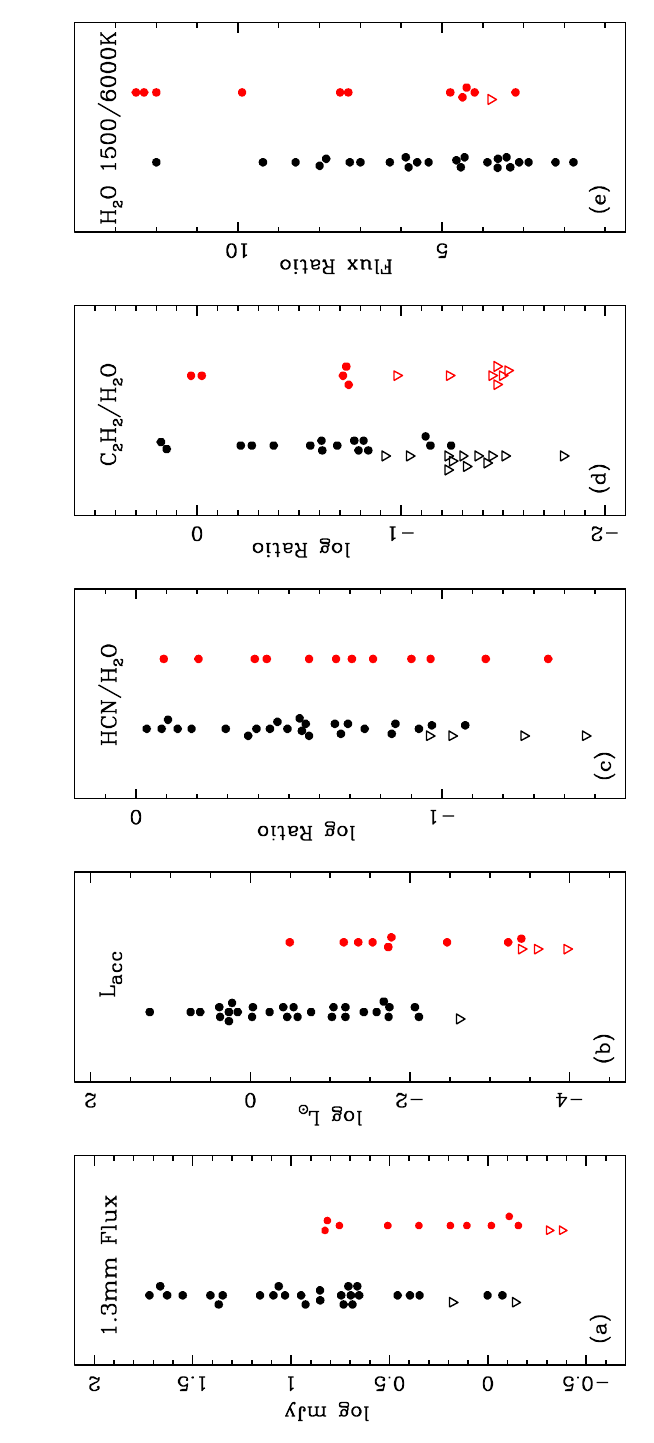}
\caption{
Distributions of IC 348 disk properties (red) compared to those of younger Taurus disks (black): 
(a) 1.4 mm flux scaled to 320 pc;
(b) accretion luminosity derived here from \ion{H}{1} emission;
(c) HCN/\water\ flux ratio; and
(d) \CtwoHtwo/\water\ flux ratio. 
Panel (e) compares the 1500/6000 K water line ratio of IC 348 disks (red) with those for younger disks from the literature (black).
In all panels, open inverted triangles indicate upper limits.
While the IC 348 disks have significantly lower disk masses and accretion luminosities than the younger disks, their hydrocarbon-to-water and cold-to-warm water ratios are similar (panels c, d, e).  
\vspace{0.5cm}} 
\label{fig:Comparing_IC348_Taurus}
\end{figure*}

\subsubsection{Disk Properties of IC 348 vs.\ Younger Systems} 

We compare in Figure~\ref{fig:Comparing_IC348_Taurus} some of the disk properties of IC 348 sources with those of younger systems.
Panel (a) compares the distributions of the distance-scaled 1.3 mm fluxes
between IC 348 (red points) and our Taurus sample (black points).
(Sub)millimeter flux density is a commonly used proxy for the disk dust mass,
because dust mass is approximately proportional to the millimeter continuum luminosity.
The median 1.3 mm flux for the IC 348 sample is 5 times lower compared to Taurus,
and the maximum flux is  $\sim 7$ lower. This difference is in accord with the 
lower disk masses for IC 348 compared to Taurus shown
by \citet{Rodriguez18}.

Our derived accretion luminosities for IC 348 and Taurus are compared in Figure~\ref{fig:Comparing_IC348_Taurus}b.
The median accretion luminosity for the IC 348 disk sample is a factor of 40 lower than the Taurus median. 
Half of the Taurus disks have accretion luminosity greater than the highest value in the IC 348 sample, 
while several of the IC 348 disks have accretion luminosities significantly lower than those measured in the Taurus sample.
The lower accretion luminosities in IC 348 are consistent with the known decline of accretion rate with age.

The HCN/\water\ and \CtwoHtwo/\water\ flux ratio distributions are shown in Panels (c) and (d) of Figure~\ref{fig:Comparing_IC348_Taurus}.
This repeats the information shown in a 2-D format in
Figure~\ref{fig:C2H2.HCN.H2Oratios}
and reiterates our conclusion that there is no measurable difference in these flux ratios between the IC 348 and Taurus samples. Further details on the dependence of the molecular ratios on submillimeter flux and accretion rate are provided in Appendix~\ref{sec:appendix:ratios}.

The last panel (Fig.~\ref{fig:Comparing_IC348_Taurus}e) compares our measured 1500/6000 K water ratios for IC 348 to those published for younger disks. 
For the young disks, we use the sample of disks in the study by \citet{Krijt25} plus additional full disks from \citet{Mallaney26}.
The latter paper uses an updated flux calibration for the JDISCS reduction pipeline, 
which increases the 1500/6000 K water ratio by 1.1 times on average. 
Hence, we used the fluxes published in \citet{Mallaney26}, but for 5 stars in \citet{Krijt25} that do not appear in Mallaney et al.\ we multiply the ratios by a factor of 1.1.
We excluded the peculiar molecular emission source MY Lup \citep{Salyk25}.
Note that the disk IRAS 04385+2550, with a 1500/6000 K ratio of 19.4, is off the top of the plot in Figure~\ref{fig:Comparing_IC348_Taurus}e.
IC 348 and the younger disk sample show a similar range in the 1500/6000 K water flux ratio,
with no evidence for a significant difference in their distributions.
As per the KS test, the water ratios are consistent with the two samples being drawn from the same distribution

Thus, despite having both lower disk masses and lower accretion rates than Taurus disks,
the IC~348 sample is characterized by
HCN/\water\ and \CtwoHtwo/\water\ 
flux ratios similar to those of younger disks,
with no sign of evolution over the age interval between the samples. 
Similarly, the values and distribution of the 1500/6000 K water flux ratios in IC 348 have not evolved from those in younger disk samples.
We discuss the implications of these results in the next section.

\section{Discussion}    \label{sec:discussion}

As shown in section~\ref{sec:results}, JWST/MIRI spectroscopy of CTTS in IC 348 reveals inner disk molecular emission properties similar to those seen in younger disk populations. 
A key finding is that the measured distributions of the HCN/\water\ and \CtwoHtwo/\water\ flux ratios among T Tauri stars in IC 348 are indistinguishable from  those observed in Taurus, for sources in the spectral type range K5--M2 ($\sim 0.4–0.8\,\Msun$). The results indicate a similar level of disk chemical diversity in both populations and, more importantly, imply that there is little evolution in the range of inner disk C/O ratios between the age of Taurus (1--2 Myr) and IC 348 (2--5 Myr). 
We also find
no measurable evolution in the distribution of cold-to-warm water emission ratios between Taurus and IC 348.
The inner disk chemical diversity (and implied C/O ratios) among the IC 348 sample also shows no relation to the relative amounts of cold and warm water emission.

Multiple indicators point to a meaningful age difference  between IC 348 and Taurus.  
As described in Appendix~\ref{sec:appendix:ages}, isochronal ages indicate that IC 348 is 
older than Taurus by at least 1 Myr and  perhaps as much as 2--3 Myr. 
In IC 348, the fraction of sources with an optically thick (Class II) disk is $\sim 40$\% for approximately solar mass stars, compared to a disk fraction of $\sim 65$\% for similar mass stars in Taurus
\citep{Luhman08,Luhman10,Fedele10,Ribas14}.
The fraction of stars undergoing detectable stellar accretion is lower for IC 348 than Taurus 
\citep[$\sim 33$\% vs.\ $\sim 60$\%;][]{Fedele10}.
Moreover, the millimeter fluxes of IC 348 sources are fainter than those in Taurus, with a flux distribution more similar to that of the 3--5~Myr old $\sigma$~Ori cluster 
\citep{Rodriguez18}. 
The sample comparisons for 1.3 mm flux  and accretion luminosity in
Figure~\ref{fig:Comparing_IC348_Taurus}
align with these broader comparisons.
All of these properties point to a significant age difference between IC 348 and Taurus.

We discuss our results in this context, 
beginning with a brief review of ideas on the chemical evolution of inner disks  (Sec.~\ref{sec:discussion.background}). We then discuss how our finding of sustained chemical diversity and slow chemical evolution in inner disks challenges current disk evolution models, which generally predict increasing C/O ratios over time (Sec.~\ref{sec:discussion.slowchemicalevolution}). Finally, we examine the implications of our results for our understanding of the role of pebble drift in the first few Myr of a disk's life 
(Sec.~\ref{sec:discussion.pebbles}). 

\subsection{Background: MIR Molecular Emission and the Chemical Signature of Planet Formation}
\label{sec:discussion.background}

The different partitioning of carbon and oxygen in the gas and solid phases of a disk, combined with the different flow rates of the (O-rich) solid and (more C-rich) gaseous reservoirs through the disk, can induce significant changes in the delivery rates of C and O to the inner disk over time.
Disk evolution and planet formation processes that impact the solid and gas flow rates can 
therefore affect the C/O ratio of the inner disk. 
When pebble drift is efficient, icy solids will be transported quickly to small radii, bringing water and oxygen to the inner disk, thereby reducing the C/O ratio. One potential consequence of this process is the so-called ``meter-size barrier problem,'' in which meter-sized objects quickly traverse the inner 1\,au of the minimum mass solar nebula 
\citep[in $\sim 100$ yr; e.g.,][]{Adachi76, Weidenschilling77},
potentially depleting the disk of its planet-building materials. The same process operating at larger disk radii ($\sim 100$\,au) and smaller solid sizes (cm) would drain outer disks on Myr timescales and greatly reduce their millimeter flux 
\citep[e.g.,][]{TakeuchiLin05}.

Planet formation processes would have the opposite effect.  
The efficient conversion of pebbles into planetesimals and larger objects will tend to suppress the transport of solids into the inner disk \citep{Ciesla06}, because these objects are too large to migrate under gas drag 
\citep{Weidenschilling77}.
Pressure bumps (i.e., gaps and rings) in the disk
can also suppress the inward transport of pebbles \citep[e.g.,][]{Pinilla12, Birnstiel24}.
Inhibiting the inward flow of O-rich solids by either mechanism allows the inflow of C-rich gas to raise the C/O ratio of the inner disk. 
The two scenarios may also be closely interrelated: pressure bumps are great places to convert pebbles into planetesimals and planets \citep[e.g.,][]{Jiang23, Morbidelli20, Birnstiel24}.
As a result, the chemical signature of planet formation processes might be characterized as the tendency toward an increasing C/O ratio in the inner disk.

We previously interpreted the spread in molecular ratios seen in {\it Spitzer} data as a chemical signature of active planet formation at Taurus age
\citep[][N13]{CN11}.
Thermal-chemical models of disk atmospheres at the time generally required enhanced C/O ratios ($> 0.4$) to account for typical hydrocarbon emission properties (emission column density, temperature, and emitting area; e.g., \citealt{N11}; see also \citealt{NA17}). The models further suggested that a factor of $<$ 2 variation in the C/O ratio of inner disk atmospheres could produce a factor of $\sim 10$ variation in the warm emission column density ratios of HCN/\water, similar to the range observed among Taurus CTTS. 
As a result, if oxygen is sequestered beyond the snow line to a greater extent in some disks, raising the C/O ratio of their inner regions, the consequent increase in the inner disk HCN abundance, combined with the decrease in the \water\ abundance, would yield dramatic increases in the HCN/\water\ ratio.   
We therefore hypothesized that if Taurus disks have experienced a range of planet formation activity, 
inducing inner disk C/O ratios that range from $\sim$ stellar
to $< 1,$ 
we could account for the observed warm column ratios.

The C/O ratios inferred for CTTS disks at $\sim 1$ Myr age  
in principle limit the efficiency of pebble drift, and the ``meter-size barrier problem'' as obstacles to planet formation.
Indeed, the apparent trend that higher HCN/\water\ ratios tend to be measured for disks with higher masses led us to
hypothesize that by Taurus age, more massive disks have typically been more efficient or more advanced than lower mass disks in forming planetesimals and planets 
\citep[][N13]{CN11}.

Taking this trend as a launch point, \citet{Banzatti20} found results similar to N13 but interpreted them differently.
They focused on the inverse quantity, \water/HCN, and its anti-correlation with disk size (which is equivalent to the N13 trend of increasing HCN/\water\ with disk mass because of the disk mass-size correlation, e.g.,
\citealt{Tripathi17,Hendler20}). 
The HCN emission was taken to be a reference quantity, and
the observed \water/HCN ratio was then interpreted as a measure of the water abundance and the extent of pebble drift, with smaller disks having experienced more pebble drift into the inner $\sim 1$~au. 
This approach did not account for the chemical impact that the water delivery rate (and the O/H abundance) would have on the hydrocarbon abundances.
Extending this idea, 
more recent papers speculate that the stronger cool water emission observed from compact disks, compared to the weaker emission from spatially extended disks, is the result of very high rates of pebble drift 
\citep[up to $400\, M_\oplus$ per Myr;][]{Romero24},
which both reduces the radial size of the millimeter continuum emission and delivers water to the inner au \citep{Banzatti23, Romero24}. 
Alternatively, cool water emission may indicate large pebble drift rates ($1-1000\, M_\oplus$ per Myr) that are related to the location of the innermost dust gap \cite[][]{Krijt25}.

In parallel, a vibrant series of theoretical papers explores pebble drift and/or its interplay with disk substructures and their resulting effect on inner disk water and its chemical composition 
\citep[e.g.,][]{Booth17, Booth19, Krijt25, Kalyaan21, Kalyaan23, Easterwood24, Schneider21, Mah24, Lienert24, Williams25, Houge25smuggling, Sellek25,Houge2026leaky}.
The models generally predict that disks can become O-rich at early times if pebble drift is efficient and the C/O ratio of the inner au rises with time as C-rich gas accretes into the inner disk. 
Spectral features indicating very carbon-rich chemistry are indeed detected from very low mass stars \citep[e.g.,][]{Pascucci09,Pascucci13,Tabone23,Arabhavi24,Arabhavi25,Kanwar24,Long25}. Whether, when, and how T Tauri disks develop such features and/or evolve into the (more modestly) C-rich systems predicted by theory is, observationally, an open question.

\subsection{Sustained Diversity and Limited Chemical Evolution}
\label{sec:discussion.slowchemicalevolution}

Perhaps surprisingly, we do not observe the expected trend toward increasing C/O ratio with age in our study, which compares
disks in two star forming regions separated in age by $\sim 1-3$ Myr,
and spanning the same range of stellar spectral types.
Instead, the two samples display the same range in C/O ratios over this timespan,
despite having markedly different disk masses, and stellar accretion rates.
To compare our results with one set of model predictions, we can look at those of \citet{Mah24}, who explore the possible impact of gaps (deep or shallow) on the chemical composition of inner disks over time and provide graphical illustrations of the range of possible outcomes for a broad range of model parameters.

In their models, a planet located at 3--10\,au creates a gap of depth $f_{\rm gap},$ where $f_{\rm gap} = 0$ when the gap is completely depleted of gas, and $f_{\rm gap} = 1$ when a gap is not present.
In the absence of a substantial gap 
($f_{\rm gap} > 0.7$; Regime I), pebbles readily drift inward, superhydrating the inner disk and reducing its gaseous C/O ratio to highly substellar values as icy pebbles evaporate (see their Figure 1).
After pebbles are depleted and water vapor in the inner disk is accreted onto the star,
the C/O ratio of the inner disk gas rises as C-rich gas continues to flow into the inner disk. As shown in Figure 1 of \citet{Mah24}, which depicts the reference case of  a $1\,M_\odot$ star surrounded by a disk with $\alpha = 3 \times 10^{-4}$
and a gap at 10 au,
the C/O ratio of the gaseous inner disk rises to the stellar C/O value of 0.55 after $\sim 4$ Myr and becomes superstellar (C/O $\sim 1$)
thereafter out to 8 Myr.

In contrast, a deep gap ($f_{\rm gap} < 0.4$; Regime III) blocks icy pebbles at larger radii from drifting inward.
The phase of high oxygen abundance is shortened, 
and the C/O ratio of the inner disk rises more quickly.
For the reference case,
the C/O ratio of the inner disk gas reaches
$\sim 0.5$ after 2 Myr and C/O $\sim 1.6$ within 4 Myr.
At intermediate gap depths  ($f_{\rm gap} \simeq 0.5-0.7$; Regime II, their ``traffic jam''),
the inner disk C/O remains substellar at all times as drifting pebbles trickle in over time.

In the context of these models, if pebble drift always dominates T Tauri disks,
we would always be in Regime I of \citet{Mah24} and all 1--2 Myr old disks would show low C/O ratios, with limited diversity. 
In contrast, if all disks have gaps with $f_{\rm gap} < 0.4$ (Regime III), they would all have approximately stellar C/O ratios at 2 Myr, again with little diversity.
In contrast, real T Tauri disks show a wide range of C/O ratios, as probed by their organics-to-water ratios.

To account for this diversity, 
we might attempt to select a mixture of cases from \citet{Mah24}, 
with different gap strength, gap formation time, and disk viscosity,
to account for the wide range in C/O ratios of disks at $\sim$1--2\,Myr of age. 
The results presented here restrict any ensemble of model disks to those that would
{\it maintain the same range in values of C/O ratios}
from the age of Taurus out to the age of IC 348.
That is, real disks have diverse C/O ratios that show little evolution from 1--2 Myr to 2--5 Myr.
Perusing the array of models in \citet[][Appendix B, D]{Mah24} illustrates the challenge of meeting these goals:
most of these models trend to increasing C/O ratios on this timescale,
a characteristic also shown in the models in \citet{Sellek_vanDishoeck_25}
and \citet{Williams25}.

Among the parameters explored in Mah et al., a few do appear to show the observed trends. 
For example, if a gap forms early ($t=0$\,Myr) and close in (at 3\,au), models with $\alpha = 3\times 10^{-4}$ and $f_{\rm gap} < 0.4$ (Regime III and II)
produce a range of C/O ratios that show little evolution over 1--5 Myr of age \citep[Fig.~B.1 of][]{Mah24}.
Another example is that of a gap that forms later (0.3 Myr) at 10\,au with $\alpha = 10^{-3}$, where models with $f_{\rm gap} < 0.15$ (Regime II) also show a broad range of C/O ratios with little evolution with time \citep[Fig.~D.2 of][]{Mah24}. 
In both examples, the slow evolution results (at least in part) from the shorter duration of early water enrichment
and a subsequent phase of reduced pebble drift through the gap.
In contrast, we would not be able to construct  an ensemble from the Regime I models alone  that shows a diverse range in C/O and maintains constant values over 1--5 Myr.
That is, in the context of Mah et al., our results imply that most disks are not in Regime I and are effective in blocking pebbles to some extent.

Recent modeling efforts raise a new challenge in suggesting that gaps are more leaky than previously believed
\citep{Stammler23, Houge2026leaky}.
Using more sophisticated dust evolution models, \citet{Houge2026leaky} find that
most dust traps are highly permeable (i.e., roughly equivalent to Regime I of Mah et al.). 
Because our results disfavor such highly permeable gaps, 
the preferred cases in \citet{Houge2026leaky} are those in which gaps  are moderately leaky or highly blocking. 
We return to this point in Sec. 5.3.

Modeling that explores these kinds of ideas \citep[and others, e.g.,][]{Houge25smuggling, Williams25} could help account for the results presented here, 
reproducing not only the broad range in molecular ratios (and inferred C/O ratios) seen in young clusters, but also their apparent slow evolution with age.

\subsection{Limited Role of Pebble Drift? }
\label{sec:discussion.pebbles}

Our results also bear on the inferred magnitude of pebble drift in protoplanetary disks. 
Recent work has interpreted the mass of observable cold water vapor and the 1500/6000 K water line ratio as probes of pebble drift rate \citep{Romero24, Banzatti25, Krijt25},
with high drift rates derived for many T Tauri disks.
The idea that cold water is delivered by migrating icy pebbles arises from the observation that the inferred gas temperatures (170-250 K) probed by the the 1500 K lines \citep{Banzatti25} are too low for efficient {\it in situ} gas-phase formation of \water.
Hence, water must be transported to the cold gas emission region
(see \citealt{Vlasblom25}). 
In the pebble drift scenario for cold water,
water ice is transported inward in the disk midplane and sublimated inside the snowline, with the liberated water vapor  mixing upward and diffusing outward to cold
disk surface layers.
A higher rate of pebble drift is hypothesized to produce an enhanced and more extended reservoir of cold surface water vapor.

Assuming that the cold water observed in $\sim 1$~Myr old T Tauri disks is supplied through pebble drift, \citet{Romero24}
inferred the current delivery rate of water ice using the derived mass of observed cold water vapor along with  assumptions about the \water\ gas lifetime and the ratio of total to observable water vapor.
Given their adopted mass fraction of water ice to total pebble mass (0.16), they estimated pebble drift rates ranging from 
$<10$  M$_\oplus$\ Myr$^{-1}$ to $\sim 400$  M$_\oplus$\ Myr$^{-1}$.
Adopting a similar approach,  \citet{Krijt25} derived an empirical relation directly connecting the 1500/6000~K ratio of $\sim 1$~Myr old T Tauri disks to an observable cold water mass. They then used the inferred mass to estimate a similar range of pebble drift rates in a larger sample of sources.

The similarity between the 1500/6000~K ratios of the IC 348 sources and those of younger ($\lesssim 1$~Myr old) T Tauri stars
(Section 4.3.4, Fig.~\ref{fig:Comparing_IC348_Taurus}e) is perhaps surprising in this context.
Applying the methodology of \citet{Krijt25} to the IC 348 sample implies icy pebble fluxes of 30 to 460 M$_\oplus$\ Myr$^{-1}$ with a median value of $\sim 200$ M$_\oplus$\,Myr$^{-1}$, comparable to the rates reported in Krijt et al.
Thus, despite the greater age of IC 348, there is no difference in the implied pebble fluxes,
in contrast to the general expectation of lower pebble mass flux,
and reduced water enrichment,
at later times
\citep[e.g.,][]{Birnstiel12,Sellek25,Krijt25,Kalyaan23}.
Xie et al.\ (2026, accepted) find a similar result: the cool-to-hot water ratio of sources in the even older (5--10 Myr)  Upper Sco region spans the same range of values found for 1--2 Myr old T Tauri stars.
If cold water mass is a probe of pebble drift, it predicts that very high pebble drift rates are sustained for millions of years.

Such high sustained pebble drift rates seem unlikely,  
because of the finite initial solid reservoir of disks.
While high pebble drift rates seem challenging to sustain over the 1--2 Myr ages of the sources in the Romero-Mirza et al.\ and Krijt et al.\ studies, the 2--5\,Myr old IC 348 sources present an even greater challenge because of their greater age. Maintaining a pebble flux rate of $200\ {\rm M}_\oplus\,{\rm Myr}^{-1}$ over the 2--5 Myr lifetime of IC 348 would require an initial solid mass of $400-1000\ {\rm M}_\oplus$, corresponding to an initial gas mass of $> 0.12-0.3\ {\rm M}_\odot$.
Such high masses are at or above the gravitational instability limit and the likely total solid mass in the disk.

The impact of pebble drift on inner disk chemistry can 
also constrain the pebble drift rate. In disks with higher rates of icy pebble drift, the inner disk receives more water and oxygen, which tends to lower its C/O ratio and therefore its HCN/\water\ and \CtwoHtwo/\water\ ratios \citep[e.g.,][]{N11,Anderson21,Arabhavi26}. 
This suggests an observational chemical test of the proposed connection between pebble drift and cold water emission.
Using the results from Section 4.3.3,
if the 1500/6000 K water line ratio
is a measure of pebble drift rate,
the disks labeled ``Coldest'' in Figure~\ref{fig:C2H2.HCN.H2Oratios}
would tend to have lower C/O ratios and fall in the lower left of the plot, and the disks labeled ``Warm'' would fall in the upper right.
This trend is not observed.
Similarly, the expected trend in Figure~\ref{fig:C2H2.HCN.H2Oratios2} 
would be higher HCN/\water\ and \CtwoHtwo/\water\ ratios for low 1500/6000 K,
and low molecular flux ratios for high 1500/6000 K ratios.
This trend is also not seen; instead the same range in HCN/\water\ and \CtwoHtwo/\water\ is measured in  disks with high and low 1500/6000 water ratios.
This result is yet more surprising given that the inferred rates of water and oxygen delivery span nearly 2 orders of magnitude \citep{Krijt25}.
Here, the lack of a
correlation between the relative strength of cold water emission and the hydrocarbon-to-water ratios in IC 348
appears to question the suggestion that cold water emission is a measure of pebble drift rate.

We can also roughly estimate the typical C/O ratio implied by the nominal pebble drift rates and our measured accretion rates. 
A typical pebble drift rate of 200 M$_\oplus$\ Myr$^{-1}$, the median value inferred for the IC~348 sample, corresponds to an oxygen inflow rate of 28 M$_\oplus$\ Myr$^{-1}$ in water ice 
(under same assumptions as in  \citealt{Romero24} and \citealt{Krijt25}). 
This pebble inflow rate of water is balanced by disk accretion at a median rate of $10^{-9}\,\Msun\,{\rm yr}^{-1}$ for the IC 348 sample.
Conservatively assuming that gas accreting from beyond the snowline is completely depleted in water and that the gas phase reservoir of C and O is dominated by CO
(i.e., neglecting icy grains coupled to the gas), and adopting a CO abundance of
$n_{\rm CO}/n_{\rm H} = 1.4 \times 10^{-4}$ characteristic of inner disk atmospheres \citep[e.g.,][]{NA17},
the mass influx rates of C and O via gas accretion are 
0.6 M$_\oplus$\ Myr$^{-1}$ and 0.8 M$_\oplus$\ Myr$^{-1},$
respectively.
Hence, pebble drift dominates the mass inflow rate of oxygen.
This picture  
assumes that the material transported to the inner disk by accretion and pebble drift 
mixes vertically to produce the C/O ratios described, similar to the assumptions of 1D models (e.g., \citealt{Mah24}).
The combined inflows of O and C from pebble drift and gas accretion correspond to a very low C/O = 0.03.
For an accretion rate at the upper end of the IC 348 sample of $10^{-8}\,\Msun\,{\rm yr}^{-1}$ (Table \ref{tab:HI_Lacc}),
the corresponding 
C/O is 0.21.
These simple estimates are in line with C/O ratios from \citet{Mah24} for drift dominated inner disks.

At these low C/O ratios, the abundances of hydrocarbons such as \CtwoHtwo\ and HCN are likely to be greatly reduced relative to \water,
and possibly undetectable spectroscopically.
For example, \citet{N11} found that commonly reported HCN/\water\ column density ratios could be reproduced with C/O ratios in the range 0.4 to $< 1.0$. When C/O declines from 0.4 to 0.2 (from an O/H increase of 0.3 dex), the column density ratios of HCN/\water\ and \CtwoHtwo/\water\ decrease by an order of magnitude.  
Similarly, for a 0.5 dex increase in O/H,
\citet[][their Fig. C.2]{Arabhavi26} predict a decrease 
in the HCN/\water\ and \CtwoHtwo/\water\ flux ratios 
by factors of 5 and 13, respectively,
compared to the already moderately low flux ratios in their reference model (see their Fig. 3).
The fact that HCN and \CtwoHtwo\ emission are commonly observed in T Tauri spectra suggests that highly O-rich inner disks 
produced by high pebble fluxes
are uncommon among T Tauri stars. If disks commonly experience a phase of  high oxygen enrichment (that produces C/O $<$ 0.1--0.2), it may occur very early, before the T Tauri phase. 

Moderately leaky disk gaps offer a potential way 
for disks to avoid excessive oxygen enrichment 
while also enjoying slow chemical evolution. 
In \citet{Houge2026leaky}, 
gaps that are moderately leaky or highly blocking 
produce lower pebble mass fluxes ($< 1-10\ M_\oplus$/Myr)
that are roughly constant over 1--5 Myr of age.
These lower pebble drift rates would  better balance disk gas accretion rates, 
yielding more moderate C/O ratios.
Reduced pebble drift of this kind at T Tauri ages
opens the door to a picture in which disk gas accretion plays a major role in setting 
the C/O ratio of the inner disk, with low level pebble drift
(regulated by a combination of gaps and planetesimal formation)
modifying the C/O ratio and driving chemical diversity. 
Declines in both the pebble reservoir and disk accretion rate with time 
(e.g., as in Fig. 10),
may then yield a slow rate of chemical evolution, similar to the observed situation.

Interestingly,  \citet{Houge2026leaky} find that planetesimal formation significantly reduces the permeability of gaps. As a result, in their models, gaps with higher blocking efficiency also tend to be sites of planetesimal formation.
This result suggests that the efficiency of planetesimal formation could be one factor in determining inner disk C/O ratios, similar to the suggestion in \citet{N13}.
This scenario may generate other benefits as well. Planetesimal formation in pressure bumps in gaps may also explain the origin of planetesimal belts in the Solar System and those that generate debris disks \citep[e.g.,][]{Najita22}.

\section{Summary and Conclusions}\label{sec:conclusions}

As reported in earlier literature, the diverse molecular ratios of T Tauri disks suggest that their inner regions are chemically diverse, likely the result of diverse planet formation histories. 
To probe the relative importance and time evolution of the planet formation and disk evolution processes that drive this diversity, we investigated the chemical properties of an older population of T Tauri disks, using JWST/MIRI to measure the MIR spectra of 12 disks in the 2--5 Myr old cluster IC 348. 
Compared to the 1--2 Myr old regions that have been primarily studied to date, IC 348 is an evolutionary ``sweet spot,'' offering the opportunity to study an older population of stars, among which a significant fraction ($\sim 40$\%) retain a detectable disk.
Older clusters 
may
show the effects of greater time evolution, 
but their smaller disk fractions make it more challenging to infer the evolution of typical disks.

We measured, for both the IC 348 disks and a comparison sample of younger Taurus disks (1--2 Myr), the fluxes of hydrocarbon features (HCN, \CtwoHtwo), \water\ lines, 
and \ion{H}{1} lines that probe the stellar accretion rate. 
Like most younger disks, 
IC 348 disks do not show the 
chemical signature (a very low organics-to-water flux ratio) 
that might be expected
of disks dominated by rapid pebble drift
and the ``meter-size barrier problem''. 
Such scenarios would rapidly remove planet-building solids,  potentially presenting a significant obstacle to planet formation.

Interestingly, the distributions of HCN/\water\ and \CtwoHtwo/\water\ molecular ratios of IC 348 disks are indistinguishable from those of younger Taurus disks 
over the same range of stellar spectral type (K5--M2; $\sim 0.4–0.8\,\Msun$).
The similar distributions imply little to no evolution in the inner disk C/O ratio 
between 1--2 Myr and 2--5 Myr,
even as disk masses decline by a factor of 5 on average
and accretion luminosities by order of magnitude. 
The limited chemical evolution we infer differs from predictions of  models of disk chemical evolution, which generally predict a measurable rise in the C/O ratio of inner disks between the ages of Taurus and IC~348 \citep[e.g.,][]{Mah24,Williams25}. 
The discrepancy indicates the need for a better understanding of the interplay between (and efficiencies of) planet formation and disk evolution processes in order to account for both the sustained diversity in the molecular ratios of T Tauri disks and their apparent limited evolution with age over the first few Myr of disk evolution.

Our results also constrain the role of pebble drift in disk evolution, both because of the greater ages probed in our study and because strong pebble fluxes should reduce the C/O ratio of inner disks. 
Earlier analyses  attempted to infer the current delivery rate of pebbles (and water ice) from 
the mass of observed cold water vapor \citep{Romero24, Krijt25}.
These ideas imply large pebble fluxes (30--460 $M_\oplus {\rm Myr}^{-1}$) into the inner disk that would be difficult to sustain over the disk lifetime, particularly at the older age of IC~348 (2--5 Myr). 
We find no evidence for 
a decline in the cold-to-warm water flux ratio distribution from Taurus to IC 348 age, which would typically be expected  
given the finite initial solid reservoirs of disks, 
and the large pebble fluxes implied.

The inferred high pebble drift rates would also drive inner disks to low C/O ratios (with extremely low C/O ratios of $\sim 0.03$ at the highest delivery rates), given the modest accretion rates measured for the IC~348 systems. 
Such low C/O ratios would suppress the presence of hydrocarbon emission in the spectra.  
The common presence of HCN and \CtwoHtwo\ in the spectra of IC~348 disks (as well as younger disks) places a chemical limit on the rate of pebble drift into inner disks and 
suggests that pebble drift rates may be 
overestimated.
We also find no evidence from our IC 348 spectra for 
a relation between the hydrocarbon-to-water ratios and the relative strength of cold water emission,
which would be expected
if the latter is a measure of pebble drift. Even if pebble drift rates are overestimated by a large, roughly constant factor, we would still expect to see such a trend. 
Overall, these results question the use of cold water emission as an indicator of rapid pebble drift. 

Reduced pebble drift could help us understand the apparent slow chemical evolution and sustained chemical diversity of inner disks. Disk gaps with moderate to low levels of leakage  \citep[as in][]{Houge2026leaky}, would both significantly reduce the pebble flux to inner disks and flatten out the pebble drift rate over 1-5 Myr of age, a combination that could deliver water to inner disks at a rate better matched to T Tauri gas accretion rates. Similarly declining pebble fluxes and accretion rates may then sustain a moderate C/O ratio for inner disks over Myr timescales, similar to the observed situation.

As a caveat to these inferences, it is important to note that our IC~348 results provide only a snapshot at one cluster age, and are based on a modest number of disks. Thus, it is important to study a larger number of T Tauri disks, both younger and older than the 1--2 Myr old populations that have primarily been studied to date. 
These data sets would help us explore how young (e.g., Taurus) and intermediate age (e.g., IC 348) disks connect to older and younger populations, 
whether disks experience a very early phase of rapid pebble drift (Section 5.3), and whether, when, and to what extent T Tauri disks 
develop the very carbon-rich chemistry seen among very low mass stars (Section 5.1).
Future surveys of such disks will likely provide valuable insights.

\begin{acknowledgments}
We thank Adrien Houge, Joe Williams, Andrea Banzatti, Ilaria Pascucci, and Chengyan Xie for sharing results in advance of publication and thoughtful comments on the manuscript.
This work is based on observations made with the NASA/ESA/CSA James Webb Space Telescope. The data were obtained from the Mikulski Archive for Space Telescopes at the Space Telescope Science Institute, which is operated by the Association of Universities for Research in Astronomy, Inc., under NASA contract NAS 5-03127 for JWST. These observations are associated with JWST Cycle 2 program 2826.
The authors acknowledge support from NASA/Space Telescope Science Institute grant JWST-GO-02826.
\end{acknowledgments}

\begin{contribution}

The two authors contributed equally.


\end{contribution}

%
\facilities{JWST(MIRI)}





\appendix

\section{Relative ages of IC 348 and Taurus}\label{sec:appendix:ages}
\restartappendixnumbering

Several lines of reasoning argue that IC 348 is evolutionarily older than Taurus, by at least 1 Myr. Here we discuss the evidence from isochronal ages, disk fractions, disk masses, and stellar accretion rates. 

{\bf Isochronal Ages.} A simple strategy for estimating cluster ages is to compute the median value of the absolute $K$-band magnitude $M_K$ for cluster stars as a function of spectral type and compare this sequence to the corresponding sequence for a reference cluster of known age. Given the luminosity fading expected from pre-main-sequence isochrones, one can then estimate the age of the cluster of interest \citep[e.g.,][]{Luhman16}.
A recent study applies this method to non-full disks in Taurus in the spectral type range K4--M4 \citep{Luhman23}.
After calculating their $M_K$ offsets 
to the median sequence for Upper Centaurus Lupus and Lower Centaurus Crux (UCL/LCC) 
and assuming UCL/LCC has an age of 20 Myr 
and that $\Delta\log L/\Delta\log {\rm age} = -0.6$, as typically found for pre-main-sequence isochrones in the relevant temperature range,
ages were determined for the Taurus subgroups.
The uncertainties are large enough that most of the subgroups could be coeval, with a mean age of 2.5 Myr \citep[][his Table 6]{Luhman23}.
The same strategy, when applied to the IC 348 sources in comparison to Upper Sco (for an age of 11 Myr), leads to an estimated age of $\sim 5$ Myr \citep{Luhman24},
older than the Taurus sources.
Thus, these isochronal studies suggest that IC 348 is older than Taurus by $\sim$2-3\,Myr.

The $\sim 5$ Myr age reported by \citet{Luhman24} is larger than the ages reported in other studies of IC 348  (e.g., 2--4 Myr; \citealt{Luhman03,Muench07}) 
and Taurus (1--2 Myr; e.g., Mohanty et al.\ 2005). This is perhaps not surprising in that studies often use different methodologies and isochrones. Because the two Luhman papers use the same methodology and assumptions, their finding that IC 348 is older than Taurus appears robust (i.e., the Taurus sources lie above the IC 348 sources in an HR diagram). 
Thus it seems reasonable to assume that IC 348 is evolutionarily older than Taurus, by at least 1 Myr and as much as $\sim$2--3 Myr.

{\bf Disk fractions and accretion rates.} Other properties of the YSO populations in the two regions also suggest that IC 348 is evolutionarily older than Taurus. 
IC 348 has a smaller disk fraction, defined as the fraction of stars with a NIR-MIR excess.
\citet{Fedele10}
compiled disk fractions for associations based on the MIR slope, and reported
$\sim$62\% for Taurus and $\sim$ 47\% for IC 348 for stars in the spectral type range K0--M4.
This agrees with the 69\% and 44\%, respectively, obtained by \citet{Luhman08} and \citet{Luhman10} for spectral types K0--M3.5.
Using different criteria,
\citet{Ribas14}
obtained similar relative disk fractions, 63\% for Taurus and 36\% for IC 348, in a study of disk frequency with age.
These results align with the broader trend of decreasing disk fraction with age 
\citep{Haisch01, Ribas14},
a trend that is consistent with the general picture that Class II sources with disks evolve into diskless Class III sources.

Similarly, the fraction of stars undergoing stellar accretion 
(at a rate $> 10^{-11} \Msun {\rm yr}^{-1}$) 
is $\sim$ 60\% for Taurus, 
higher than the $\sim$ 33\% for IC 348 in the spectral type range K0--M4 
\citep{Fedele10}.
As in the case of disk fractions, these results, when interpreted in the context of the broader trend of declining accretion 
fraction with age \citep{Fedele10},
also suggest that IC 348 is older than Taurus.

{\bf Millimeter fluxes and disk masses.} 
\citet{Rodriguez18} compared the millimeter flux distributions of the young stellar population in IC 348 with those of other young clusters and associations. As they show, disks in IC 348 have fainter millimeter fluxes, on average, than disks in Taurus (1--2 Myr), Lupus (1--3 Myr), and Cha I (2--3 Myr), 
and are much brighter than disks in Upper Sco (5--10 Myr).
Taurus disks are $\sim 4$ times brighter on average. The closest match in millimeter flux distribution is $\sigma$ Ori (3--5 Myr), again suggesting that IC 348 is evolutionarily older than Taurus.

A similar situation is found for the smaller disk samples compared in this paper. As shown in
Figure~\ref{fig:Comparing_IC348_Taurus}a,
the 1.3\,mm fluxes of our selected IC 348 sources are $\sim 5$ times lower on average than those of the comparison Taurus sample.  Figure~\ref{fig:sample_selection_submm}
illustrates how our IC 348 sample includes sources with a broad range of 1.3\,mm fluxes (0.7--10 mJy) which are 
in the upper half
of the IRAC flux distribution (Fig.~\ref{fig:sample_selection_IRAC}),
and primarily in the upper half of the millimeter flux distribution in its spectral type range (Fig.~\ref{fig:sample_selection_submm}). 
Similarly, the Taurus sample is also drawn from sources with higher IRAC fluxes (Fig.~\ref{fig:sample_selection_IRAC}) and the upper half of the millimeter flux distribution of Taurus sources. 

\bigskip

\section{MIRI Spectra of IC 348 Sources in the 24 \micron\ Region}\label{sec:appendix:cold}
\restartappendixnumbering

\begin{figure}[ht]
\centering
\includegraphics[trim = 0.5cm 5.0cm 0.5cm 1.5cm, clip, width=0.65\textwidth]{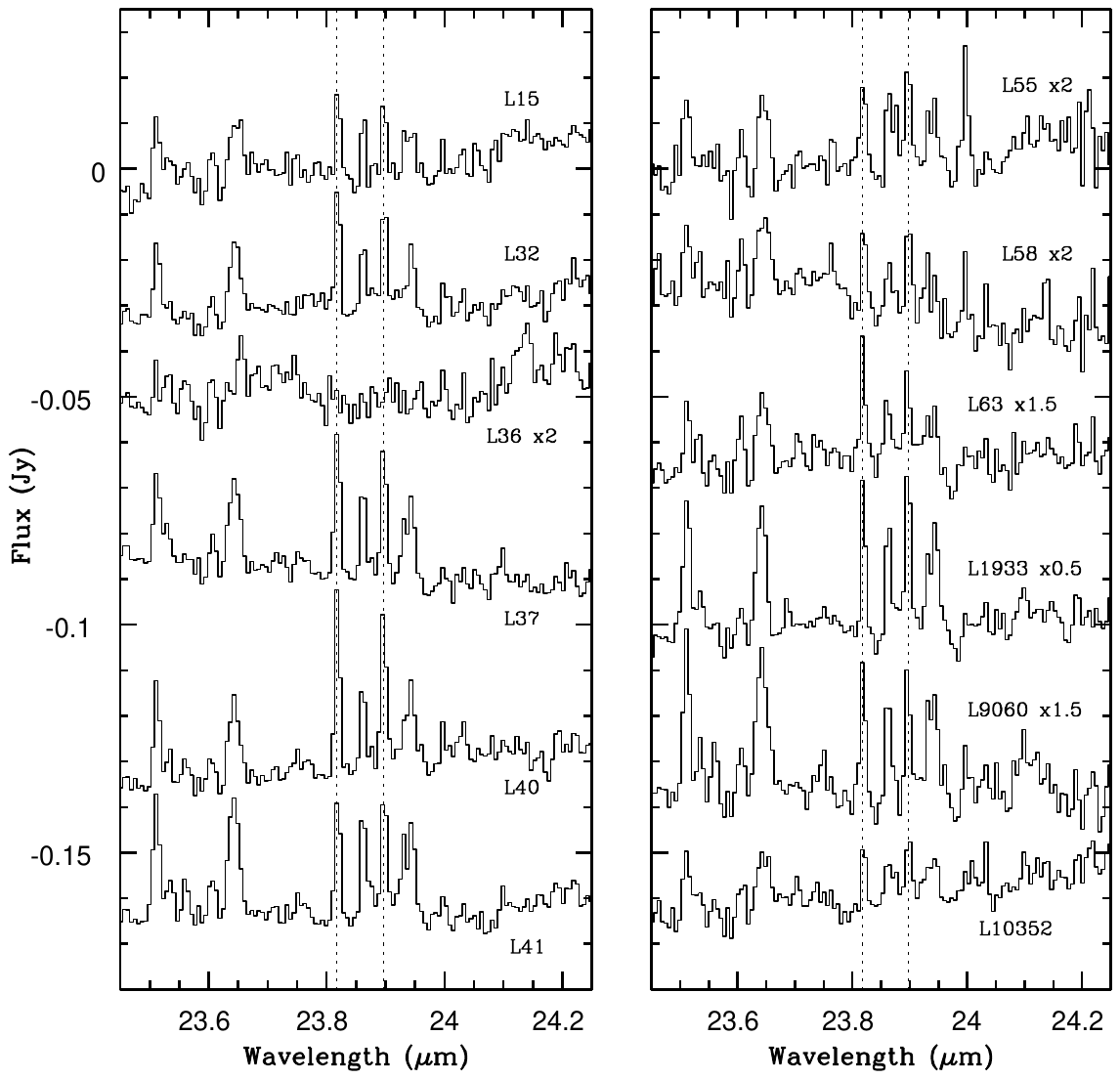}
\caption{Close-up of the MIRI spectra in the 24 \micron\ region. The low energy ($\sim$ 1500 K) \water\ lines are marked by the dotted lines.
The spectra are shifted vertically for plotting, and in some cases scaled in flux by the indicated factor. } 
\label{fig:cold_water}
\end{figure}

Figure~\ref{fig:cold_water} shows the MIRI spectra for the IC 348 targets in the 24 \micron\ region, which includes the low energy ($\sim$1500 K) \water\ lines used in the analysis.

\bigskip

\section{Defining Feature Bands for Flux Measurements}\label{sec:appendix:features}
\restartappendixnumbering

\begin{figure}[t!]
\centering
\includegraphics[trim = 0.5cm 3.5cm 1cm 2.5cm, clip, width=0.75\textwidth]{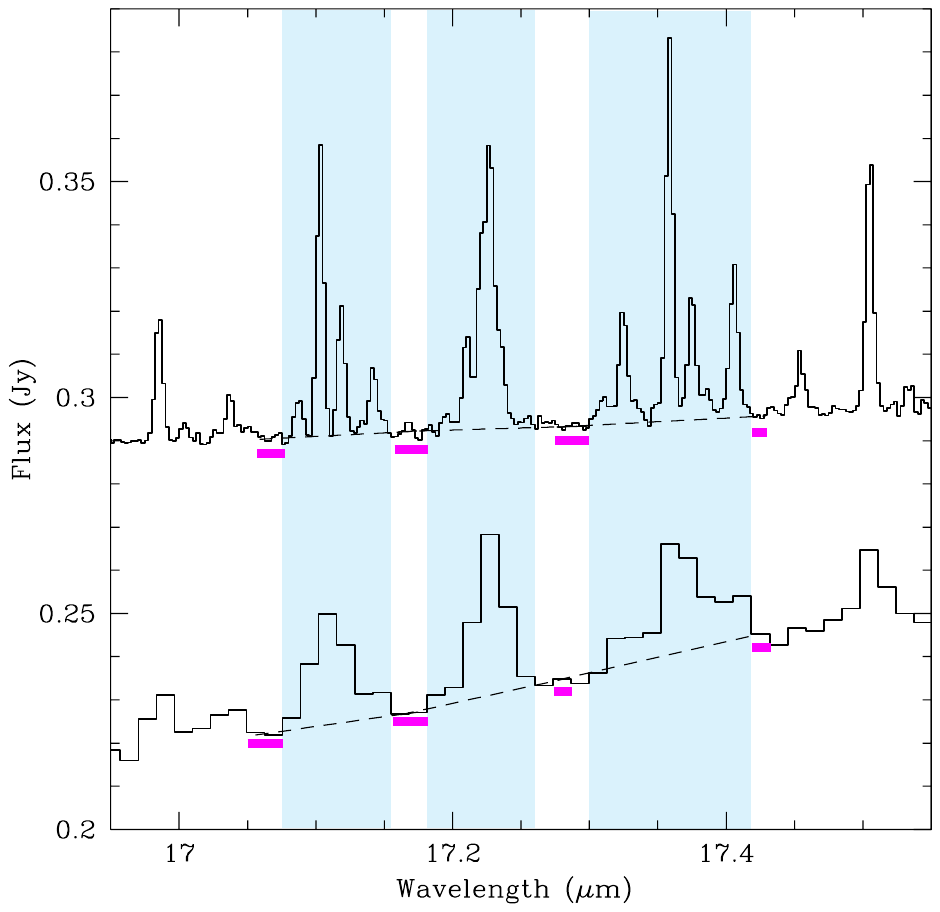}
\caption{Illustration of the measurement of the \water\ flux, showing the JWST-MIRI spectrum of LRL 1933 (top)
and the Spitzer-IRS spectrum of FM Tau (bottom).
The colored bands indicate the wavelength intervals over which the \water\ flux is measured for the MIRI and IRS spectra. 
The magneta bars indicate the intervals that define the linear continuum,
and the black dashed lines show the continuum fit for each spectrum. 
The IRS spectrum has been shifted in flux. }
\label{fig:measuring_water}
\end{figure}

\begin{figure}[ht]
\centering
\includegraphics[trim = 0.5cm 5.0cm 0.5cm 5cm, clip, width=0.65\textwidth]{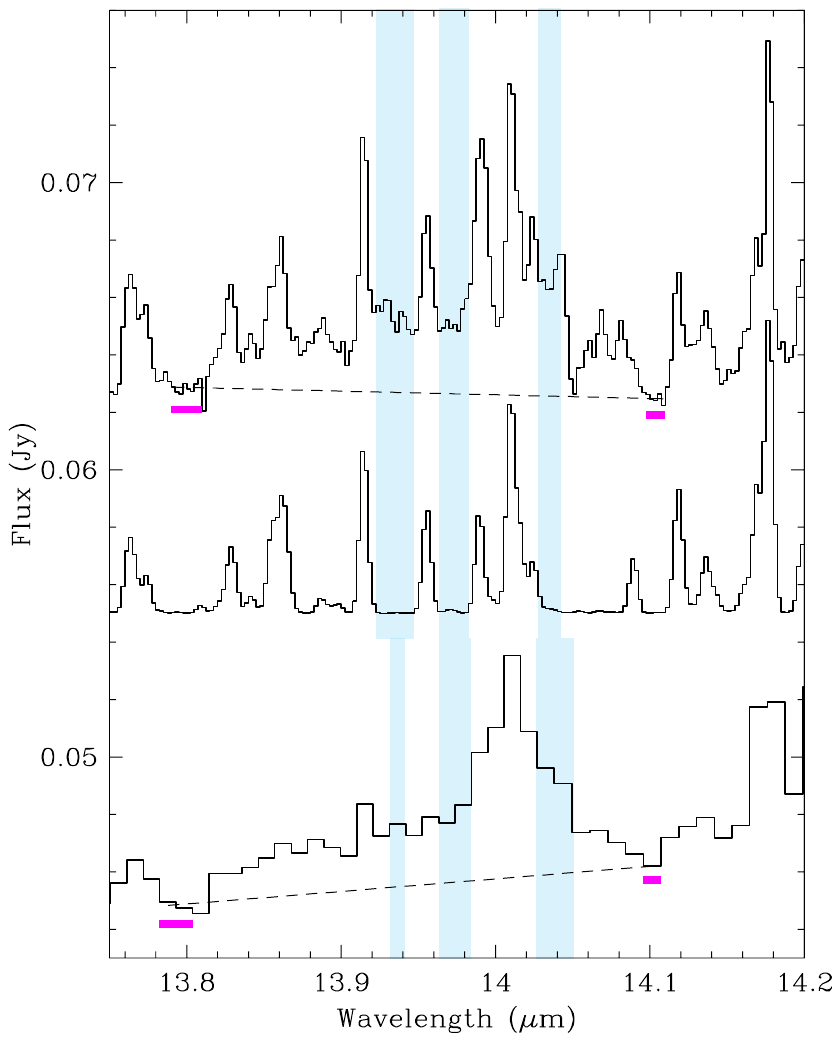}
\caption{
The same as Figure~\ref{fig:measuring_water}, but for the measurement of the HCN flux,
showing the JWST-MIRI spectrum of LRL 9060 (top),
the Spitzer-IRS spectrum of RW Aur (bottom),
and a model \water\ spectrum at MIRI resolution (middle). Colored bands show the regions used to measure the HCN emission while avoiding  nearby \water\ emission lines.
The IRS spectrum has been scaled and shifted in flux.
} 
\label{fig:measuring_HCN}
\end{figure}

\begin{figure}[ht]
\centering
\includegraphics[trim = 0.5cm 5.0cm 0.5cm 5cm, clip, width=0.6\textwidth]{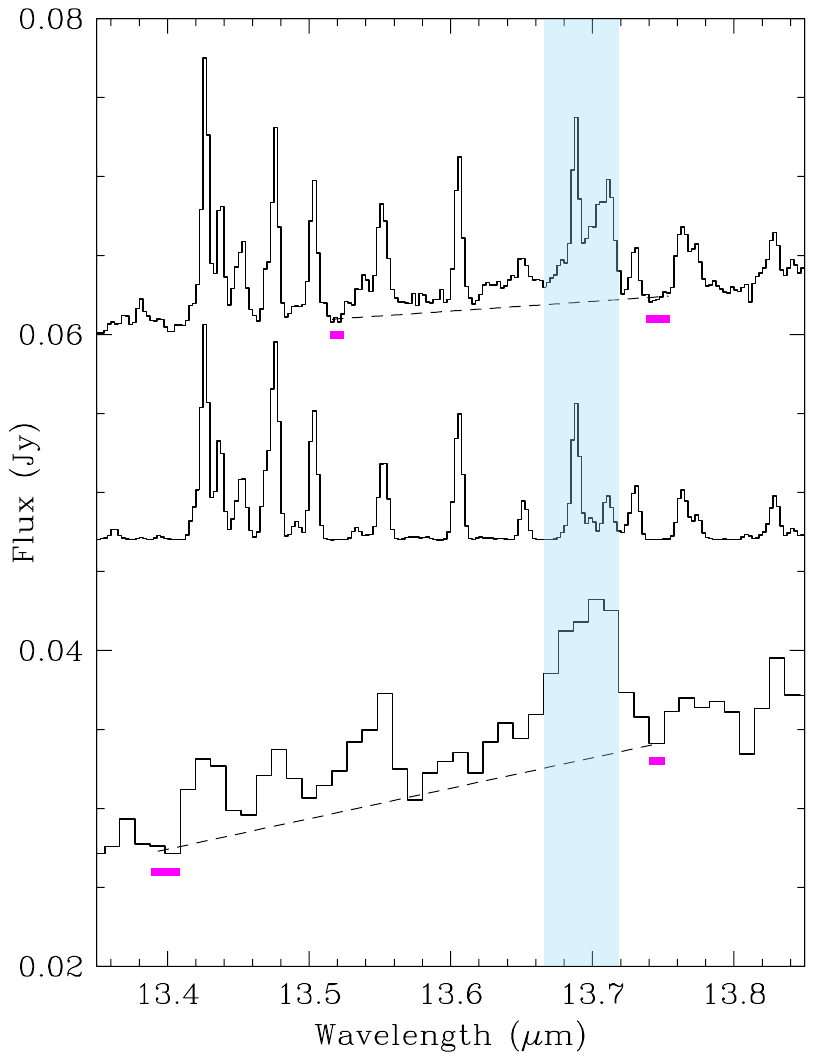}
\caption{ The same as Figure~\ref{fig:measuring_water}, but for the measurement of the \CtwoHtwo\ flux,
showing the JWST-MIRI spectrum of LRL 9060 (top), 
the Spitzer-IRS spectrum of DE Tau (bottom),
and a model \water\ spectrum at MIRI resolution (middle).
The IRS spectrum has been scaled and shifted in flux. } 
\label{fig:measuring_C2H2}
\end{figure}

\begin{figure}[ht]
\centering
\includegraphics[trim = 0.5cm 1cm 1cm 1cm, clip, width=0.65\textwidth]{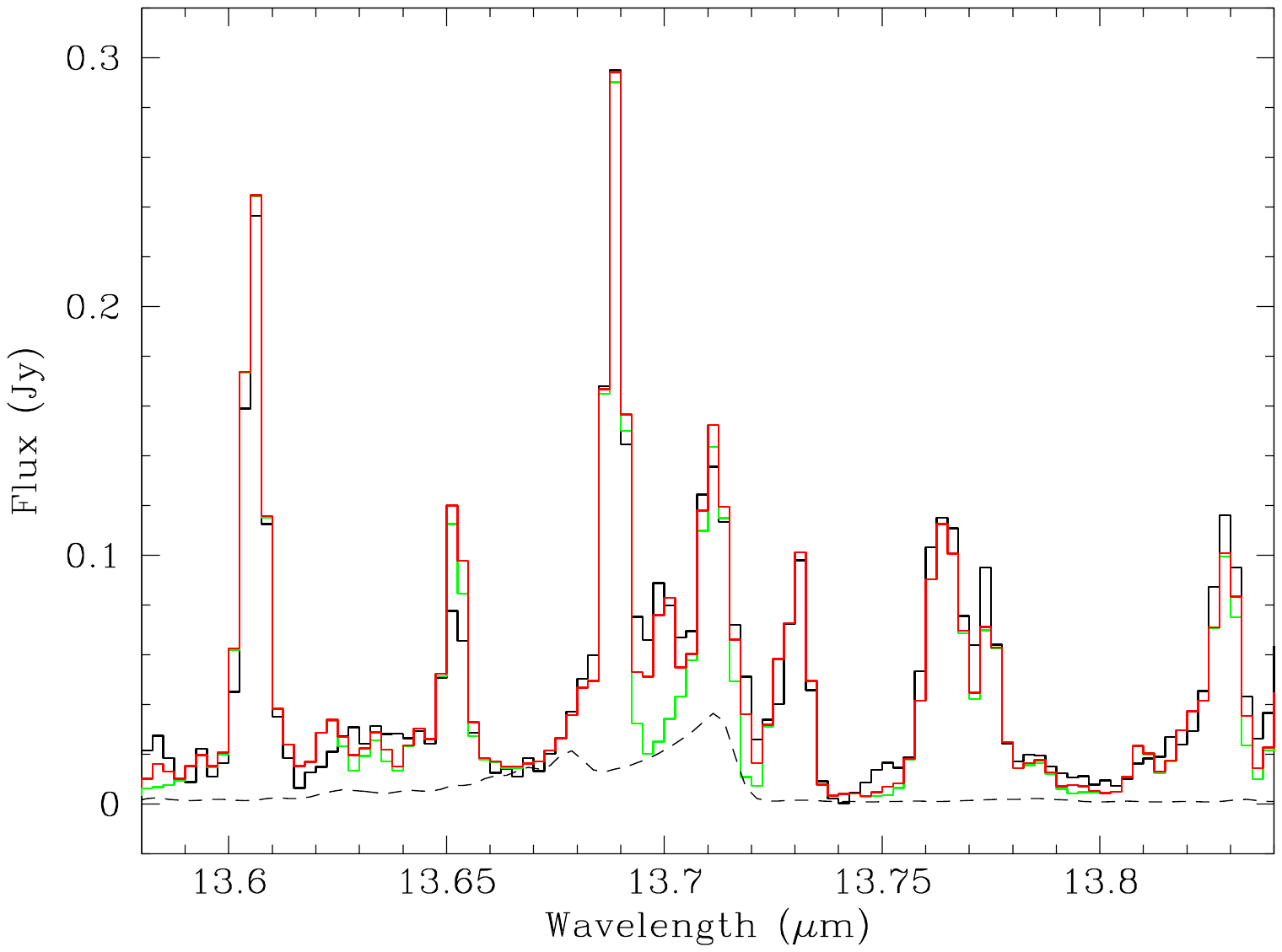}
\caption{Spectrum of FZ Tau (black histogram) compared with a model that includes \water, \CtwoHtwo, and OH using the HITRAN line list (green), and a model that includes additional \water\ lines from the HITEMP line list (red). The inclusion of the additional high energy lines is particularly important when \CtwoHtwo\ is weak. The dashed line is the model contribution from \CtwoHtwo\ emission.} 
\label{fig:FZTau.c2h2}
\end{figure}

Figures \ref{fig:measuring_water}--\ref{fig:measuring_C2H2} show the bands used to integrate the flux for the \water, HCN and \CtwoHtwo\ features studied,
and the wavelength intervals used to fit a linear continuum,
for both MIRI and IRS spectra. 
Each figure shows an example spectrum, which was chosen to be near the
sample median in terms of equivalent width and the ratio of HCN or \CtwoHtwo\
to \water. A model \water\ spectrum is also shown for the HCN and \CtwoHtwo\ regions
(the 17\,\micron\ region in Figure \ref{fig:measuring_water} is nearly pure \water\ lines). 

As in N13, model spectra were calculated  as required to help determine where to set 
the above intervals and for   making some corrections as described below. 
For \water, we employed a line list based on HITRAN2020 \citep{Gordon22}
augmented
with additional lines from HITEMP \citep{Rothman10},
in order to include higher
energy transitions that are missing from HITRAN but observed in MIRI spectra. This line list
was created by calculating an LTE slab model with the HITEMP line list for a
temperature of 1000 K and column density of $10^{19} \,{\rm cm}^{-2}$, and then eliminating lines
with an optical depth less than 0.01. Transitions that were not already in the
HITRAN line list were merged into the HITRAN list to produce the final combined
line list.
Line properties for all other molecules are from HITRAN2020.

In this work, our augmented \water\ line list is most relevant for \CtwoHtwo, where a
number of high energy (10000--12000\,K) \water\ transitions make an observable
contribution to the combined \water\ flux. This is demonstrated in Figure \ref{fig:FZTau.c2h2} for
FZ Tau, a CTTS with weak \CtwoHtwo\ and strong \water\ emission 
\citep{Pontoppidan24}.
The spectrum near 13.7\,\micron\ is compared to synthetic spectra that include two
temperature components for \water\ (following Pontoppidan et al. 2024), plus \CtwoHtwo, HCN and OH.
The model plotted in green used the straight HITRAN water line list, while the red
histogram is a model with our augmented \water\ line list, which fills in missing flux in the blended \CtwoHtwo+\water\ feature. For this particular model, the higher
energy transitions add an additional 26\% to the water flux. These additional
\water\ lines should be included in analyses of the \CtwoHtwo\ Q-branch and are important
for lower \CtwoHtwo\ to \water\ ratios.

The \water\ flux is the sum of the three bands centered at 17.12\,\micron, 17.22\,\micron, and 17.36\,\micron\ (Fig.~\ref{fig:measuring_water}),
similar to N13 and \citet{Banzatti20}.
The transitions within these bands cover a range of upper level energy from 2400\,K to 6300\,K.
For HCN, three sub-bands are defined in the Q-branch that avoid \water\ lines,
OH at 14.05\,\micron, and a combined \COtwo\ plus \CtwoHtwo\ feature at 13.88\,\micron\ (Fig.~\ref{fig:measuring_HCN}).
There still exists a low level of \CtwoHtwo\ within the sub-bands, which is negligible
in most cases; however, if the flux from \CtwoHtwo\ is much greater than HCN, this
simple procedure will overestimate the HCN flux. 

Due to the limited contamination-free region within the \CtwoHtwo\ Q-branch,
we corrected for the \water\ emission that falls within the defined \CtwoHtwo\ band by
searching for measurable \water\ emission features that could serve as a predictor
of the \water\ emission with the flux bandpass (Fig.~\ref{fig:measuring_C2H2}).
Building upon \citet{Banzatti25}, we calculated a series of two-component \water\ model spectra, consisting of a base model with a temperature of 850\,K and a column density of $10^{18}\,{\rm cm}^{-2}$, plus
a warm 400\,K component of progressively larger area relative to the hot component.
These model series were repeated with a hot component of 750\,K or 950\,K, or with a
column density twice or half that of the base model.
We found that the \water\ feature at 12.45\,\micron\ (a blend dominated by transitions with $E_{\rm up}$ of 4212\,K, 3629\,K and 7640\,K, with smaller contributions from 9025\,K to 14,000\,K lines) was a good predictor of the \water\ flux within the \CtwoHtwo\ bandpass for this
series of models, with flux of the 12.45\,\micron\ feature equal to the contaminating water flux to within 12\%.
In addition, a correction for HCN was applied, although in all cases this was less
than that of \water.  Model HCN spectra were calculated for column densities of
$10^{15}\,{\rm cm}^{-2}$ and $10^{17}\,{\rm cm}^{-2}$ at temperatures of 700\,K to 900\,K. The HCN flux in the \CtwoHtwo\
bandpass was found to be 0.10 times the sum of the HCN sub-bands, to within 13\%.

For the measurements made from the IRS spectra, some adjustments and corrections
were made to account for the lower resolution and spectral sampling.
For the 17\,\micron\ region, bleeding of \water\ emission into continuum regions at IRS resolution lowers the measured flux slightly.  While this effect is small,
model spectra were used to determine a correction factor of 1.04, which was applied
to the measured IRS water fluxes.
For HCN, the sub-bands were adjusted to better match the water spectrum at IRS
resolution. To account for the different interval widths, we used model HCN spectra 
 to determine correction factors to return the same flux at MIRI resolution.
For \CtwoHtwo, a different interval was used for the short wavelength end of the
continuum, because the 13.52\,\micron\ interval in the MIRI spectra is no longer a good continuum point at IRS resolution (Fig.~\ref{fig:measuring_C2H2}). The 12.45\,\micron\ water feature used to correct for \water\ contamination required a correction factor of 1.29 to
account for the continuum dilution by neighboring water lines; a correction
was also applied to the \CtwoHtwo\ flux to account for emission that falls outside
the \CtwoHtwo\ bandpass at IRS resolution.

As a test of the IRS measurements, the MIRI spectra for the IC 348 sample were
smoothed and rebinned to match that of the IRS spectra, and the IRS measurement
procedures were applied and compared with the fluxes measured directly from the
MIRI data. For \water, the MIRI flux was recovered to within 4\% (10/12 stars), while for
HCN the flux was recovered to within 5\%. The flux of the 12.45\,\micron\ water feature
(used to correct \CtwoHtwo) had a scatter of 8\%.
For \CtwoHtwo, only five disks are marked as detections (Table \ref{Table.IC 348.Results}). 
For four of the detections, the scatter in the corrected \CtwoHtwo\ flux is 2.5\% but with an offset in the mean of 4\%, 
while a fifth (LRL 58) disagrees with the MIRI measurement by 30\%.
In general, our measurement of the molecular fluxes from the IRS data does not appear to produce a significant systematic difference compared to  measurements from MIRI data.

\begin{figure}[ht]
\centering
\includegraphics[trim = 2cm 4.5cm 1cm 4cm, clip, width=0.55\textwidth]{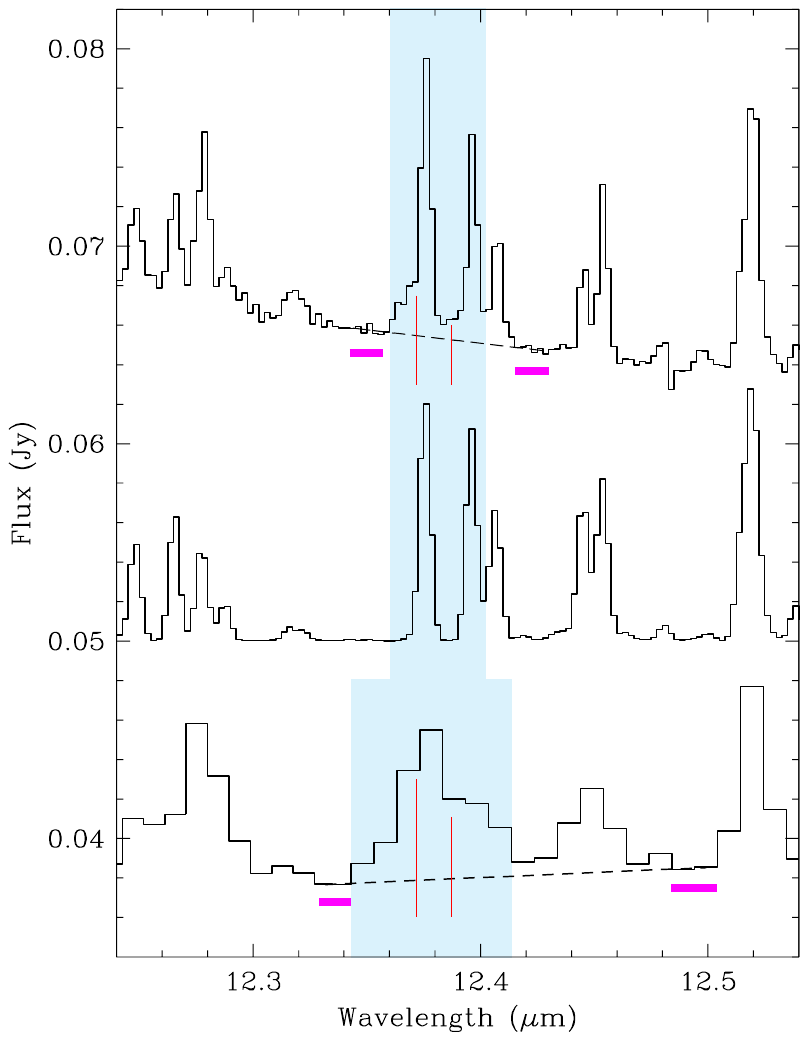}
\caption{The same as Figure~\ref{fig:measuring_water}, 
but illustrating measurement of the \ion{H}{1} flux, showing the JWST/MIRI spectrum of LRL 9060 (top),
the Spitzer/IRS spectrum of DP Tau (bottom), and a model \water\ spectrum at MIRI resolution (middle).
The wavelengths of the
\ion{H}{1} (7--6) and (11--8) transitions are marked by the red vertical lines.
In the case of LRL 9060, the \water-corrected \ion{H}{1} flux is 29\% of the total measured flux in the band. }
\label{fig:measuring_HI}
\end{figure}

Figure \ref{fig:measuring_HI} shows the wavelength intervals used for the continuum and flux measurements for \ion{H}{1} at 12.37\ \micron. Our measurements include both the
\ion{H}{1}(7--6) and \ion{H}{1}(11--8) transitions. The flux measurements for MIRI spectra
must be corrected for the two contaminating \water\ transitions at 12.3756 and 12.3962 \micron.
We used the same series of \water\ emission models to search for a good predictor of the water emission within the flux bandpass and adopted a linear combination of the \water\ transitions at 13.13\,\micron\ and 14.43\,\micron. 
Specifically, the correction subtracted from the measured blended flux is
$({\rm H_2O\ Correction}) = 2.09\,F_{13.13} + 0.43\,F_{14.43}$. 

Over the series of \water\ models, this estimate predicted the \water\ flux in the bandpass to within 2\%.
The corrected \ion{H}{1} fluxes were converted to line luminosity and then to accretion
luminosity using Equation (1) in \citet{Rigliaco15}. As previously noted, this equation
applies to the combined (7--6)+(11--8) luminosity.
As a comparison, we also calculated the accretion luminosities using only the
\ion{H}{1}(7--6) line and the relation to accretion luminosity in \citet{Tofflemire25},
where our correction for the single contaminating \water\ line is
$ 0.85\,F_{13.13} + 0.43\,F_{14.43}$.
The two sets of accretion luminosities were in excellent agreement with no systematic offset.

For the IRS data, the \water\ correction of the flux in the \ion{H}{1} band must also include a
third contaminating \water\ transition at 12.407 \micron\ (Fig.~\ref{fig:measuring_HI}).
\citet{Rigliaco15} used the 13.13\,\micron\ \water\ line to correct for contaminating water
emission in the blended feature. Our \water\ models give an average of 3.58 for the ratio between
the contaminating flux and the 13.13\,\micron\ transition, in agreement with Rigliaco et al..
This shows that, regardless of the wording of the text in Rigliaco et al., their correction
actually accounts for all of the contaminating \water\ lines and not only the 12.39 \micron\ line.
We did not use the 13.13\,\micron\ line, because we found that confusion from HCN and \CtwoHtwo\ 
lines, and blended \water\ lines, makes it difficult to measure at IRS resolution in many sources. 
Instead, we used a combination of water features at 11.64 and 11.72  \micron, each of which is a blend of water transitions.
The spectra and models were integrated over wavelength 
intervals of 11.628--11.656 \micron\ and 11.713--11.742 \micron, for the two
features, respectively.
The water correction used was $({\rm H_2O\ Correction}) = 1.15\,F_{11.64} + 0.58\,F_{11.72}$,
which predicted the model \water\ emission in the bandpass to within 10\%.

\bigskip



\section{Molecular Flux Ratios vs.\ Disk mass and Accretion rate}\label{sec:appendix:ratios}
\restartappendixnumbering

As described in Section 4.3.4,
IC 348 and Taurus have similar molecular flux ratios, despite IC 348 having lower stellar accretion rates and disk masses. 
Here, we plot the molecular ratios against these two quantities. 

\begin{figure}[t]
\centering
\includegraphics[trim = 0.5cm 5.0cm 1cm 3cm, clip, width=0.45\textwidth]{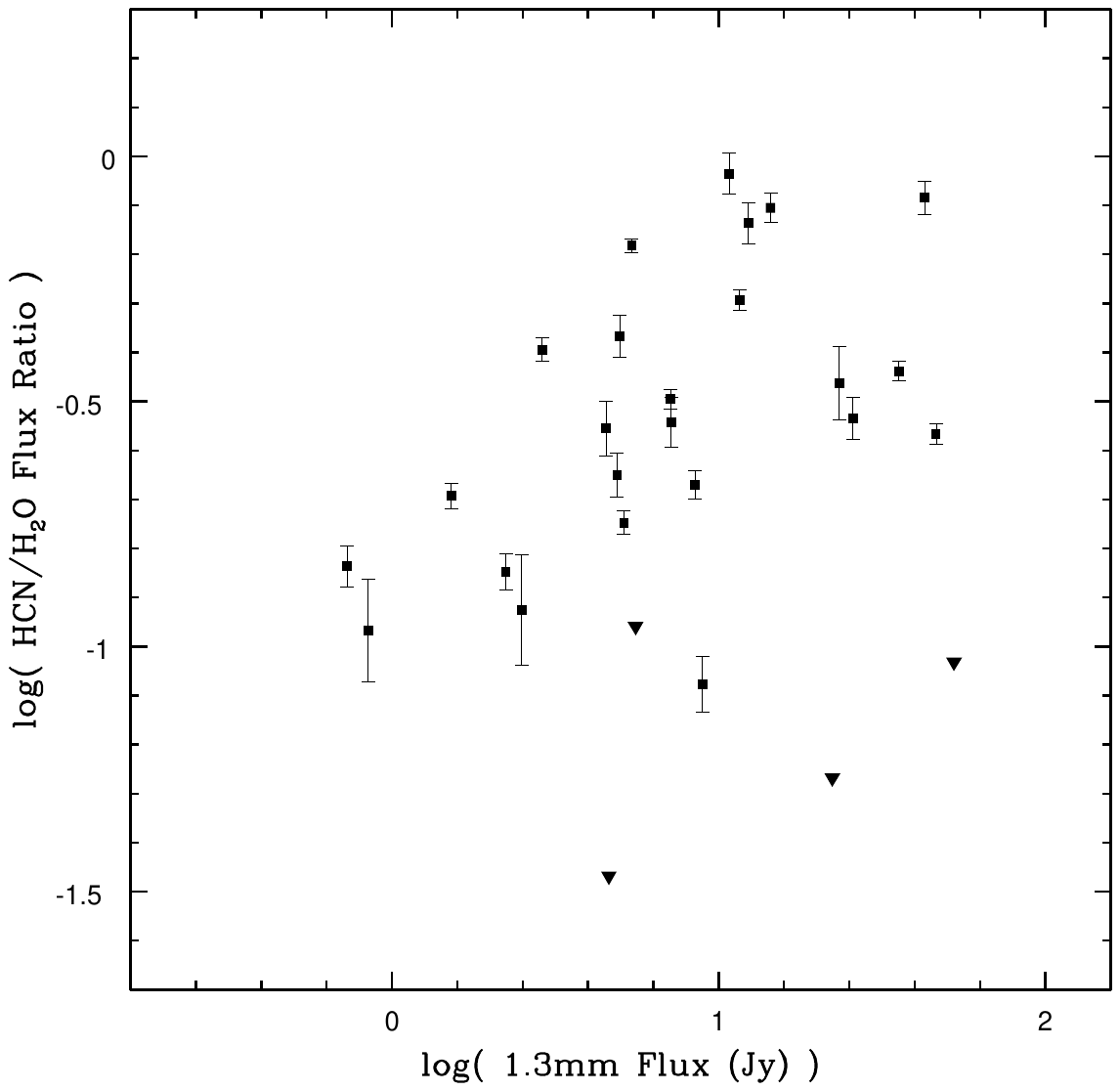}
\includegraphics[trim = 0.5cm 5.0cm 1cm 3cm, clip, width=0.45\textwidth]{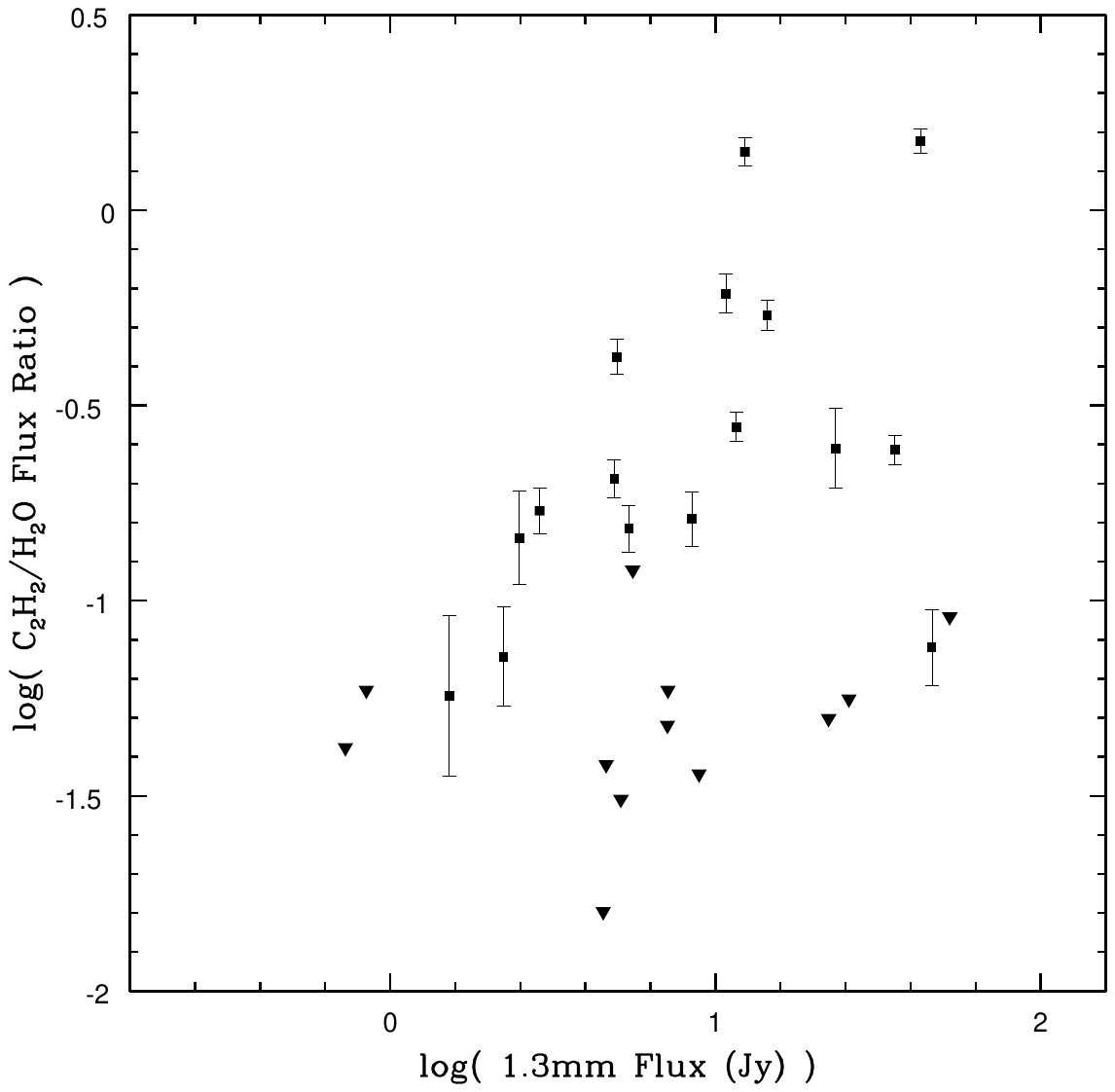}
\caption{Flux ratios of the HCN (left) and \CtwoHtwo\ (right) relative to \water\ features in the spectra of Taurus disks as a function of 1.3~mm flux.
In both panels, the millimeter fluxes are scaled to the IC 348 distance of 320 pc to facilitate comparison with IC 348.
Inverted triangles indicate upper limits.
} 
\label{fig:FluxRatios_Fmm_Tau}
\end{figure}

Figure~\ref{fig:FluxRatios_Fmm_Tau}, which plots only the Taurus sample, illustrates possible trends in  molecular flux ratios with 1.3\,mm flux. 
Both HCN/\water\ (left) and \CtwoHtwo/\water\ (right) show an upper envelope in flux ratio that increases with millimeter flux. High molecular flux ratios are not found for lower disk masses, while the highest flux ratios are seen in the higher mass disks. At the same time, high mass disks show a range of flux ratios down to upper limits. The HCN/\water\ trend is similar to results reported in the literature. N13 reported a trend between HCN/\water\ and submillimeter disk mass, and \citet{Banzatti20} found a trend between the inverse of these ratios (e.g., \water/HCN) and disk dust radius (which is correlated with mass; e.g., \citealt{Tripathi17}).

\begin{figure}[ht]
\centering
\includegraphics[trim = 0.5cm 4.5cm 1cm 3cm, clip, width=0.45\textwidth]{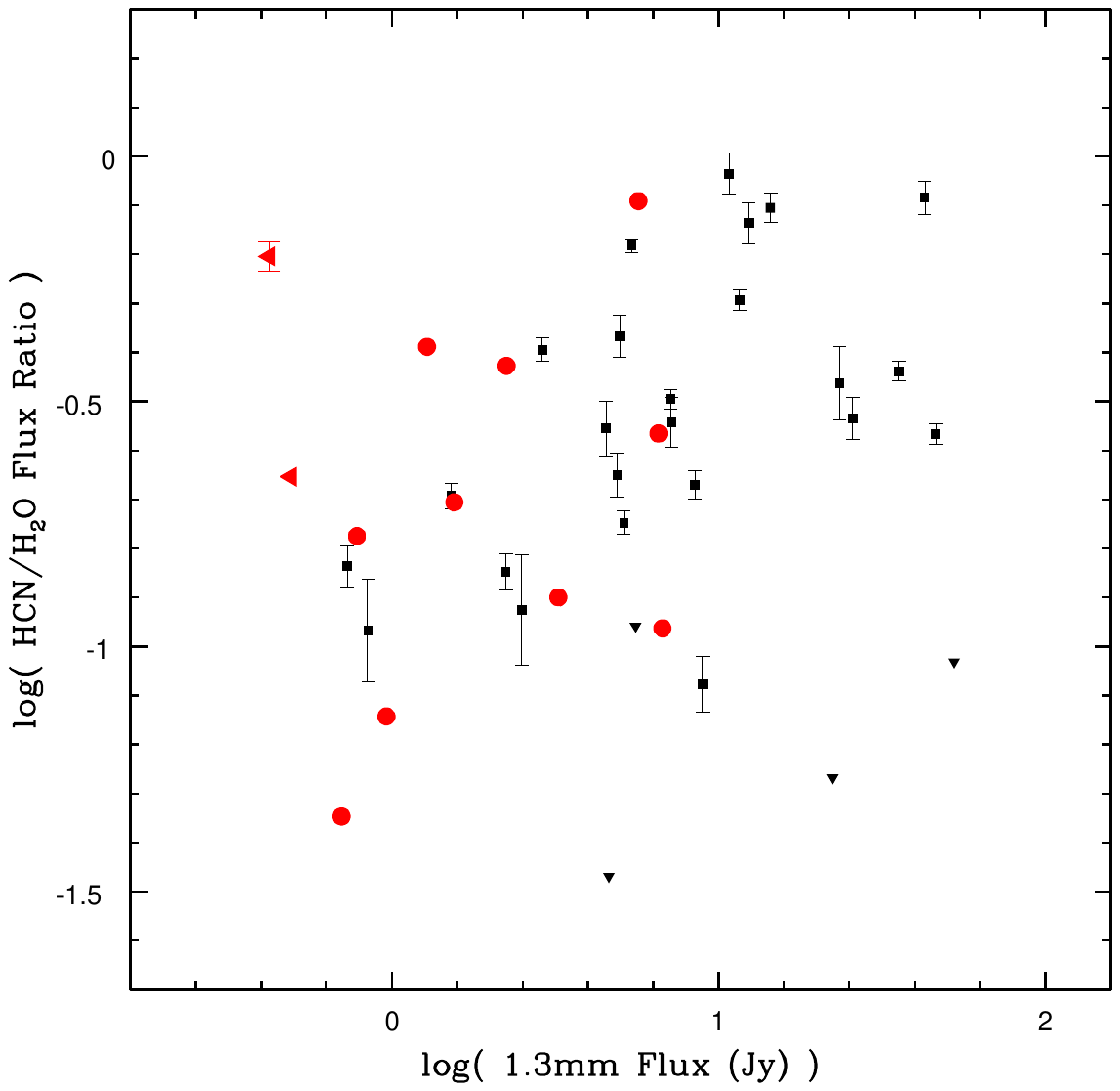}
\includegraphics[trim = 0.5cm 4.5cm 1cm 3cm, clip, width=0.45\textwidth]{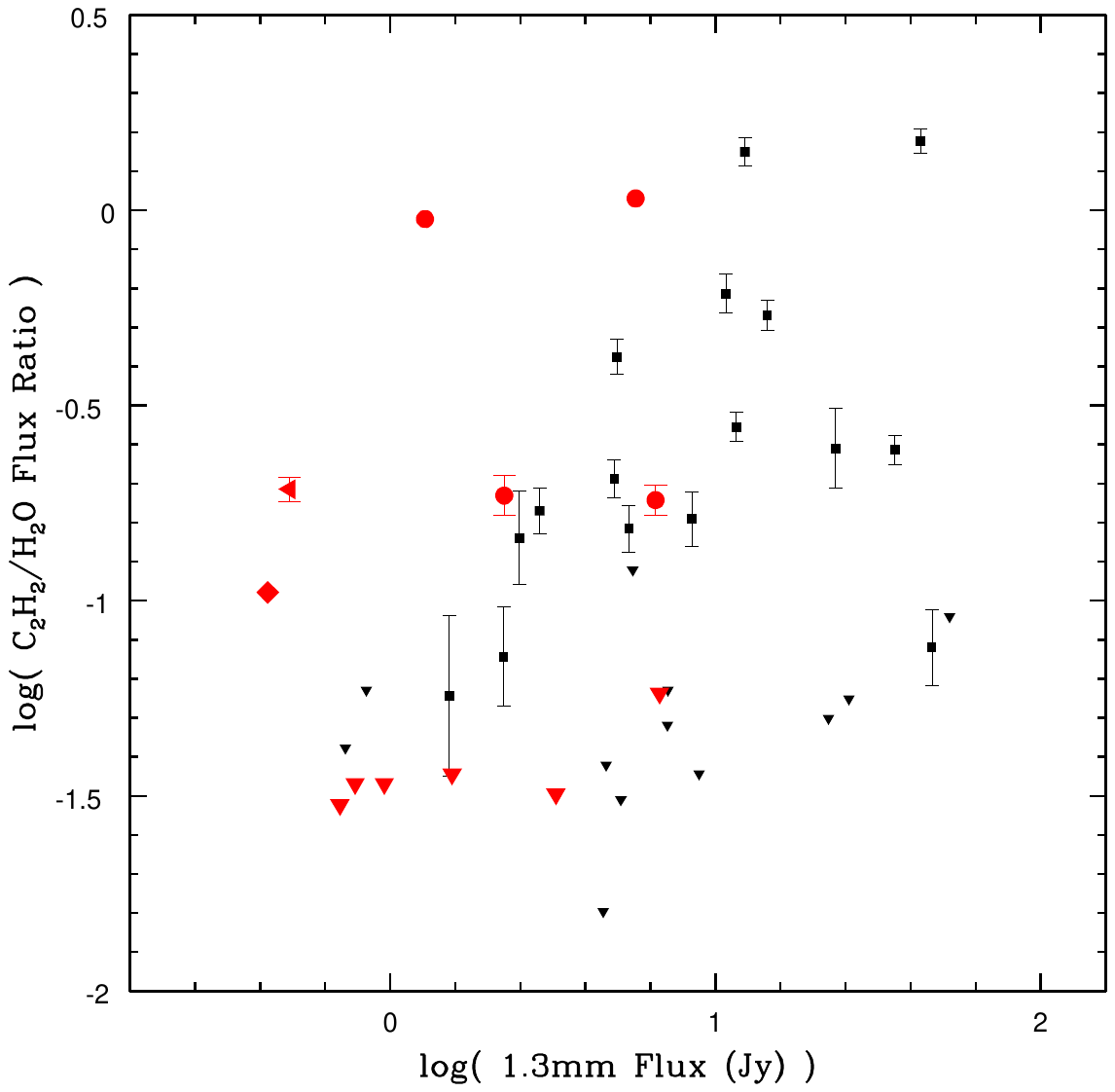}
\caption{Flux ratios of HCN (left) and \CtwoHtwo\  (right) relative to \water\ features in the spectra of IC 348 (red points) and Taurus (black points) sources as a function of 1.3~mm flux, scaled to a common distance of 320~pc. Although the IC 348 disks are 5 times weaker in 1.3~mm flux, their molecular flux ratios have the same distribution as the Taurus disks.
Triangles indicate upper limits on the molecular ratio or 1.3~mm flux, and the diamond indicates an upper limit on both.
} 
\label{fig:HCN.to.H2O.v.Fmm}
\end{figure}

The trends for the Taurus sample, which covers a restricted range in stellar masses $0.4–1\, \Msun$,
contrasts with results for lower mass ($< 0.2\, \Msun$) stars and their correspondingly lower mass disks.  
As first shown by 
\citet{Pascucci09, Pascucci13}, 
the mid-infrared spectra of disks around low-mass young stars show higher ratios of \CtwoHtwo/HCN and hydrocarbons to water, indicative of higher C/O ratios. 
The results reported by \citet{Grant25} show a dichotomy in the \CtwoHtwo/\water\ ratios between their very low mass star (VLMS) and T Tauri star (TTS) samples,
with systematically higher ratios in very low-mass stars. 
When plotted against disk mass over this very large range in stellar mass (0.02--1.5 $M_\odot$), the combined sample indicates a declining trend in \CtwoHtwo/\water\ with increasing disk mass, in the opposite sense of that in Figure~\ref{fig:FluxRatios_Fmm_Tau}.
The relation among the hydrocarbon-to-water ratios, disk mass, and stellar mass (or luminosity) is an area for further research.

As shown in Figure~\ref{fig:HCN.to.H2O.v.Fmm},  the IC 348 molecular flux ratios have a similar distribution to those in Taurus, but their disk masses are lower (Ruiz-Rodriguez et al.\ 2018). The HCN/\water\ distribution is similar to that of Taurus, i.e., the red points follow the black points, with the exception of one outlier at low submillimeter flux. The situation for \CtwoHtwo/\water\ is similar in that the sources with the highest \CtwoHtwo/\water\ ratios have higher submillimeter fluxes. The small sample size of IC 348, makes it difficult to see trends in that sample alone. 

\begin{figure}[ht!]
\centering
\includegraphics[trim = 0.5cm 5.0cm 1cm 3cm, clip, width=0.45\textwidth]{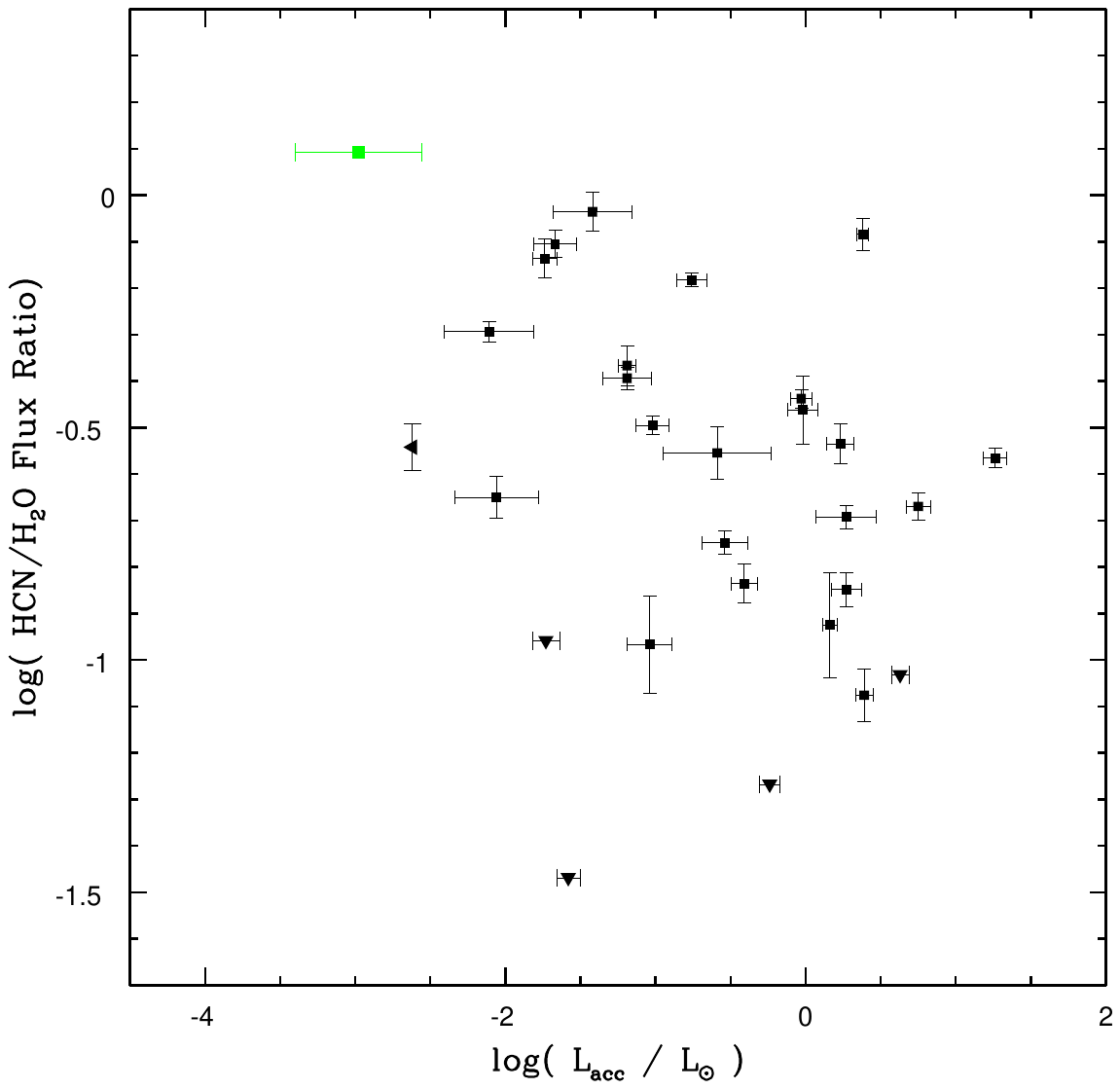}
\includegraphics[trim = 0.5cm 5.0cm 1cm 3cm, clip, width=0.45\textwidth]{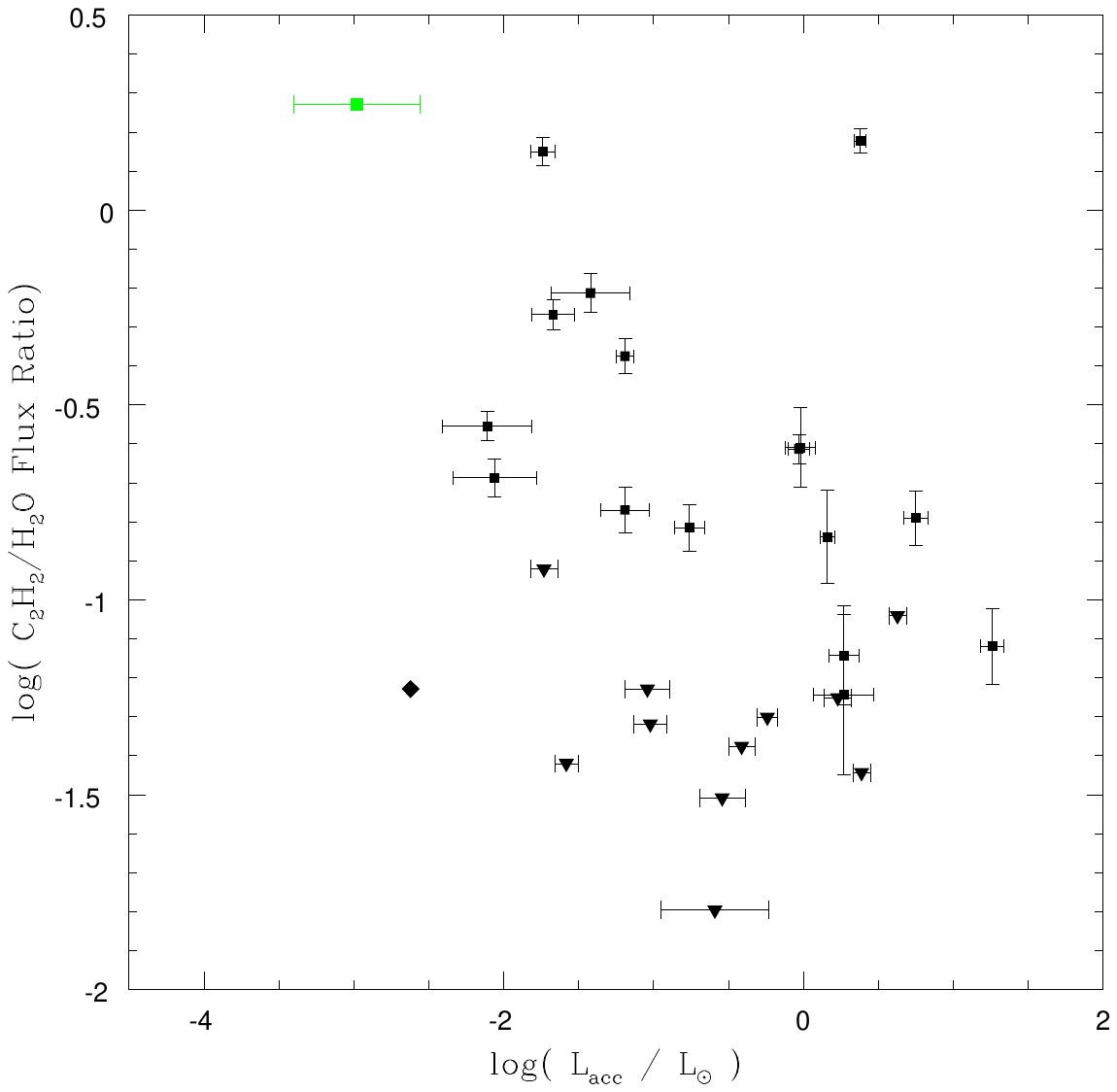}
\caption{Flux ratios of HCN (left) and \CtwoHtwo\ (right) relative to \water\ features in the spectra of Taurus disks as a function of accretion luminosity. 
Values of $L_{\rm acc}$ are from simultaneous \ion{H}{1} measurements. 
The green square indicates DoAr 33 ($L_{\rm acc}$ from B. M. Tofflemire et al. 2025). 
Triangles indicate upper limits on either the flux ratio or $L_{\rm acc}$, and diamonds indicate upper limits on both.
} 
\label{fig:FluxRatios_Lacc_Tau}
\end{figure}

\begin{figure}[ht]
\centering
\includegraphics[trim = 0.5cm 5.0cm 1cm 3cm, clip, width=0.45\textwidth]{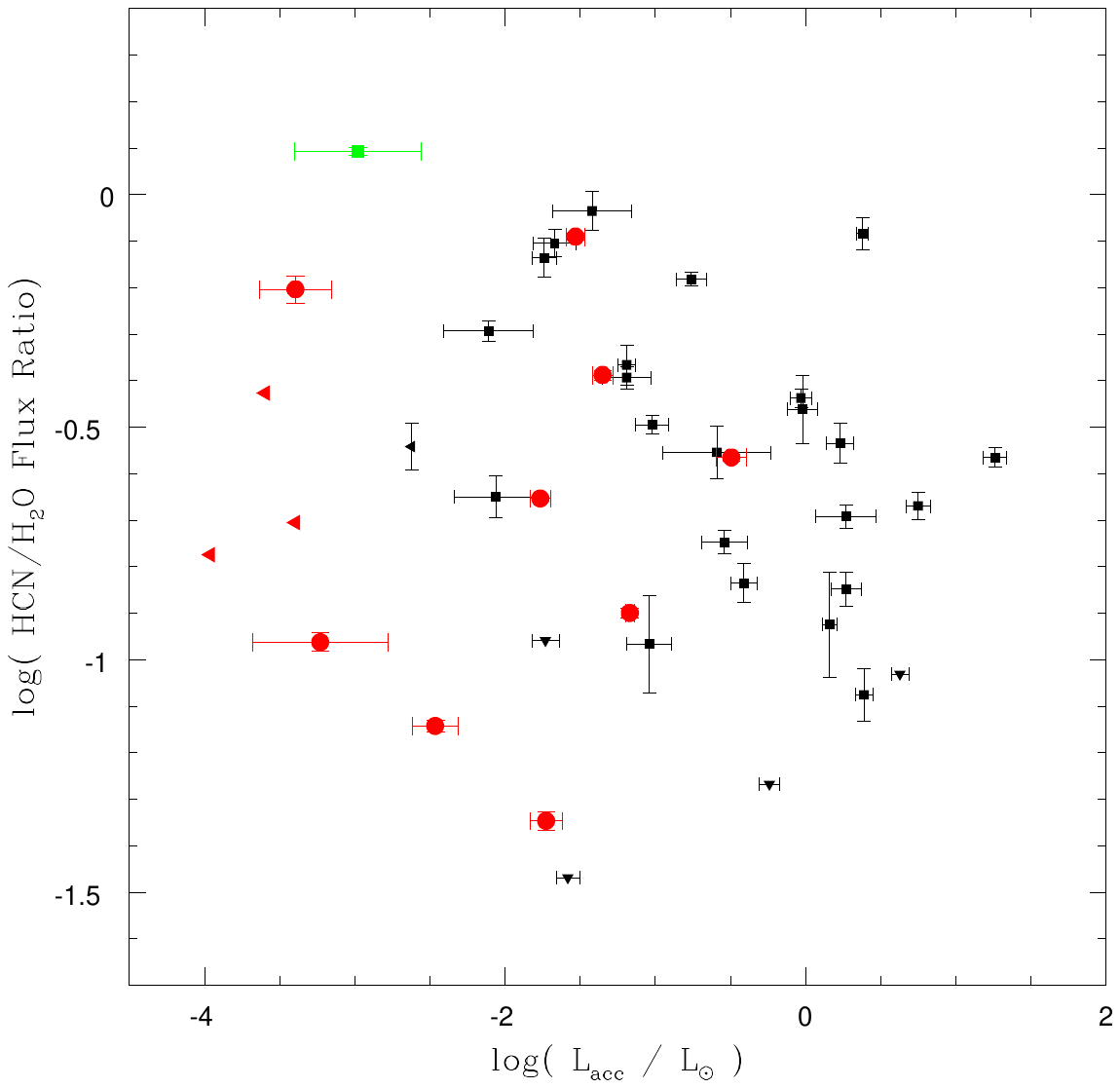}
\includegraphics[trim = 0.5cm 5.0cm 1cm 3cm, clip, width=0.45\textwidth]{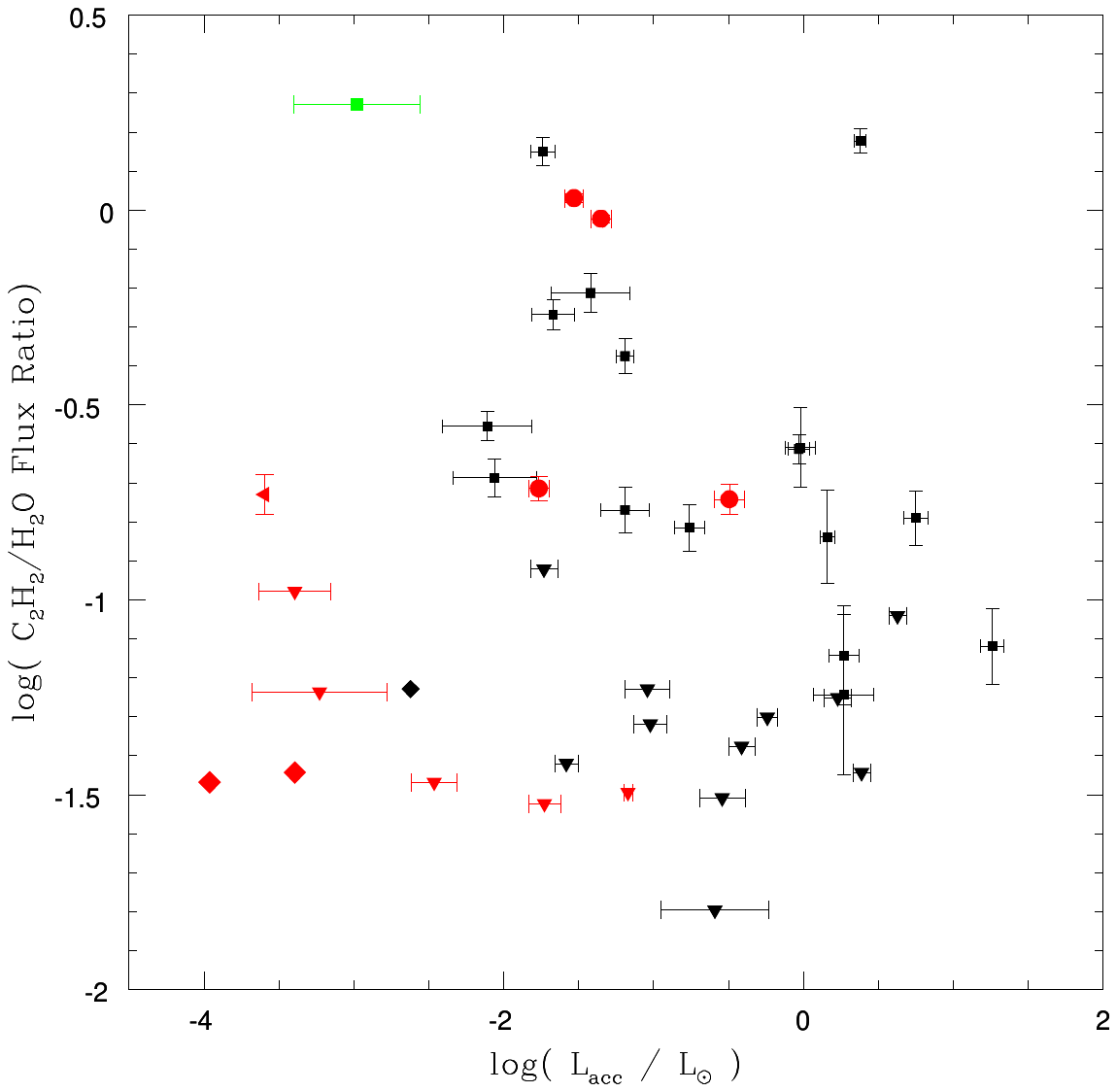}
\caption{Flux ratios of HCN (left) and \CtwoHtwo\ (right) relative to \water\ as a function of accretion luminosity $L_{\rm acc}$ for IC 348 (red points) and Taurus (black points). 
Values of $L_{\rm acc}$ are from simultaneous \ion{H}{1} measurements for both samples.
Although the values of $L_{\rm acc}$ are more than a order of magnitude lower in IC 348 than in Taurus, their molecular flux ratios have the same distribution.
The green square indicates DoAr 33 ($L_{\rm acc}$ from \citealt{Tofflemire25}).  Triangles indicate upper limits on either the flux ratio or $L_{\rm acc}$, and diamonds indicate upper limits on both.
}
\label{fig:FluxRatios_Lacc}
\end{figure}

The Taurus sample also suggests a decreasing upper envelope of molecular flux ratios with increasing accretion luminosity. As shown in Figure~\ref{fig:FluxRatios_Lacc_Tau}, the upper envelope of both the HCN/\water\ and \CtwoHtwo/\water\ distributions appears to rise toward lower accretion rate. With the exception of one point in the upper right of both plots (DL Tau), high flux ratios are not seen at high $L_{\rm acc}$ ($> 0.3 L_\odot$). 
High flux ratios are found at lower $L_{\rm acc}$ ($< 0.1 L_\odot$), and lower flux ratios are seen at all $L_{\rm acc}$.

The location of DoAr~33 in
Figure~\ref{fig:FluxRatios_Lacc_Tau}
(green point) is consistent with the upper envelope trend of increasing molecular flux ratios at lower accretion rate.
\citet{Colmenares24} have suggested that the high column densities of 
\CtwoHtwo\ and ${\rm C_4H_2}$
in DoAr 33 is a consequence of its very low accretion rate. 
That is, molecules created near the inner disk edge, in a carbon-rich chemistry driven by sublimation of carbon grains at high temperature, are less readily accreted onto the star at low accretion rate. 
While the position of DoAr~33 in Figure~\ref{fig:FluxRatios_Lacc_Tau} is consistent with this idea, we also see that many IC~348 disks have similarly low accretion rates but lower molecular flux 
(Figure~\ref{fig:FluxRatios_Lacc}). 
Hence, while low accretion rates could contribute towards low C/O in inner disks, it cannot be the only factor.

The results in \citet{Grant25} also show a decline in the \CtwoHtwo/\water\ ratio with increasing stellar accretion rate between the VLMS to TTS regimes. However, values for several of the IC 348 disks will fill in the lower left region of their Figure 3b, and the result will be less of an overall trend but rather a dichotomy between the \CtwoHtwo/\water\ ratios of the lower and higher mass TTS disks.

As shown in Figure~\ref{fig:FluxRatios_Lacc}, the IC 348 molecular flux ratios have  distributions similar to that of the Taurus sample and may also show a decreasing upper envelope of molecular ratios with accretion luminosity. However, the size of the IC 348 sample is too small to distinguish a trend in that sample alone.


\bibliography{main}{}
\bibliographystyle{aasjournalv7}



\end{document}